\documentclass[onecolumn]{aastex701}
\usepackage{subcaption}
\usepackage{array}
\usepackage{amsmath,bm}
\usepackage{natbib}
\usepackage{xcolor}
\usepackage{booktabs}
\usepackage{gensymb}
\usepackage[T1]{fontenc}
\usepackage{newtxtext,newtxmath}

\usepackage{tabularx}
\usepackage{booktabs}
\usepackage{makecell}
\usepackage{longtable}
\usepackage{pifont}
\newcommand{\cmark}{\ding{51}}
\newcommand{\xmark}{\ding{55}}

\newcounter{subfigglobal}

\DeclareRobustCommand{\VAN}[3]{#2}
\let\VANthebibliography\thebibliography
\def\thebibliography{\DeclareRobustCommand{\VAN}[3]{##3}\VANthebibliography}

\begin{document}
	
	
	\title[]{Local Heavily Obscured Active Galaxies Observed With \textit{Chandra} (I):\\ Spectral Properties of the Nuclear and Extended X-ray Emission}
	
	\correspondingauthor{Riccardo Middei}
	\email{riccardo.middei@inaf.it}
	
	\author{R. Middei}
	\affiliation{INAF Osservatorio Astronomico di Roma, Via Frascati 33, 00078 Monte Porzio Catone, RM, Italy}
	\affiliation{Center for Astrophysics | Harvard \& Smithsonian, 60 Garden Street, Cambridge MA 02138, USA}
	\email{riccardo.middei@inaf.it}

	\author{G. Fabbiano}
	\affiliation{Center for Astrophysics | Harvard \& Smithsonian, 60 Garden Street, Cambridge MA 02138, USA}
	\email{gfabbiano@cfa.harvard.edu}
	
	\author{A. Trindade Falc\~ao}
	\affiliation{NASA Goddard Space Flight Center, Code 662, Greenbelt, MD 20771, USA}
	\affiliation{Center for Astrophysics | Harvard \& Smithsonian, 60 Garden Street, Cambridge MA 02138, USA}
	\email{annatrindadefalcao@gmail.com}

	\author{D.~{\L}.~Kr\'{o}l}
	\affiliation{Center for Astrophysics | Harvard \& Smithsonian, 60 Garden Street, Cambridge MA 02138, USA}
	\affiliation{Astronomical Observatory of the Jagiellonian University, Orla 171, 30-244 Krak\'ow, Poland}
	\email{dominika.krol@cfa.harvard.ed}
	
	\author{W. P. Maksym}
	\affiliation{NASA Marshall Space Flight Center, Huntsville, AL 35812, USA}
	\email{walter.p.maksym@nasa.gov}

	\author{M. Elvis}
	\affiliation{Center for Astrophysics | Harvard \& Smithsonian, 60 Garden Street, Cambridge MA 02138, USA}
	\email{melvis@cfa.harvard.edu}

	\begin{abstract}
		The \textit{Chandra} X-ray Observatory has revealed that kiloparsec-scale extended X-ray emission is commonly associated with heavily absorbed ($\log (N_{\rm H}$/cm$^{-2})\,>\, 23$) Active Galactic Nuclei (AGN) in the local Universe. We present a homogeneous spatially resolved spectral analysis of  a sample of nearby obscured AGN observed with deep ($>100$~ks) {\it Chandra}/ACIS exposures and all associated with extra-nuclear extended emission. Exploiting the sub-arcsecond resolution of {\it Chandra}, we separate unresolved nuclear emission from physically distinct extended regions and model their spectra using photoionized and thermal plasma components. The extended emission is spectrally complex, requiring a combination of photoionized and collisionally ionized components in most regions. The spectral parameters exhibit systematic radial variations, with nuclear regions typically hosting higher-column and higher-ionization components, while extended regions span similarly broad but overlapping ranges without a simple monotonic radial progression. Hard ($>3$ keV) X-ray emission in excess of nuclear PSF spillover is detected in 19/20 sources, while extended Fe~K$\alpha$ emission is observed in 12/20 systems. These results establish a uniform baseline for the X-ray properties of heavily obscured AGN and provide constraints on the geometry and physical conditions of AGN-driven feedback in the local Universe.
	\end{abstract}
	
	\keywords{}

	
	\section{Introduction}
	\label{sec:intro}
	
	Simulations of galaxy formation and evolution within the $\Lambda$CDM framework \citep[e.g.,][]{benson_what_2003} have long required external sources of energy input to reproduce the observed galaxy luminosity function. In the absence of such feedback, models predict a power-law distribution of galaxy masses, whereas observations reveal a broken power law \citep{schechter_analytic_1976}. Agreement is reached only near the inflection point, with significant discrepancies at both the faint and bright ends: fewer galaxies are observed than predicted \citep[e.g.,][]{keres_galaxies_2009}. At the low-luminosity end, supernova feedback is thought to expel gas from shallow potential wells, suppressing further star formation. At the high-luminosity end, active galactic nuclei (AGN) are the most likely drivers of galaxy suppression. Energy released by accreting supermassive black holes (SMBHs) can exceed the binding energy of their host galaxies’ interstellar medium (ISM), with ratios E$_{\rm BH}$/E$_{\rm gal} > 80$ \citep{fabian_observational_2012}. This process, known as AGN feedback, is now recognized as a fundamental ingredient of galaxy evolution.
	
	AGN feedback operates through three primary channels: radiation, winds, and jets \citep{fabian_observational_2012}. Radiative feedback is traced by kiloparsec-scale photoionized bicones seen in extended narrow-line regions (ENLRs), marked by enhanced [O~III]/H$\alpha$ or [O~III]/H$\beta$ ratios \citep{unger_extended_1987, bianchi_soft_2006}, and by spatially coincident soft X-ray emission detected with {\it Chandra} \citep[][Section 6.4.3]{fabbiano_interaction_2024}. {\it Chandra} has also revealed extended hard ($>$3~keV) X-ray emission and 6.4~keV Fe~K$\alpha$ fluorescence in nearby AGN \citep[e.g.,][Section 6.5 of this review]{fabbiano_discovery_2017, fabbiano_interaction_2024}, likely produced by reflection off distant molecular clouds. Relativistic jets constitute another feedback mode: while large-scale radio jets are present in only $\sim$10\% of AGN \citep[e.g.,][]{kellermann_vla_1989}, most AGN contain embedded low power radio jets \citep[e.g.,][]{falcke_hubble_1998, nagar_radio_1999}. Comparison of radio data and {\it Chandra} images in selected energy bands has shown shocked thermal X-ray emission associated with the termination of the jet/radio lobes \citep[][]{wang_deep_2011,paggi_cheers_2012,Krol_2025_High_Resolution}. Moreover, simulations suggest that jet–ISM interactions generate cross-jet shocked X-ray emitting outflows \citep{mukherjee_jet-ism_2018}  that may explain the cross-cone X-ray emission detected in some AGN with {\it Chandra} \citep[e.g.,][]{travascio_agn-host_2021, fabbiano_jet-ism_2022}.

	Winds represent a third feedback channel. Though slower ($v\sim$1,000~km~s$^{-1}$) and less collimated than jets, they may be nearly ubiquitous in radio-quiet AGN. By shocking the ISM, winds can heat gas to $T \sim 10^7$K, giving rise to $\sim$1~keV X-ray emission. Structures such as LINER-like cocoons around high-ionization regions may trace this phase as the wind expands into the host galaxy, though their origin remains debated \citep{maksym_mapping_2016, ma_spatially_2021, trindade_falcao_mapping_2025,Domi2026}.
	
	X-ray imaging spectroscopy provides a direct probe of all three feedback modes. In particular, {\it Chandra}/ACIS-S has enabled detailed studies of nearby, radio-quiet (but typically containing a few $\sim$100 pc extended low-power radio jets), and heavily obscured AGN \citep[e.g.,][]{wang_deep_2011-2, fabbiano_deep_2018, trindade_falcao_deep_2023}. In these systems, circumnuclear obscuration acts as a natural coronagraph, suppressing direct nuclear emission and revealing extended emission and AGN–ISM interactions \citep[e.g.,][]{fabbiano_interaction_2024}.  Individual galaxies have been studied in detail revealing extended soft and hard X-ray emission and complex emission spectra, requiring a mix of photoionized and collisionally ionized plasmas \citep[see][]{fabbiano_deep_2018,fabbiano_interaction_2024}. However, a uniform, quantitative characterization of the extended X-ray emission across the broader sample of highly obscured AGN observed by {\it Chandra} is still lacking.
	
	In this paper, we present the results of such a spectral  analysis. We have constructed an analysis pipeline to merge individual observations of each AGN, simulate the nuclear point-spread function (PSF), to establish the unresolved and extended emission, evaluate spectral extraction regions based on the azimuthal and radial properties of the images, and then perform a multi-component spectral analysis.  The paper is organized as follows: in Section~\ref{sec:sample} we describe the sample selection; in Section~\ref{sec:data_analysis} we detail the data reduction and analysis; in Section~\ref{sec:results} we present results for our AGN sample, including the selection of spectral extraction regions and the results of spatially resolved spectroscopy ($\sim$0.2" resolution achieved via sub-pixel analysis). In Section~\ref{sec:discussion} we examine the spectral properties of the sample, and we summarize our conclusions in Section~\ref{sec:conclusions}. In a follow-up paper we will compare luminosities and spectral properties of nuclear and extended region to extract sample properties that may constrain the physics of AGN feedback. In a third paper in this series we will perform a uniform analysis of the spatial properties of the emission as a function of energy.
	
	\section{Sample Selection}
	\label{sec:sample}
	
	To investigate the interaction between AGN and their host galaxies in a systematic manner, we have compiled a sample of 20 nearby sources exhibiting extended X-ray emission as observed by \textit{Chandra}. This sample is primarily drawn from the hard X-ray-selected BASS catalog \citep{koss_bass_2022}, supplemented with the [O~III]-selected sample of \citet{risaliti_distribution_1999}. The selection is restricted to heavily obscured sources with column densities $\log (N_{\text{H}}/\text{cm}^{-2}) > 23$ and local objects within a distance of $D < 250$~Mpc. We have expanded this sample by including Mrk\,3, Mrk\,34, and Mrk\,78. Although these three sources were not selected via the BASS/Risaliti approach, they were added due to their well-established nature as heavily obscured AGN. Their inclusion is further justified by the availability of deep ($>$100\,ks) archival {\it Chandra} ACIS-S exposures \citep{Garmire2003} that are essential for resolving the faint, extended X-ray features targeted in this study. These criteria ensure a a well-characterized sample of obscured, moderately low-redshift AGN ($z<0.06$) suitable for studying feedback processes. {\it Chandra} studies of the extended X-ray emission require heavily obscured or Compton-thick (CT) systems, where the bright nuclear continuum is suppressed. This natural attenuation minimizes pile-up effects and facilitates the detection of faint circumnuclear structures. Our sample also includes NGC~1167, which, although probably not highly obscured, is a weak AGN, allowing the study of its diffuse X-ray emission \citep{fabbiano_jet-ism_2022}.
	
	The final sample is listed in Table~\ref{tab:log}, ordered by right ascension. For each source we report R.A. (J2000), Dec (J2000), redshift, distance (from NED, corrected to the CMB frame), parsec-to-arcsecond scale, line of sight absorbing nuclear column density ($\log N_{\rm H}$), black hole mass ($\log M_{\rm BH}$), and the total {\it Chandra} exposure used in this work. All sources are covered by deep \textit{Chandra} observations ($\gtrsim$100~ks). For the variable AGN NGC~1365 \citep[][]{risaliti_occultation_2007, risaliti_variable_2009}, we restrict our analysis to observations obtained while the source was in a CT state \citep[][]{nardini_chandrahetg_2015}.
	
	\begin{table}[t]
\caption{Properties of CT AGNs analyzed in this study.}
\centering
\begin{tabular}{lccccccccccc}
	\hline
	\textbf{Galaxy} & \textbf{R.A.} & \textbf{Dec} & \textbf{{\it z}} & \textbf{Distance} & \textbf{Scale} & \textbf{$\log N_{\rm H}$} & \textbf{$\log M_{\rm BH}$} &  \textbf{Total Exp} & \textbf{Reference}\\
	& (J2000) & (J2000) &  & (Mpc) & (pc/$''$)& (cm$^{-2}$) & ($M_{\odot}$) & (ks) & ($\log N_{\rm H}$), ($\log M_{\rm BH}$), (extended) \\
	\hline

	Mrk\,573 & 25.9907 & 2.3498 & 0.0172 & 72 & 348 &$>$24.5$^{\Delta}$ & 7.6  & 120 & (1), (8), (15,16,17,18) \\

    NGC\,835 & 32.3525& -10.1358 & 0.0128 & 55 & 264 &23.66 &- & 102 & (2), (-), (19)\\

	NGC\,1068 & 40.6696 & -0.0133 & 0.0038 & 16 & 78&$>$25$^{\Delta}$ & 6.9-8.0 &101 & (3), (8,9,10), (20,21,22)  \\

	NGC\,1167 & 45.4265 & 35.2057 & 0.01653 & 72 &336  & 21.5 & 7.9  & 193 & (4), (8), (-) \\

	NGC\,1365 & 53.4015 & -36.1404 & 0.0055 & 23 & 113& $>$24.1$^{\Delta}$ &7.3-7.8  & 200 & (5), (11), (-)\\

	NGC\,1386 & 54.1924 & -35.9994 & 0.003 & 12 & 62& $>$24.5$^{\Delta}$ & 7.4 &102& (3), (8), (-)  \\

    Mrk\,3    &93.9015 & 71.0375    & 0.01351 & 60   & 306  & 24.1  &  8.7 & 426  & (25), (26), (28)\\

	ESO\,428-G014 & 109.1301 & -29.3247 & 0.0057 & 23 & 117 &23.3 & 7.0-7.5 &156& (6), (12), (-) \\

    Mrk\,78     &115.6738   & 65.1771   & 0.0379 & 174  & 820  &$>$24.1$^{\Delta}$  & 7.9 &  101 & (27), (26), (29)\\

	NGC\,3081 & 149.8730 & -22.8263 & 0.0081 & 25 & 166 & 23.8 & 7.1 & 268 & (3), (8), (-) \\

	NGC\,3079 & 150.4908 & 55.6798 & 0.0037 & 16 & 76 & 24.6$^{\Delta}$ & 7.6 & 126 & (2), (8), (-)\\

    Mrk\,34     & 158.5356   & 60.0305   & 0.0509 & 236   &  1108 & 24.3$^{\Delta}$  & 7.8  & 299  & (24), (24), (30)\\

	NGC\,3393 & 162.0977 & -25.1621 & 0.0125 & 61 & 256 &24.4$^{\Delta}$&7.5 & 880 & (2), (13), (-)\\

	NGC\,4945 & 196.3645 & -49.4682 & 0.0019 & 4 & 39 & 24.8$^{\Delta}$ & 6.0 & 420 & (2), (14), (-)\\

	NGC\,5252 & 204.5665 & 4.5426 & 0.0231 & 112  & 466 & 22.7 & 8.0 &220 & (7), (8), (-)\\

	NGC\,5643 & 218.1699 & -44.1746 & 0.004 & 16 & 83 & 25$^{\Delta}$ & 6.1-7.0 & 172 & (2), (23), (-) \\

	NGC\,5728 & 220.5997 & -17.2532 & 0.0093 & 43 &191 & 24.1 & 8.2 &242 & (2), (8), (-) \\

	ESO\,137-G034 & 248.8088 & -58.08 & 0.009 & 42 &184 & 24.4$^{\Delta}$ & -  &215 & (2), (-), (-) \\

	IC\,5063 & 313.0098 & -57.0688 & 0.0114 & 52 &233 &23.6 &7.7 &271 & (2), (8), (-) \\

	NGC\,7212 & 331.7554 & 10.2311 & 0.0266 & 120 & 534& $>$24.5$^{\Delta}$ &7.5 &148 & (3), (8), (-) \\
    
	\hline
\end{tabular}

\tablecomments{
Columns (1)–(4) list each galaxy, their coordinates, and redshift; Column (5) is the cosmology-corrected scale from NED; Column (6) lists the line-of-sight column densities. Sources marked with $\Delta$ satisfy the criteria for CT ($\log$(N$_{\rm H}/cm^{-2})\,>\,24.1$ \citep{comastri_compton-thick_2004}; Column (7) lists the black hole masses; Column (8) lists the total exposure time of the observations analyzed in this work; Column (9) gives the references for the line-of-sight column densities, black hole masses listed previously and the first studies analyzing the extended X-ray emission of these sources.  
References:  
(1) \citealt{bianchi_high-resolution_2010};  
(2) \citealt{ricci_compton-thick_2015};  
(3) \citealt{bassani_three-dimensional_1999};  
(4) \citealt{akylas_xmm-newton_2009};  
(5) \citealt{nardini_chandrahetg_2015};  
(6) \citealt{feruglio_multiphase_2020};  
(7) \citealt{cappi_is_1996};  
(8) \citealt{marinucci_link_2012};  
(9) \citealt{lodato_non-keplerian_2003};  
(10) \citealt{hure_origin_2002};  
(11) \citealt{risaliti_xmmnewton_2009};  
(12) \citealt{fabbiano_deep_2019};  
(13) \citealt{greene_precise_2010};  
(14) \citealt{greenhill_distribution_1997};  
(15) \citealt{paggi_cheers_2012};  
(16) \citealt{jones_extended_2021};  
(17) \citealt{bianchi_high-resolution_2010};  
(18) \citealt{gonzalez-martin_soft_2010};  
(19) \citealt{gonzalez-martin_x-ray_2016};  
(20) \cite{young_chandra_2001};  
(21) \citealt{gao_extended_2025};  
(22) \citealt{wang_chandra_2001};  
(23) \citealt{poitevineau_galaxy_2025};
(24) \citealt{gandhi_nustar_2014};
(25) \citealt{guainazzi_nature_2016};
(26) \citealt{woo_active_2002};
(27) \citealt{zhao_properties_2021};
(28) \citealt{sako_chandra_2000};
(29) \citealt{fornasini_termination_2022};
(30) \citealt{maksym_ufo_2023}
}
\label{tab:log}
\end{table}
	
\begin{figure}[h]
	\centering
	\includegraphics[width=.95\textwidth]{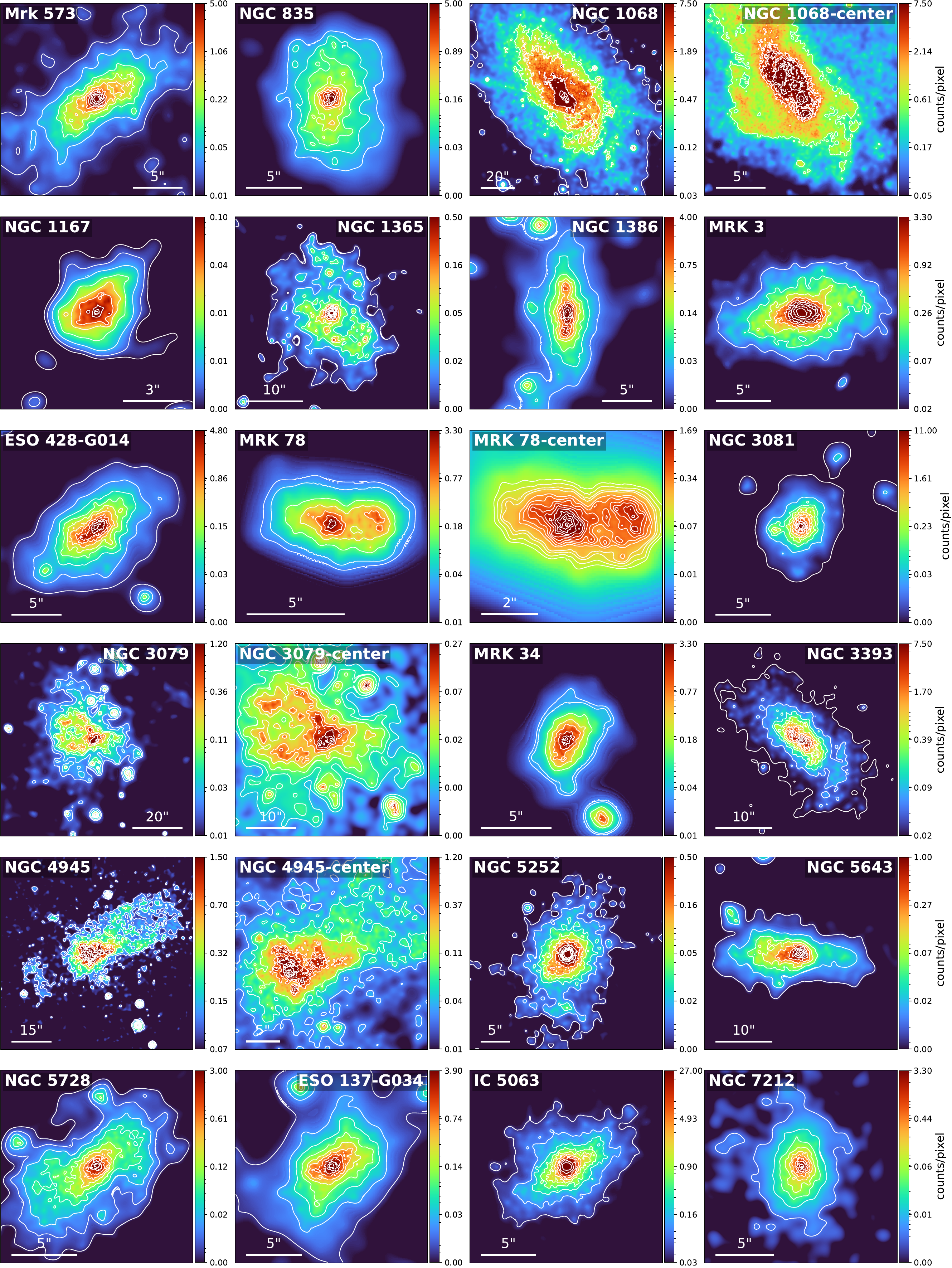}
	\caption{Adaptively smoothed {\it Chandra}/ACIS-S images of the AGN sample. All images are shown in the full 0.3-7~keV band. North is up, east is to the left. The binning and contour values are listed in Table~\ref{tab:dmadapt_params}. For NGC~1068 and Mrk~78 we show two different spatial scale images to highlight the entire emission regions and the features of the emission closer to the nucleus. Similarly, for NGC~3079 and NGC~4945, we show two different scale images to highlight both the AGN-connected bubble and its position relative to the galactic plane.}
	\label{fig:imagescool}
\end{figure}

	Figure~\ref{fig:imagescool} shows the central regions of the galaxies in our sample. Each image was produced in the full 0.3-7~keV band using the sub-pixel ACIS-S binning values listed in Table~\ref{tab:dmadapt_params}. To enhance both compact and extended structures, we applied adaptive smoothing requiring 5-9 counts within the kernel (see Table~\ref{tab:dmadapt_params}), with kernel radii spanning 0.5-15 pixels in 30 logarithmically spaced steps. For each target, the adopted sub-pixel scale, minimum counts per smoothing kernel, and contour levels are summarized in Table~\ref{tab:dmadapt_params}.
	
\begin{figure}[h]
	\centering
	\includegraphics[width=.9\textwidth]{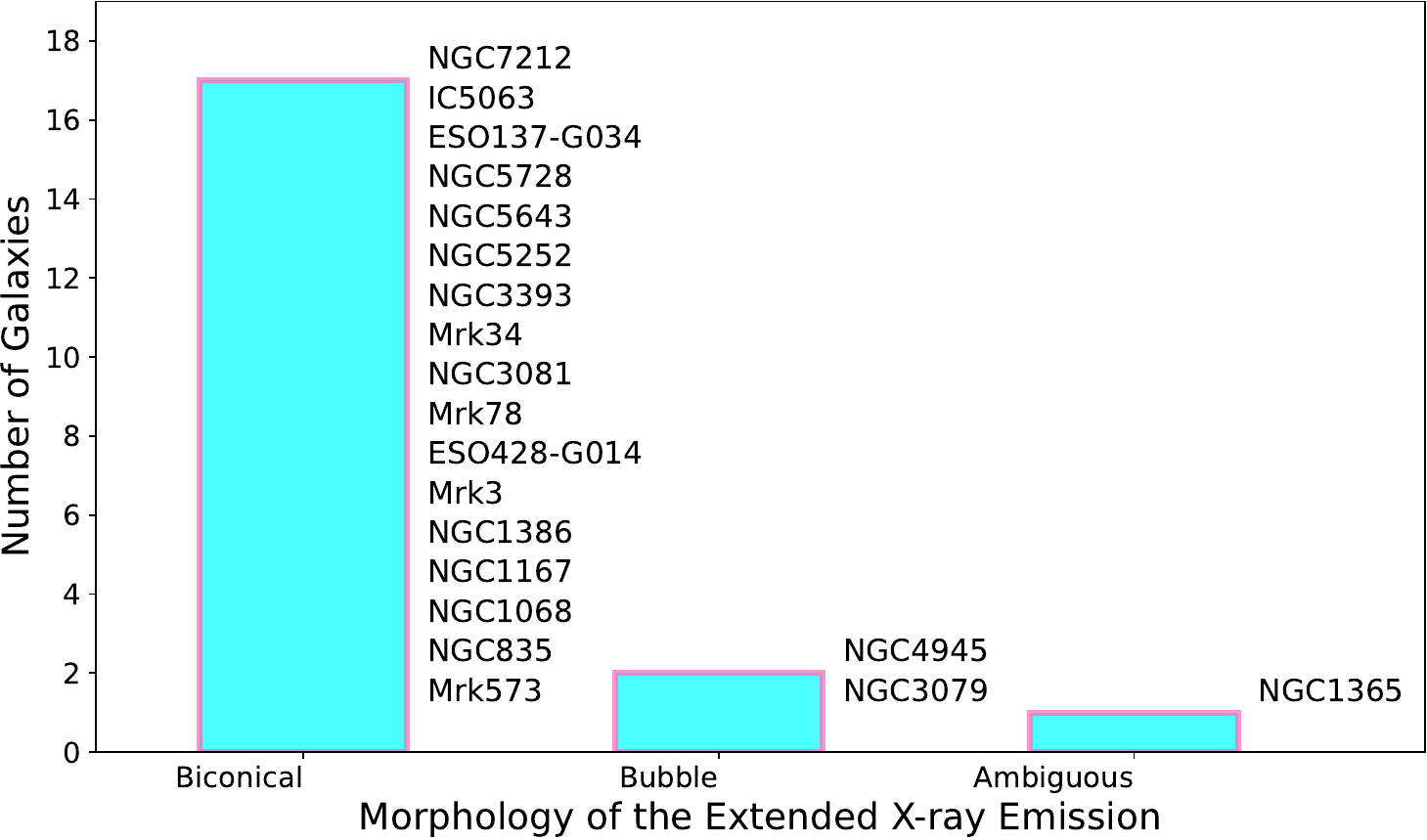}
	\caption{Distribution of X-ray morphological classifications for the 20 AGNs in our sample. 17 systems show biconical structures aligned with optical ionization cones, 2 display bubble-like morphologies suggestive of shock-heated outflows, and 1 (NGC~1365) exhibits a more complex structure influenced by both AGN activity and circumnuclear star formation. The classification is based on the radial and azimuthal profiles shown in Figures~\ref{fig:profiles1}-\ref{fig:profiles5}.}
    \label{fig:morphology}
\end{figure}

	By selection,  extended X-ray emission is present in all the AGN in our sample, as reported in previous individual studies (see Table~\ref{tab:log} for the earliest reported detections). In most cases, the extended emission is elongated along the same direction as the optical narrow-line regions (i.e., the ionization cones), with additional emission often present in the cross-cone direction as well \citep[see review by][]{fabbiano_interaction_2024}. For 17 AGN, the extended X-ray structures are approximately symmetric about the nucleus and primarily aligned with the host galaxy’s disk, in a bi-conical morphology. Three AGN, however, show different morphologies. In NGC~4945 \citep[e.g.,][]{porraz_barrera_hot_2024} and NGC~3079 \citep[e.g.,][]{li_pressure_2024}, the extended X-ray emission exhibits a bubble-like structure with enhanced brightness at large radii, oriented roughly perpendicular to the galactic plane. In NGC~1365 \citep{wang_imaging_2009}, the diffuse X-rays reflect both AGN excitation and contributions from a circumnuclear star-forming ring, resulting in a more complex or ambiguous morphology.  The full set of morphologies across the sample is summarized in Figure~\ref{fig:morphology}.
	

	\section{Data Reduction and Analysis}
	\label{sec:data_analysis}
	To ensure a uniform treatment of the AGN sample (Table~\ref{tab:log}), we developed an automated pipeline for reducing and analyzing the {\it Chandra} observations. The workflow includes four main stages: (1) reprocessing, astrometric alignment, and merging of individual observations (Section~\ref{sec:imaging_repro}); (2) PSF simulations (Section~\ref{sec:psf}); (3) extraction of radial and azimuthal surface brightness profiles (Section~\ref{sec:radial_az}); and (4) spectral extraction and modeling (Section~\ref{sec:spec_analysis}).
	
	The pipeline is implemented through shell scripts and makes use of \texttt{CIAO}~4.17\footnote{\url{https://cxc.cfa.harvard.edu/ciao/}} \citep{fruscione_ciao_2006}, in combination with \texttt{DS9} \citep{joye_new_2003}, \texttt{MARX} \citep{davis_raytracing_2012}, and \texttt{XSPEC} \citep{arnaud_xspec_1996}.

	\subsection{Image Reprojection, Astrometric Alignment, and Merging}
	\label{sec:imaging_repro}
	For each source, we retrieved all publicly available {\it Chandra} observations as of April 25, 2025. Individual datasets were reprocessed with \texttt{chandra\_repro}\footnote{\url{https://cxc.cfa.harvard.edu/ciao/ahelp/chandra\_repro.html}} to apply the latest calibration. The observation with the longest exposure was adopted as the astrometric reference. Its centroid was determined in the hard band (5.5–7~keV), binned to 1/8 of the native pixel scale, using \texttt{dmstat}\footnote{\url{https://cxc.cfa.harvard.edu/ciao/ahelp/dmstat.html}} within a $0.2''$ circular aperture. This approach, established in prior studies of obscured AGN \citep[e.g.,][]{fabbiano_chandra_2018, fabbiano_deep_2018-1}, exploits the fact that nuclear emission in CT sources is mostly compact at high energies.
	
	Centroids for all remaining observations were derived using the same procedure, and relative astrometric shifts were applied with \texttt{wcs\_update}\footnote{\url{https://cxc.cfa.harvard.edu/ciao/ahelp/wcs\_update.html}} and \texttt{dmhedit}\footnote{\url{https://cxc.cfa.harvard.edu/ciao/ahelp/dmhedit.html}}. Aspect solution files were updated accordingly to ensure internal consistency.
	
	All reprojected event files and aspect solutions were then merged with \texttt{merge\_obs}\footnote{\url{https://cxc.cfa.harvard.edu/ciao/ahelp/merge\_obs.html}}, producing deep combined images and event lists. Centroid alignment was verified in the 5.5–7~keV band. Final alignment was inspected visually across the individual observations.

	\subsection{PSF Simulations}
	\label{sec:psf}
	To assess the significance of extended X-ray emission, we generated PSF simulations in the 0.3-7 keV band. Nuclear spectra were first extracted from $1.5''$ apertures using \texttt{specextract}\footnote{\url{https://cxc.cfa.harvard.edu/ciao/ahelp/specextract.html}} and fit in \texttt{XSPEC} with models appropriate for Compton-thick or Compton-thin AGN. The best-fit spectral models were then used as inputs for ray-tracing simulations in \texttt{MARX}, adopting an \texttt{ASPECT\_BLUR} parameter of $0.19''$ (reduced to $0.07''$ for select cases; see \citealt{ma_extended_2023}). The resulting simulated event files were combined with \texttt{dmmerge}\footnote{\url{https://cxc.cfa.harvard.edu/ciao/ahelp/dmmerge.html}} to produce deep reference PSFs for direct comparison with the observed data.

	\subsection{Spectral extraction regions}
	\label{sec:radial_az}
	The definition of regions from which to extract the  data for spectral analysis was based on the spatial properties of the X-ray surface brightness of each AGN. The results of this analysis are summarized in Figures~\ref{fig:profiles1}-\ref{fig:profiles5}, where for each AGN we show the 0.3-7~keV image, the azimuthal surface brightness profile centered on the central point-like AGN source, and the radial surface brightness profile compared with the PSF. 
	
	\subsubsection{Azimuthal and Radial Profiles}
\begin{figure*}[t]
	\centering
	
	\includegraphics[width=0.96\textwidth]{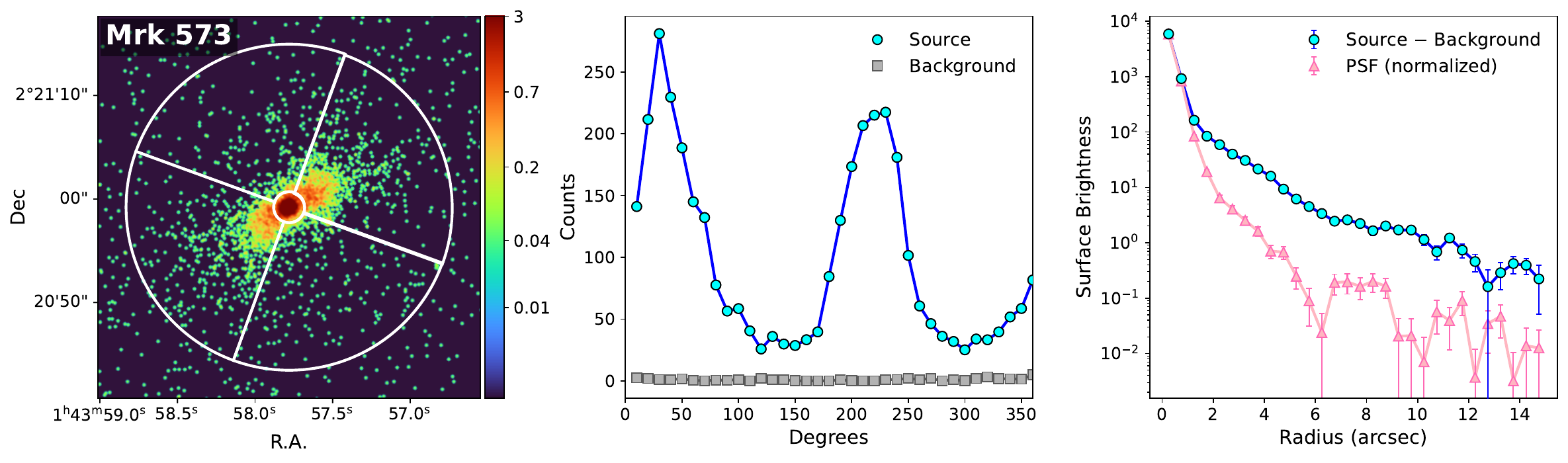}
	\vspace{0.3cm}
	
	\includegraphics[width=0.96\textwidth]{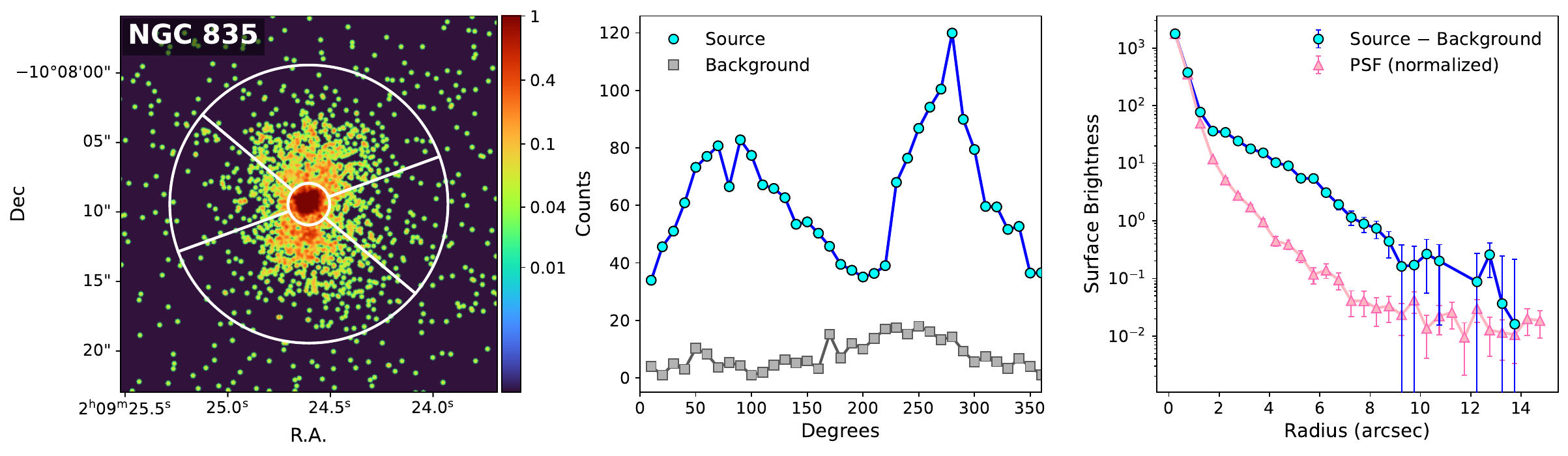}
	\vspace{0.3cm}
	
	\includegraphics[width=0.96\textwidth]{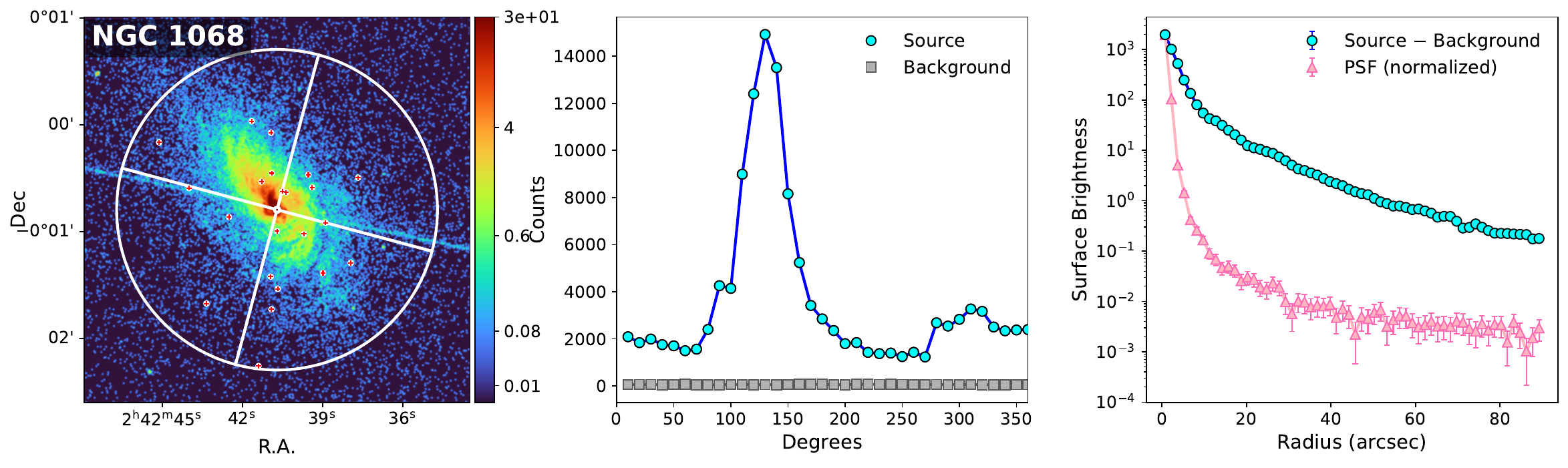}
	\vspace{0.3cm}
	
	\includegraphics[width=0.96\textwidth]{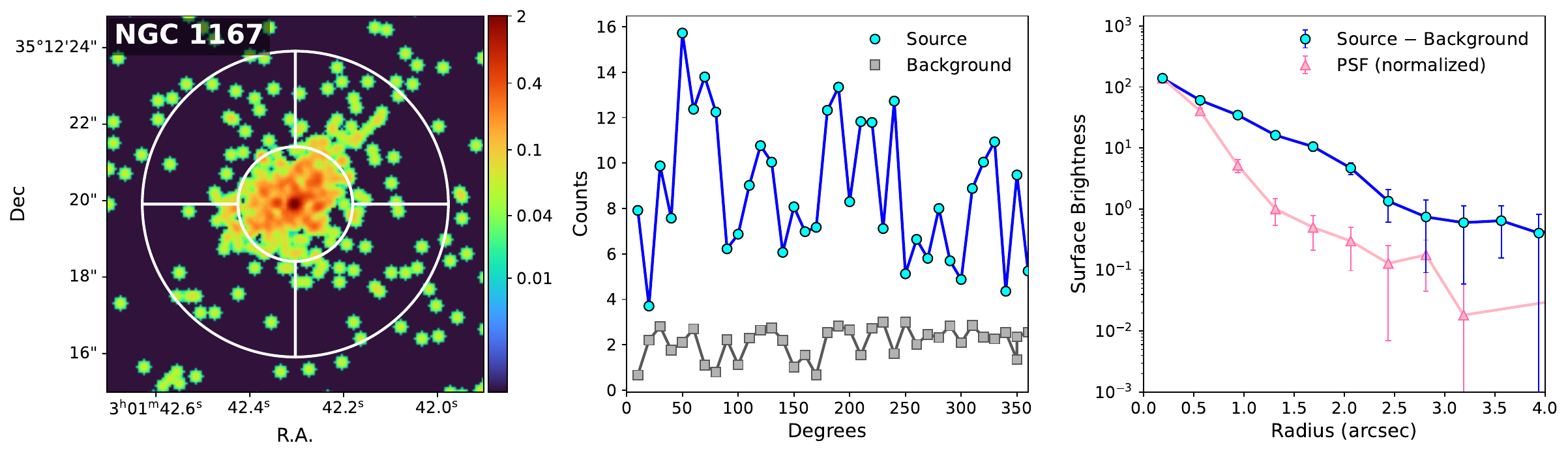}
	
	\caption{Characterization of the spatial distribution of the 0.3-7~keV X-ray emission. \textit{Left panel:} \textit{Chandra} image in the 0.3-7~keV band with the spectral extraction regions overlaid, including the nuclear (core) and extended (cone and cross-cone) components. These regions were defined based on the azimuthal and radial surface brightness profiles. \textit{Middle panel:} Azimuthal surface brightness profile of the 0.3-7~keV emission, compared with the background level, highlighting anisotropies associated with extended emission. \textit{Right panel:} Radial surface brightness profile of the the azimuthally integrated background-subtracted 0.3-7~keV emission, compared with the normalized point spread function (PSF) in the same energy band. Deviations from the PSF at larger radii indicate the presence of spatially extended emission beyond the unresolved nuclear component.}
	\label{fig:profiles1}
\end{figure*}

\begin{figure*}[t]
	\centering
	
	\includegraphics[width=0.99\textwidth]{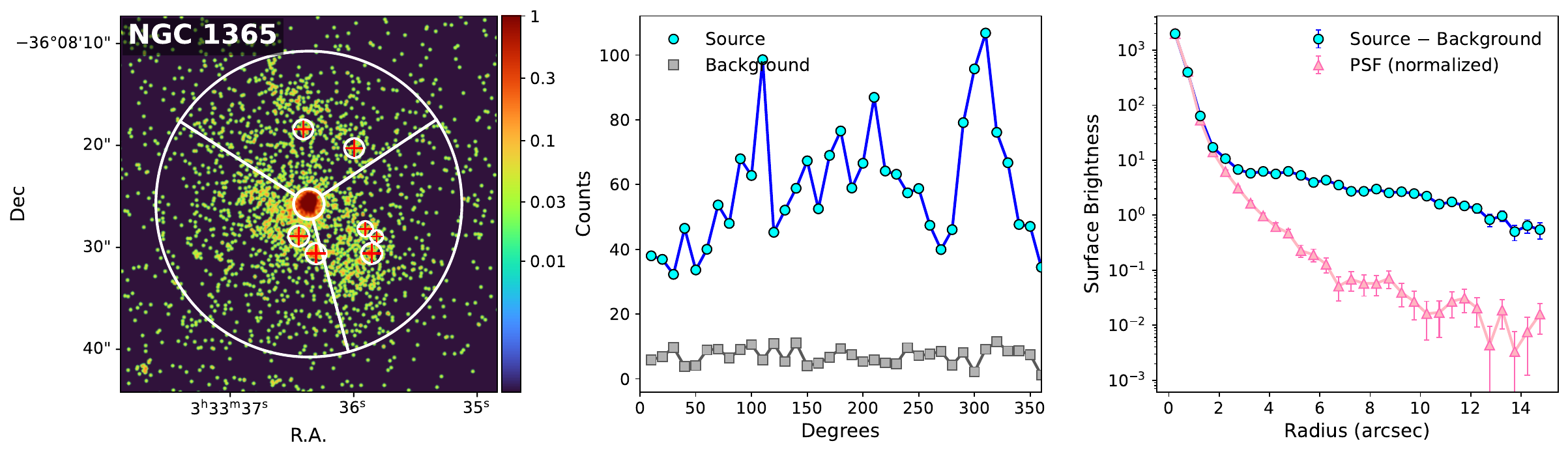}
	\vspace{0.32cm}
	
	\includegraphics[width=0.99\textwidth]{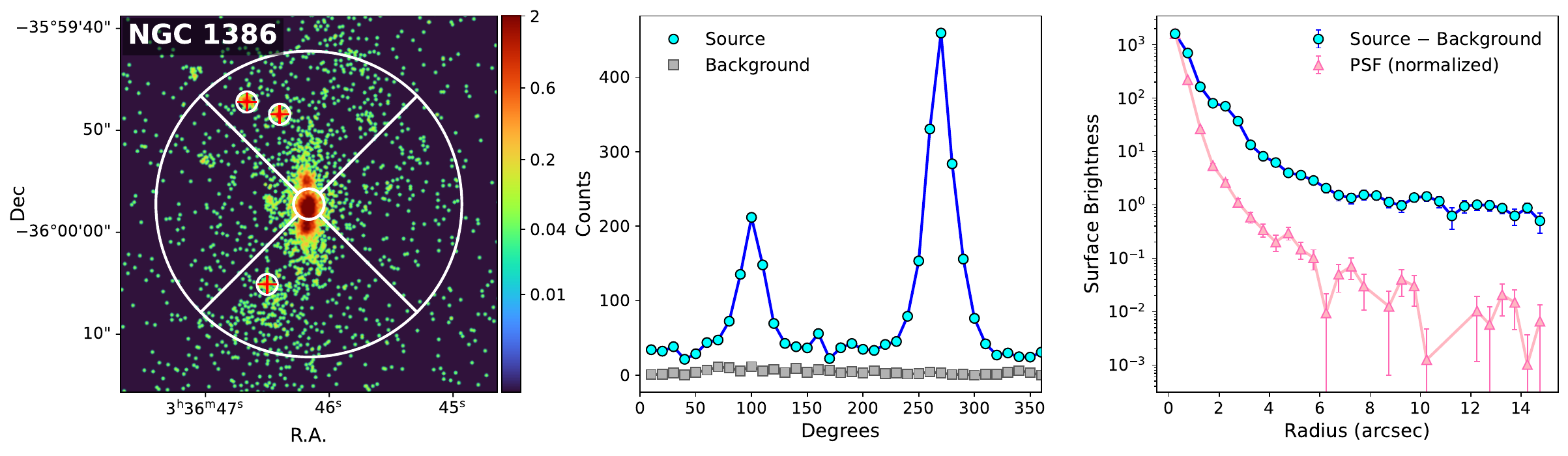}
	\vspace{0.32cm}
	
	\includegraphics[width=0.99\textwidth]{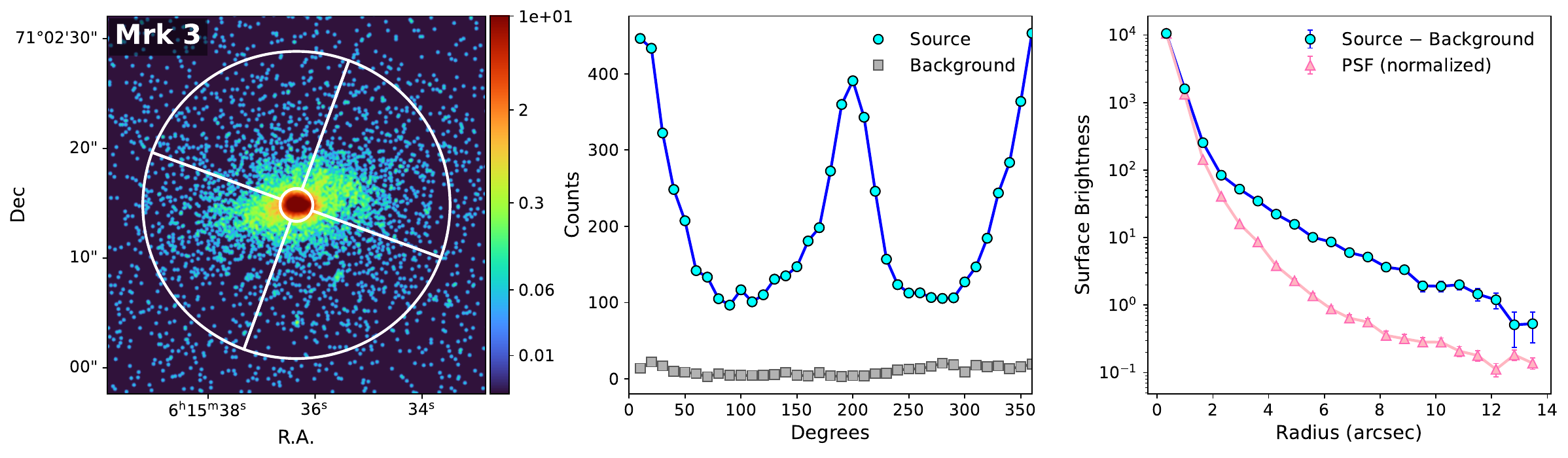}
	\vspace{0.32cm}
	
	\includegraphics[width=0.99\textwidth]{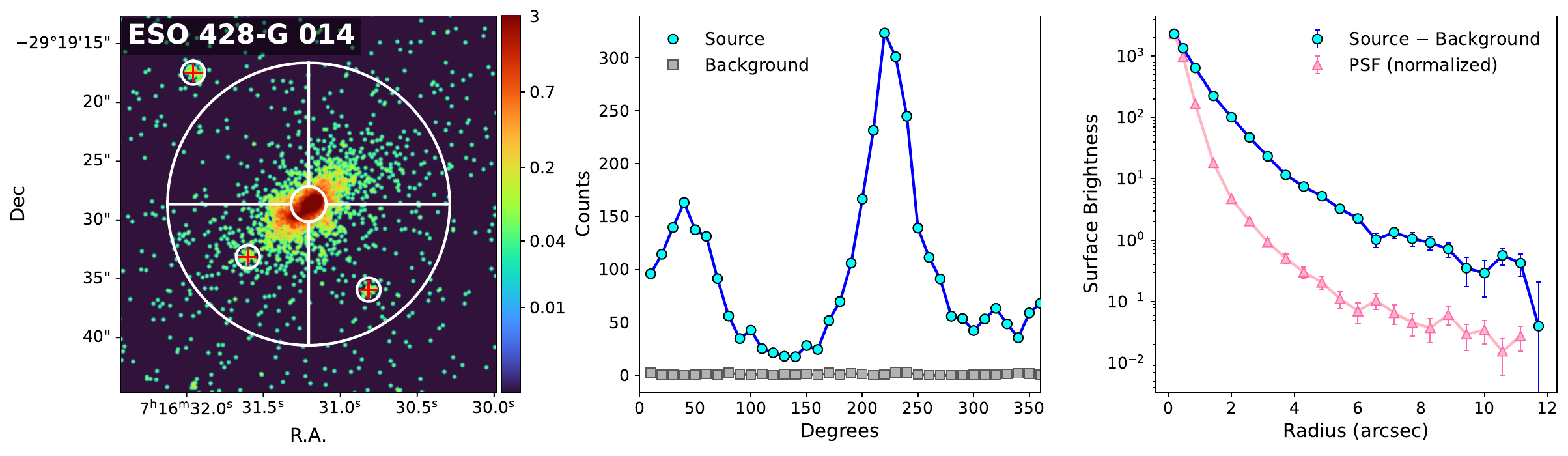}
	\vspace{0.32cm}
	
	\caption{Same as in Figure~\ref{fig:profiles1} for NGC\,1365, NGC\,1386, Mrk 3 and ESO\,428-G014.}
	\label{fig:profiles2}
\end{figure*}

\begin{figure*}[t]
	\centering
	
	\includegraphics[width=0.99\textwidth]{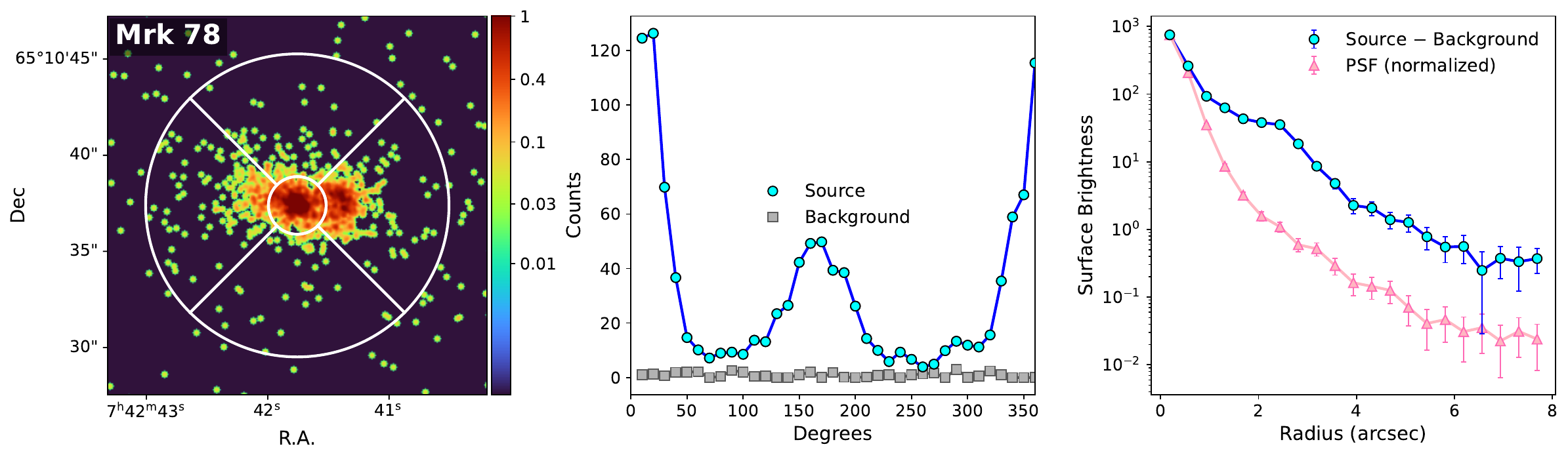}
	\vspace{0.32cm}
	
	\includegraphics[width=0.99\textwidth]{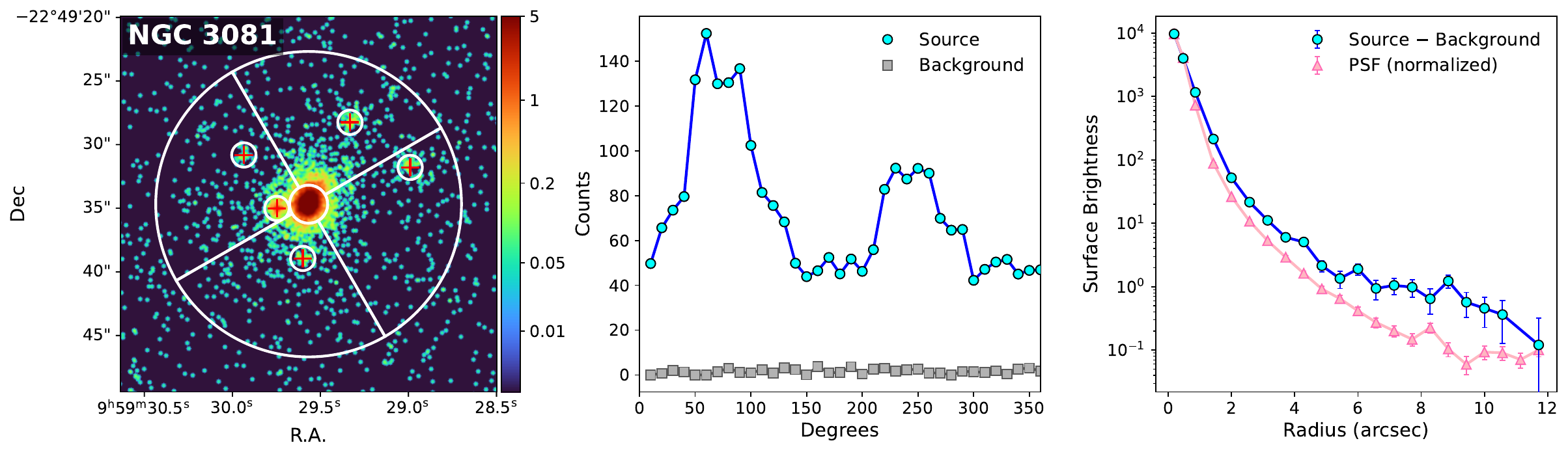}
	\vspace{0.32cm}
	
	\includegraphics[width=0.99\textwidth]{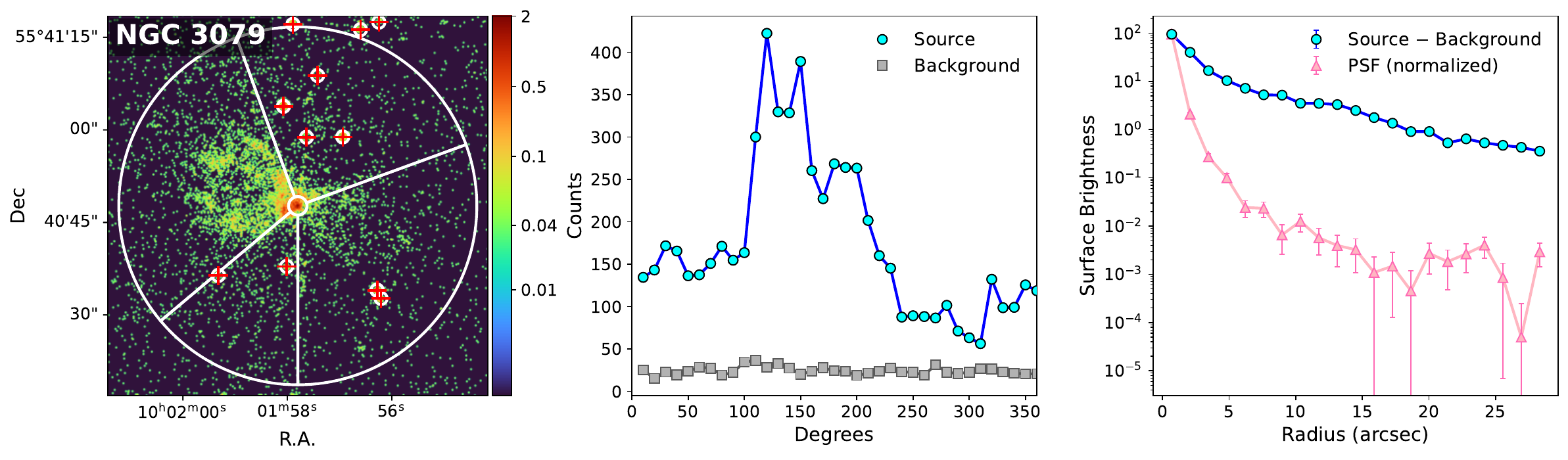}
	\vspace{0.32cm}
	
	\includegraphics[width=0.99\textwidth]{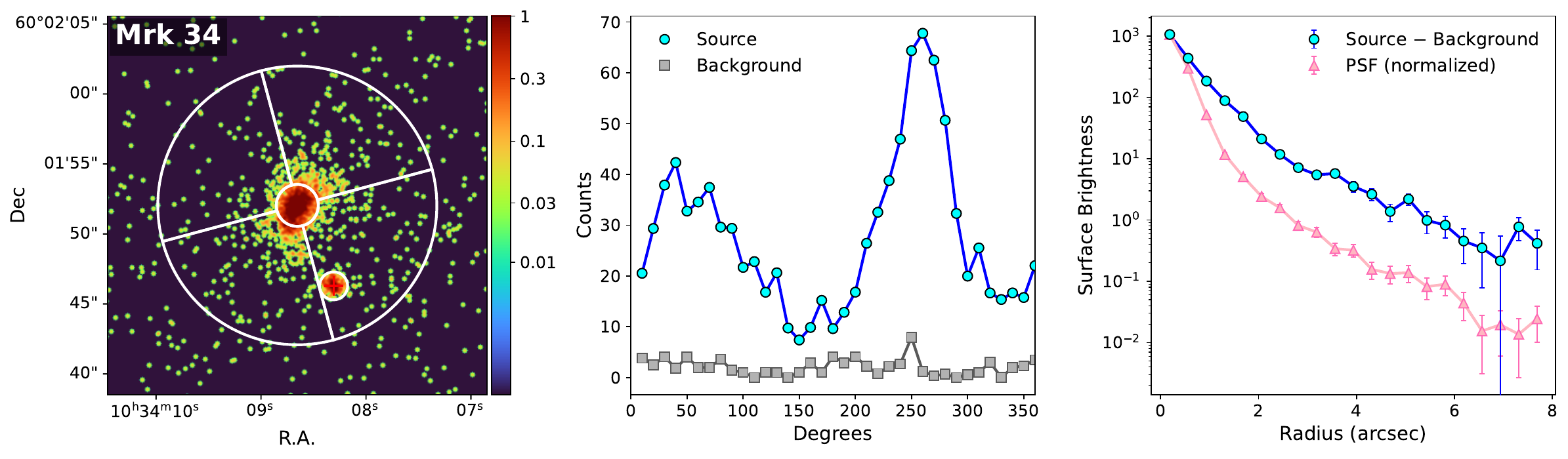}
	\vspace{0.32cm}
	
	\caption{Same as in Figure~\ref{fig:profiles1}, for Mrk\,78, NGC\,3081, NGC\,3079 and Mrk\,34.}
	\label{fig:profiles3}
\end{figure*}

\begin{figure*}[t]
	\centering
	
	\includegraphics[width=0.99\textwidth]{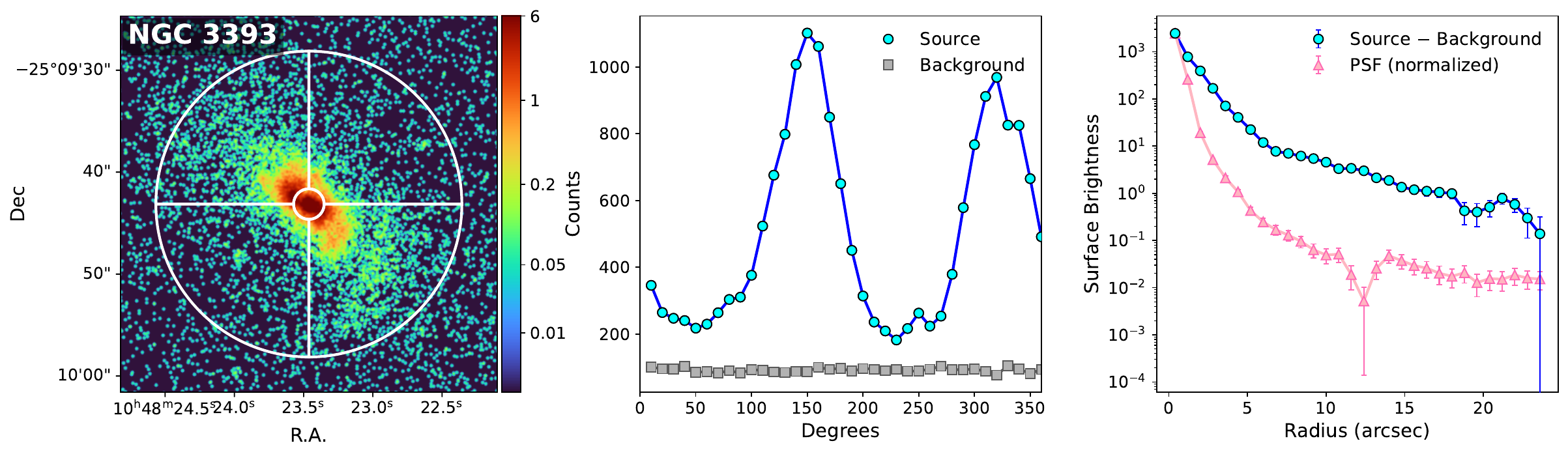}
	\vspace{0.32cm}
	
	\includegraphics[width=0.99\textwidth]{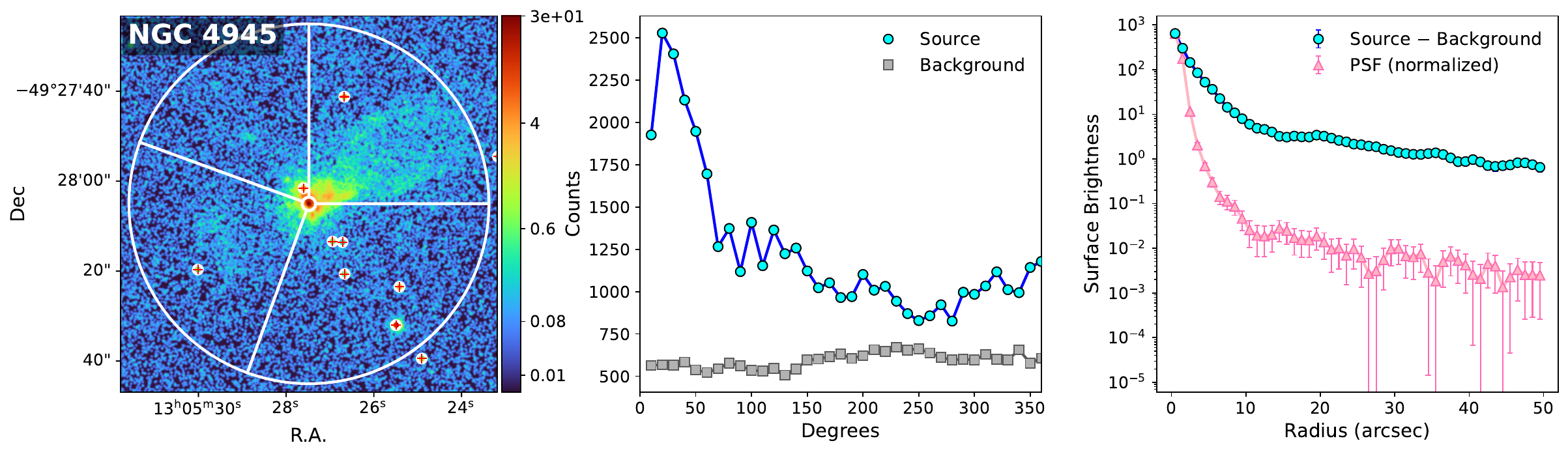}
	\vspace{0.32cm}
	
	\includegraphics[width=0.99\textwidth]{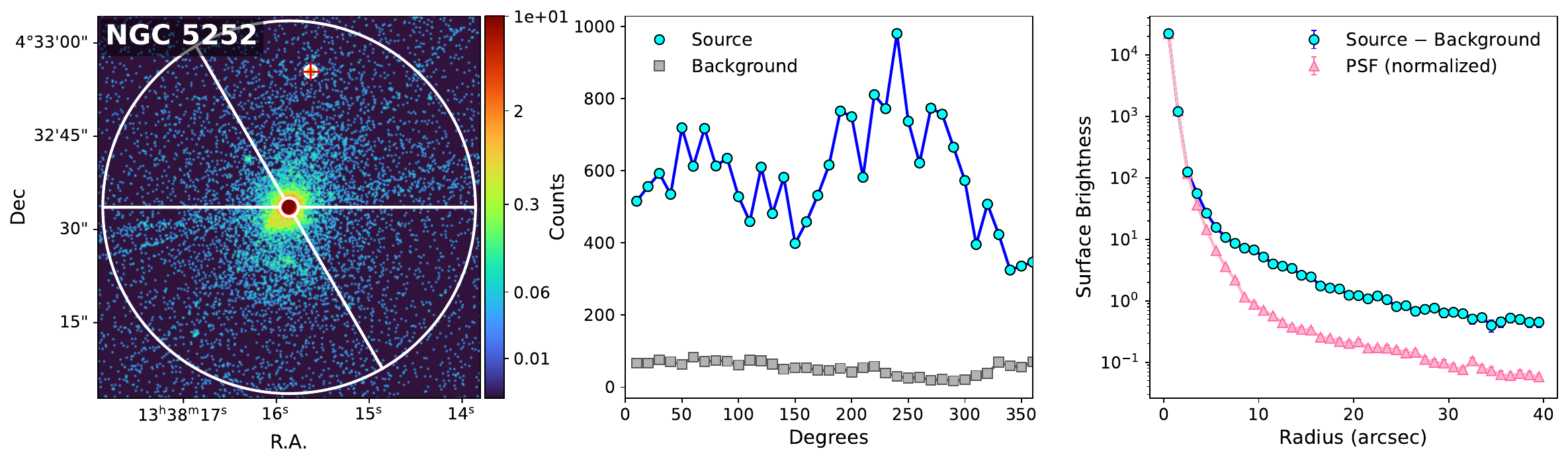}
	\vspace{0.32cm}
	
	\includegraphics[width=0.99\textwidth]{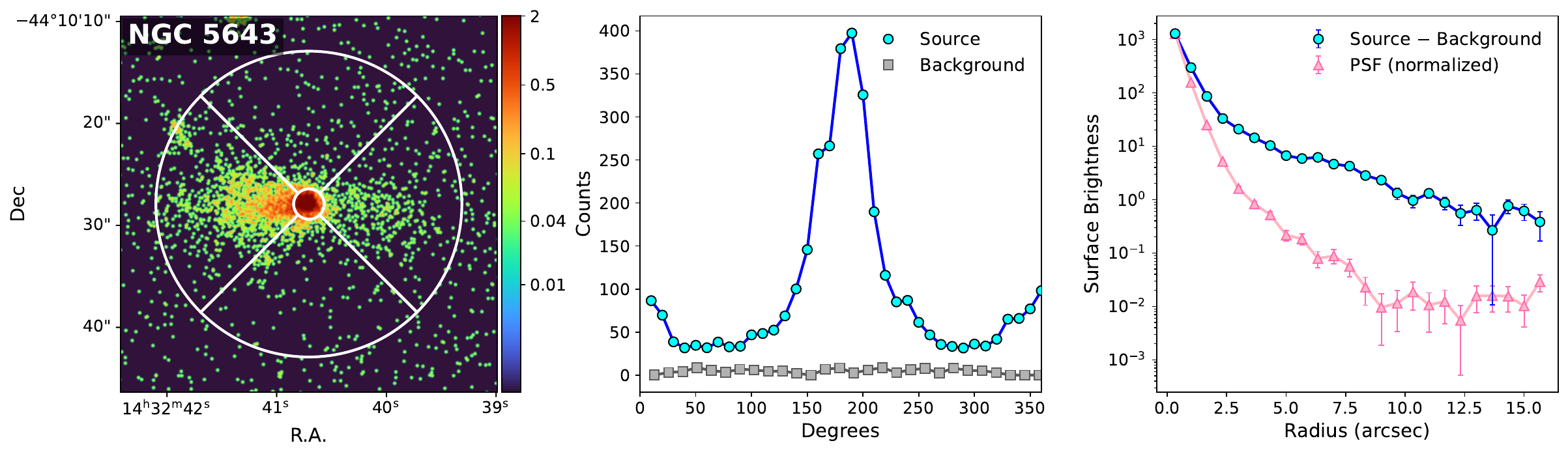}
	\vspace{0.32cm}
	
	\caption{Same as in Figure~\ref{fig:profiles1} for NGC\,3393, NGC\,4945, NGC\,5252 and NGC\,5643.}
	\label{fig:profiles4}
\end{figure*}

\begin{figure*}[t]
	\centering
	
	\includegraphics[width=0.99\textwidth]{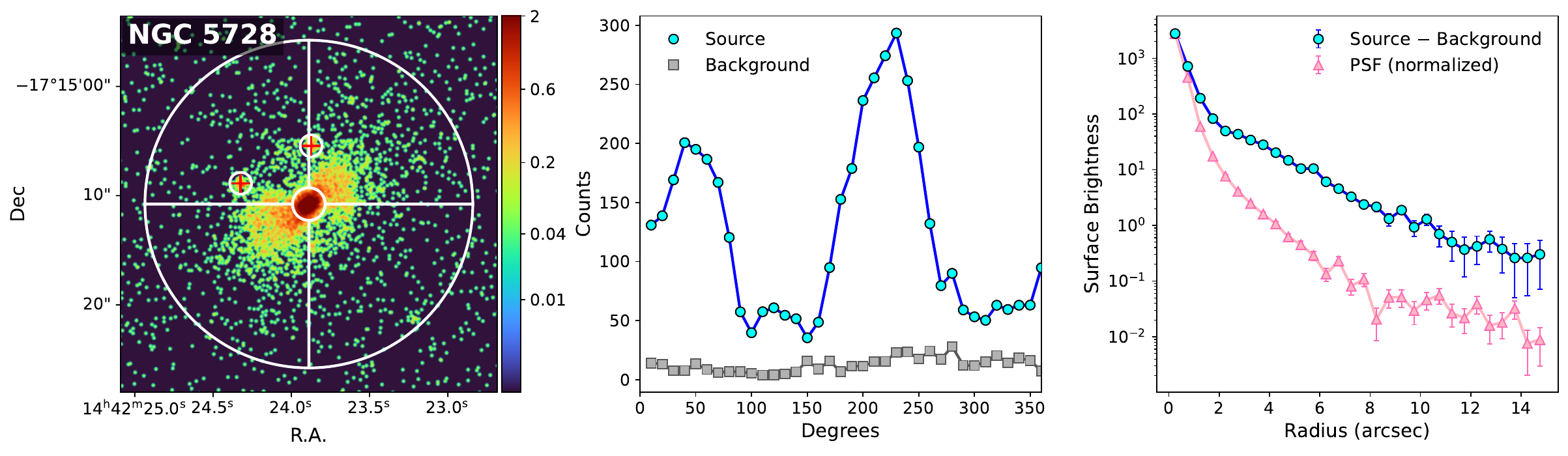}
	\vspace{0.32cm}
	
	\includegraphics[width=0.99\textwidth]{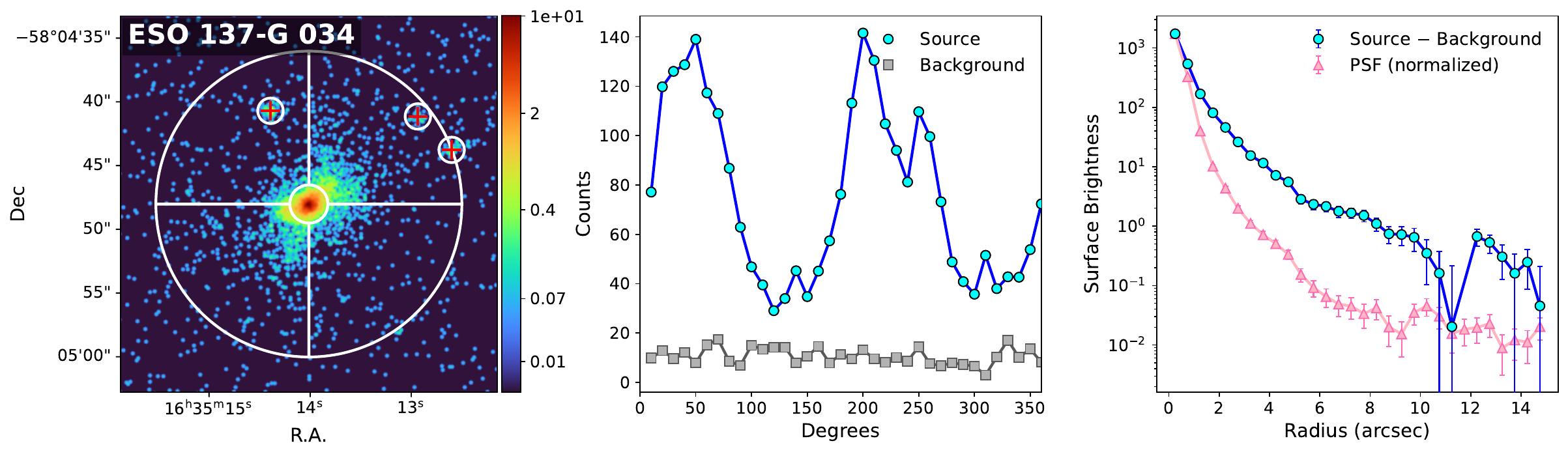}
	\vspace{0.32cm}
	
	\includegraphics[width=0.99\textwidth]{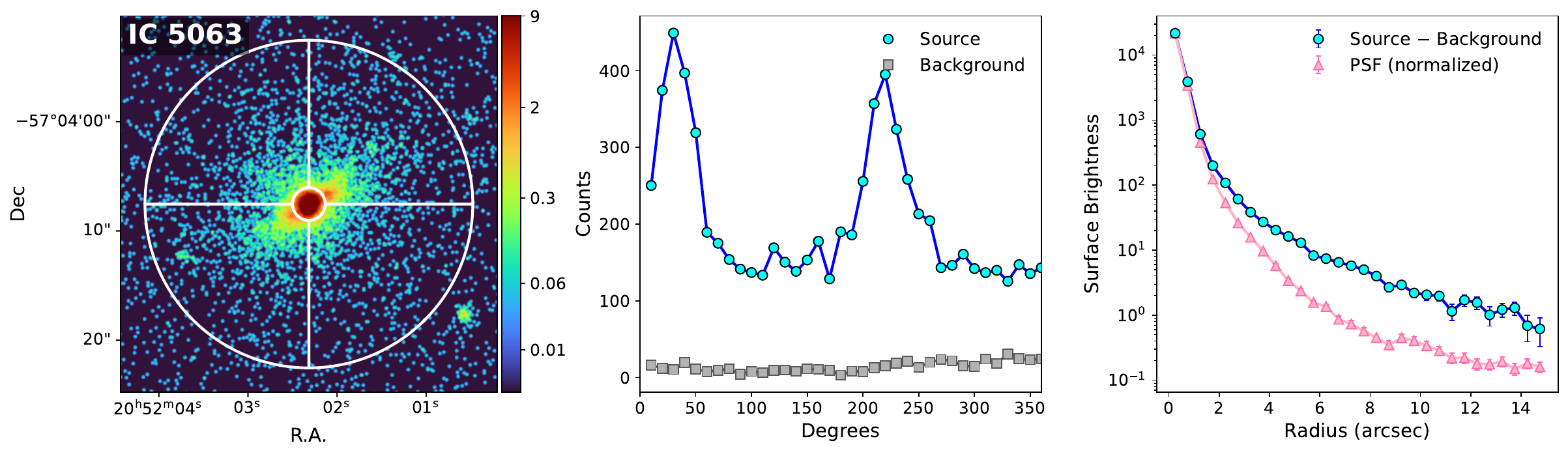}
	\vspace{0.32cm}
	
	\includegraphics[width=0.99\textwidth]{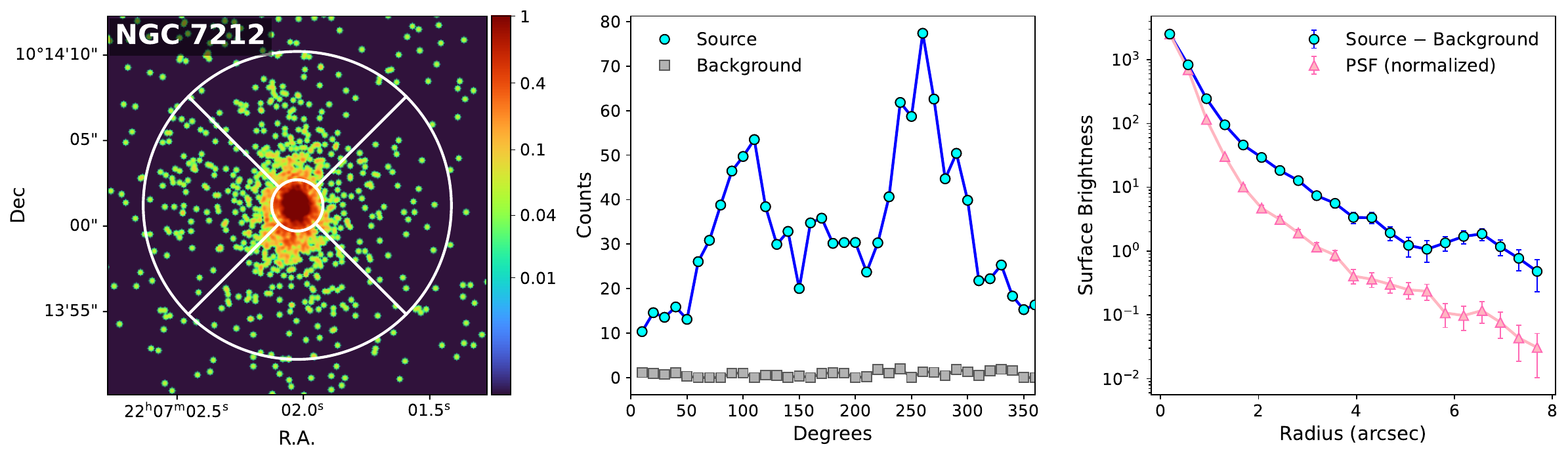}
	\vspace{0.32cm}
	
	\caption{Same as in Figure~\ref{fig:profiles1} for NGC\,5728, ESO\,137-G034, IC\,5063 and NGC\,7212.}
	\label{fig:profiles5}
\end{figure*}

	To analyze the surface brightness profiles, we first excluded point-like sources in the field-of-view. These sources were identified with \texttt{wavdetect}\footnote{\url{https://cxc.cfa.harvard.edu/ciao/ahelp/wavdetect.html}} and excluded using $1''$ (or 1.5$''$ for brighter sources) circular masks, which were applied consistently across both radial and azimuthal analyses. These source exclusion regions are marked on the 0.3-7~keV images of each AGN in Figures~\ref{fig:profiles1}-\ref{fig:profiles5}.
	
	Radial and azimuthal surface brightness profiles were extracted from the point-source-subtracted merged event files and compared with their corresponding simulated PSFs in the full 0.3-7~keV band. Azimuthal profiles were extracted in the 0.3-7~keV band using pie-shaped sectors covering $1.5'' < r < r_{\rm max}$, divided into 36 angular bins of $10^\circ$ each. 
	
	Radial profiles were constructed in concentric annuli centered on the AGN, extending from an inner radius of $r_{\rm in} = 1.5''$ to outer radii of $r_{\rm max} \in {8'', 15'', 30'', 40'', 50'', 90''}$, depending on the extent of the emission in each source. As a fiducial value to define the extent of the diffuse emission, we use the radius where the measured surface brightness emission reaches a value of 5\% of the background emission, except for three galaxies (NGC~1068, NGC~4945, NGC~5252), for which we stopped at $\sim$10\%. When the extended emission profiles reach this level, we consider it to mark the boundary of the diffuse component. Background levels were measured from identically shaped annuli placed in nearby, source-free regions. The background-subtracted per-pixel counts were then compared to those of the normalized PSF, scaled to match the innermost radial bin. 
	
	Figures~\ref{fig:profiles1}-\ref{fig:profiles5} show the full-band (0.3-7~keV) images, and the azimuthal and radial profiles for each galaxy in our sample.
	
	\subsubsection{Nuclear and Extended Regions}
	Nuclear spectra were extracted from $1.5''$ radius circular apertures centered on the hard-band centroid. The extended emission spectra were extracted from pie-shaped sectors based on the azimuthal and radial profiles. These regions are plotted on the images in Figures~\ref{fig:profiles1}-\ref{fig:profiles5}. For the 17 sources with well-defined biconical structures (see Figure~\ref{fig:morphology}), we selected “cone” and “cross-cone” sectors, as indicated by the azimuthal profiles, where by "cone" we define the most radially extended regions, which are generally co-spatial with the optical extended emission line regions (see \citealt{fabbiano_interaction_2024}). 
	
	In the case of systems with irregular morphologies (NGC~4945, NGC~3079, and NGC~1365), the extraction regions were defined based on the image morphology and radial and azimuthal profiles. 
	
	Specifically for NGC~1365, we chose three regions, two of which (E and W regions) encompass most of the emission from the star-forming ring \citep{wang_imaging_2009}. The N region includes also emission that \citet{wang_imaging_2009} ascribed to the AGN excitation. 
	
	\subsection{Spectral  Analysis}
	\label{sec:spec_analysis}
	Spectra  were extracted from the regions shown in Figures~\ref{fig:profiles1}-\ref{fig:profiles5} from individual observations and then combined with \texttt{mergespec}\footnote{\url{https://cxc.cfa.harvard.edu/ciao/ahelp/mergespec.html}}. The spectra were binned at a minimum of five counts per bin. Spectral fitting was performed in \texttt{XSPEC} over the 0.3-7~keV band using the C-statistic \citep{cash_parameter_1979}. Galactic absorption was modeled with \texttt{tbabs}, fixing column densities to values from \citet{bekhti_hi4pi_2016}. The flat \texttt{XSPEC} built-in $\Lambda$CDM cosmology was adopted, with $H_0 = 70$~km~s$^{-1}$~Mpc$^{-1}$, $\Omega_{\rm m} = 0.27$, and $\Omega_\Lambda = 0.73$.
	
	Our fitting approach employs composite photoionization \citep[\texttt{CLOUDY}, see e.g.][]{ferland_cloudy_1998} and thermal models \citep[\texttt{APEC};][]{foster_updated_2012}\footnote{https://heasarc.gsfc.nasa.gov/docs/software/xspec/manual/node134.html}, along with components describing the compact nuclear AGN emission. Since previous \textit{Chandra} spectral studies have revealed the presence of hard extended emission \citep[e.g.,][for the case of ESO~428-G014]{fabbiano_discovery_2017, fabbiano_deep_2018}, we also include a spectral component to model this hard extended emission. Following \citet{fabbiano_deep_2018}, we introduce additional spectral complexity for each spectrum only when required by the fit. The spectral models used to fit the nuclear and extended regions are described below. 
	
	Due to the relatively low ($\sim$100-150 eV, R\,=\,5--50) energy resolution of ACIS-S and possible degeneracies between model components, different parameter sets often yielded statistically equivalent fits. Following \citet{fabbiano_deep_2018}, when fit results have comparable statistical quality, we selected the solutions that minimized correlated residuals. Correlated residuals suggest a bad representation of emission line-like features, particularly in the soft X-ray band. To verify that adding multiple components to minimize these residuals does not lead to arbitrary overfitting, we performed a step-by-step benchmark test on the nuclear region of NGC\,1068. As detailed in Appendix~B, our test demonstrates how the incremental introduction of photoionized components is justified to remove structured residuals, reaching a threshold where the data quality limits us from adding new model components.
	
	\subsubsection{Nuclear Spectral Spillover}
	\label{sec:spillover}
	The circular region with a radius of 1.5$''$, used to extract the nuclear spectrum, contains approximately 95\% of the hard \textit{Chandra} PSF\footnote{see Chandra Proposers’ Observatory Guide, Section 6.16, \url{cxc.harvard.edu/proposer/POG/html/chap6.html}; see also Chandra ABC Guide to Pile Up, \url{http://cxc.harvard.edu/ciao/download/doc/pileup\_abc.ps}}. This radius was chosen to maximize the coverage of the extended regions. However, in this configuration, the extended spectra may be contaminated by nuclear emission, particularly at energies $>4$~keV. This contamination is expected to affect especially the fainter cross-cone regions in AGN with brighter nuclear emission.
	
	To account for nuclear spillover contamination in the extended regions, we adopted a simplified geometrical and spectral scaling approach. The spillover from the unresolved nuclear source at the telescope aimpoint, as in the observations used here, is known to be zimuthally distributed over $360^{\circ}$ around the point source \url{} and, according to instrument calibration documentation\footnote{\url{https://cxc.harvard.edu/proposer/POG/html/chap4.htmltth\_fIg4.18}}, primarily affects the hard X-ray band due to the slightly broader PSF wings at higher energies. We therefore assumed an average spillover contribution of $6\%$ of the intrinsic nuclear flux in the hard X-ray band. Starting from the best-fit spectral model of the nuclear region, we isolated only the hard components (i.e., the \texttt{MYTorus} tables and, when required, the absorbed power-law component), as the spillover is expected to predominantly impact this energy range. For each extranuclear extraction region, the spillover contribution was modeled by multiplying the nuclear hard-component model by a constant scaling factor corresponding to the fractional azimuthal coverage of that region. In the case of four approximately symmetric regions, we adopted a scaling factor of 0.015 (i.e., one quarter of the assumed $6\%$ nuclear spillover). When the extraction regions were not azimuthally symmetric, we adjusted the scaling factors proportionally to their relative angular extent. In practice, this resulted in spillover contributions of approximately $2\%$ for the two larger regions and $1\%$ for the two smaller regions, while maintaining the assumption of a total $6\%$ spillover in the hard band.

	\subsubsection{Spectral Models for the Core/Nuclear Emission}
	\label{sec:core_emis}
	The spectral extraction regions for the core/nuclear spectra have radii of 1.5$''$ that correspond to physical sizes in the range of 60-1660~pc for the distances of the AGN in our sample. Therefore, even with \textit{Chandra}, these regions are not fully spatially resolved and will contain a mix of extended circumnuclear emission contributing to the soft spectral band and a point-like component arising from the absorbed nuclear source that only contributes at the higher energies (see e.g., \citealt{fabbiano_deep_2018, fabbiano_deep_2018-1} in the case of ESO~428-G014). We therefore fitted the nuclear spectra with a suite of models that could describe the circumnuclear excitation (photoionization and collisionally-excitation) plus models representative of the emission of CT or heavily absorbed nuclei (depending on the nuclear $N_H $, see Table~\ref{tab:log}). 
	
	The soft X-ray emission ($<$3~keV) was modeled as a combination of thermal (collisionally-ionized) and photoionized components. Collisionally ionized plasma was represented with \texttt{APEC}, with both temperature and normalization left free during fitting. The photoionized emission was modeled using \texttt{CLOUDY} c23.01 grids, adopting a broken power-law ionizing continuum consistent with previous studies \citep[e.g.,][]{kraemer_mass_2020}. The photoionizing continuum was defined using \texttt{CLOUDY}’s built-in AGN prescription (agn 6.00 -1.40 -0.50 -1.), corresponding to a standard AGN spectral energy distribution parameterization. The hydrogen density of the photoionized X-ray emitting gas was fixed at $\log n_{\rm H}=5$~cm$^{-3}$. The ionization parameter was gridded over the range $-2 \le \log U \le 4$ in steps of 0.25~dex, and the absorption column density over $19 \le \log (N_{\rm H}/\rm cm^{2}) \le 23.5$ in steps of 0.1~dex. The normalization of each photoionized component was allowed to vary during spectral fitting. The \texttt{XSPEC} additive tables were generated with \texttt{CLOUDY} including both the transmitted and reflected spectra, similar to \citet[][]{trindade_falcao_deep_2024}. Each model component therefore self-consistently includes the reflected continuum together with the associated fluorescent and recombination line emission produced at the illuminated face of the cloud, accounting for the scattered hard X-ray continuum and the fluorescent Fe K$\alpha$ line without the need for an additional reflection component such as \texttt{pexrav} \citep[][]{Magdziarz1995}. This differs from our previous works \citep[e.g.][]{Pepi2018}, where \texttt{CLOUDY} was used only to model the photoionized emission below $\sim$2.5 keV, while the reflection continuum was described separately with \texttt{pexrav}.
	
	Nuclear spectra were modeled with the \texttt{MYTorus}\footnote{\url{https://www.mytorus.com/}} model \citep[][]{murphy_x-ray_2009}. The reflected continuum and fluorescent Fe~K$\alpha$ and Fe~K$\beta$ lines were represented by the \texttt{MYTorusS} and \texttt{MYTorusL} components, respectively. We adopted the default toroidal geometry with a half-opening angle of $60^{\circ}$, corresponding to a fixed covering factor. The normalizations of the scattered continuum and line components were tied, with a multiplicative constant included to allow relative scaling. A single column density was assumed for both the reflected continuum and fluorescent lines, but was left free to vary during the fitting. 
	
	For the low $N_{\rm H}$ AGN (see Table\,\ref{tab:log}) NGC\,5252, the data required the inclusion of an additional absorbed power-law component to represent the transmitted fraction of the nuclear continuum, in addition to the \texttt{MYTorusS} and \texttt{MYTorusL} components. The column density of this transmitted component was allowed to vary independently from that of the reflection components, consistent with a decoupled torus geometry. We adopted an edge-on line of sight (inclination angle of 90$^{\circ}$) for this forward-scattering configuration.
	
	Given the limited 0.3-7~keV bandpass and to reduce parameter degeneracies, the photon index of the primary continuum, and hence the one used in the model components accounting for the reflected flux, was fixed to $\Gamma = 1.8$. This choice is consistent with the mean slope measured in large AGN samples \citep[e.g.,][]{piconcelli_xmm-newton_2005, bianchi_caixa_2009, matzeu_supermassive_2023}.
	
	For NGC~4945, the 6-7~keV energy range shows a strong emission line at 6.7~keV, in addition to the neutral 6.4 keV Fe K$\alpha$ line, generally ascribable to highly ionized iron Fe~XXV. In this case, we fitted a Gaussian (\texttt{zGauss} in \textsc{XSPEC}) to the underlying test model. The line's energy centroid and its normalization were fitted, while its corresponding $\sigma$ was kept frozen to a value of 50~eV.

	\subsubsection{Spectral Models for the Extended Emission}
	As in the case of the nuclear spectra (Section~\ref{sec:core_emis}), we used photoionization and thermal models to describe the extended emission. The soft ($<$3 keV) spectra are generally well modelled with \texttt{APEC} and \texttt{CLOUDY} components. The hard ($>$3 keV) spectra are generally well fitted with CLOUDY components. In two cases (see Section 4.2), we added a \texttt{pexrav} reflection component \citep{magdziarz_angle-dependent_1995}, which describes a cut-off power-law spectrum reflected from neutral, slab-like material \citep[see e.g.,][]{trindade_falcao_discovery_2024}. This model allows us to account for both the continuum and the reflected flux, or specifically the latter. We used \texttt{pexrav} to model only the reflection component by fixing the reflection scaling factor at R=-1, thereby excluding any contribution from the direct continuum. Only the model normalization was fitted. For consistency with the nuclear continuum (see Section~\ref{sec:core_emis}), the photon index was fixed at $\Gamma\,=\,1.8$, and the high-energy cutoff was set to 300~keV, well outside the {\it Chandra} bandpass. Solar abundances were assumed, and the inclination angle of the reflecting slab was fixed at $60^{\circ}$.
	
	When required by the data, we included a narrow ($\sigma=50$~eV) Gaussian component to account for the neutral Fe~K$\alpha$ line when detected in the extended emission. The line energy was fixed at 6.4~keV. An additional narrow Gaussian component was added when the data showed evidence for more ionized Fe species, such as Fe~XXV at a rest energy of 6.7~keV. For the Fe~K$\alpha$ line, only the normalization was allowed to vary, whereas for the second Gaussian component, both the line centroid and the normalization were free parameters.
	
	\subsubsection{Luminosities}
	\label{sec:lumin}
	Intrinsic (unabsorbed) X-ray luminosities were computed using the convolution model \texttt{clumin} in \texttt{XSPEC}, which integrates the model flux over specified energy ranges and applies the appropriate redshift correction. Throughout this work, we report component luminosities in the 0.3-7~keV band. We note that \texttt{clumin} returns the intrinsic model luminosity after accounting for both Galactic and intrinsic absorption, as defined by the adopted spectral components. Consequently, the reported luminosities represent the unabsorbed emission from each spectral component and are directly comparable across models and energy ranges. Luminosities were calculated for all spectral components. The \texttt{MYTorus} component provides the contribution from the scattered and fluorescent nuclear emission; however, the intrinsic nuclear luminosity itself is not accessible within the \textit{Chandra} energy band (0.3-7~keV). Estimating this luminosity requires observations at higher energies, such as those provided by \textit{NuSTAR}.
	
	\begin{figure}[t]
\centering
\includegraphics[width=0.95\textwidth]{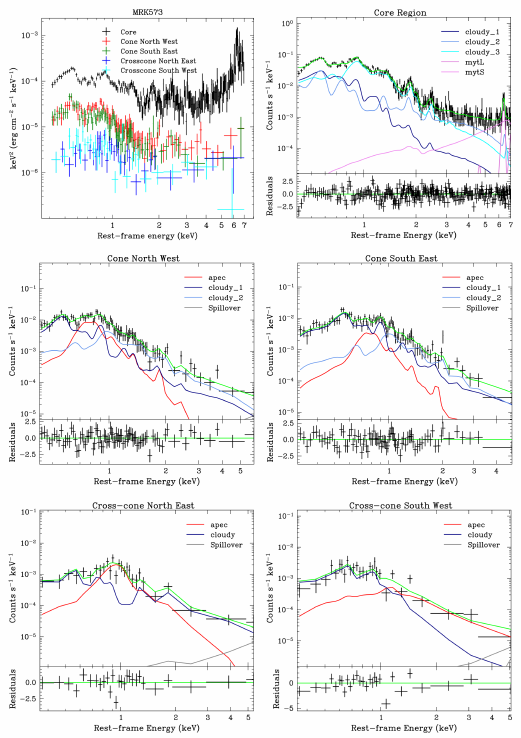}
\caption{Top-left panel: Unfolded \textit{Chandra} spectra extracted from the regions defined in the corresponding panels of Figure~3. The remaining panels show the individual spectra and corresponding best-fit models for the various regions.}
\label{fig:mrk573_spec}
\end{figure}

\begin{figure}[t]
\centering
\includegraphics[width=0.95\textwidth]{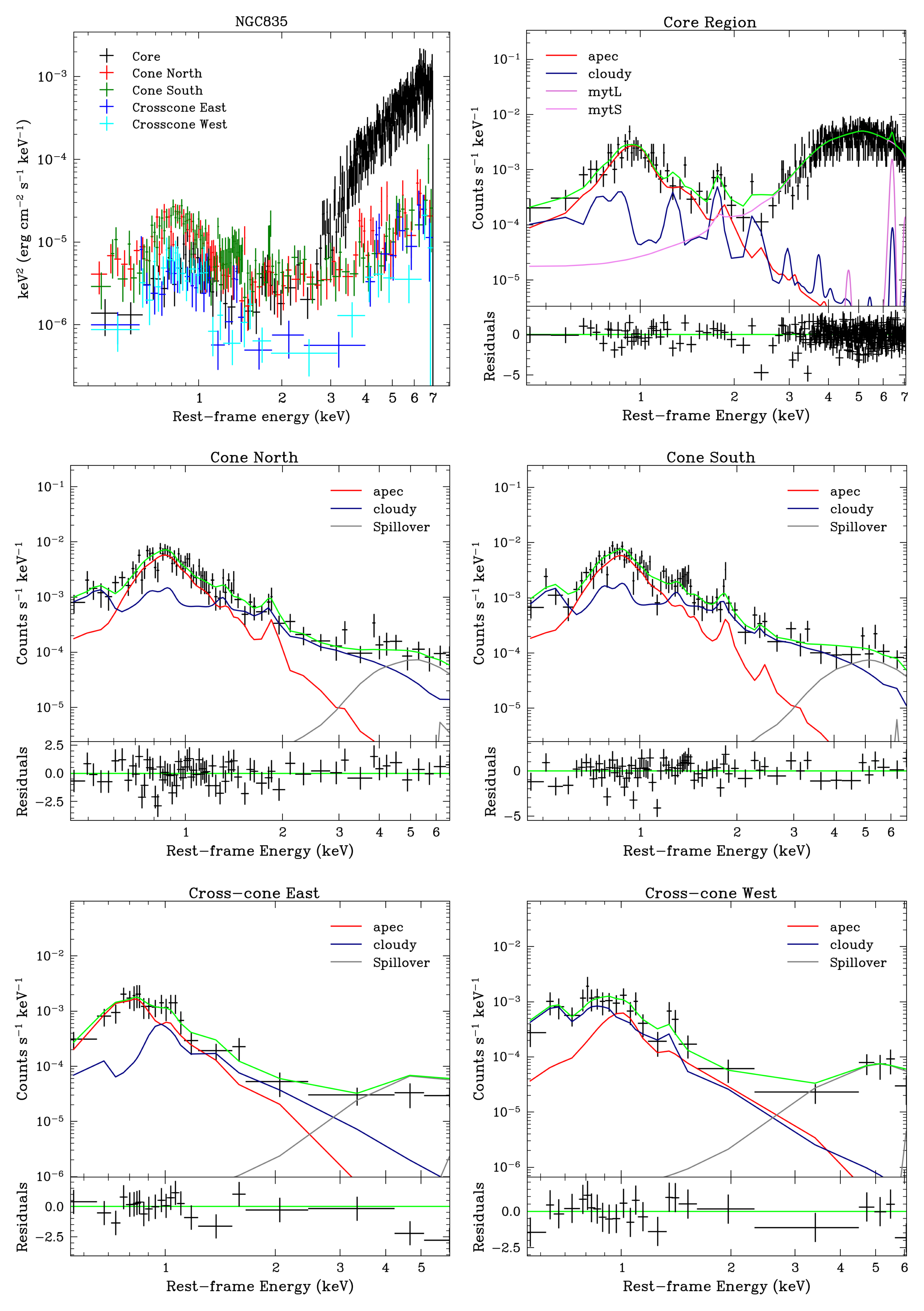}
\caption{Same as in Fig.~8, spectra for NGC~835.}
\label{fig:ngc835_spec}
\end{figure}

\begin{figure}[t]
\centering
\includegraphics[width=0.95\textwidth]{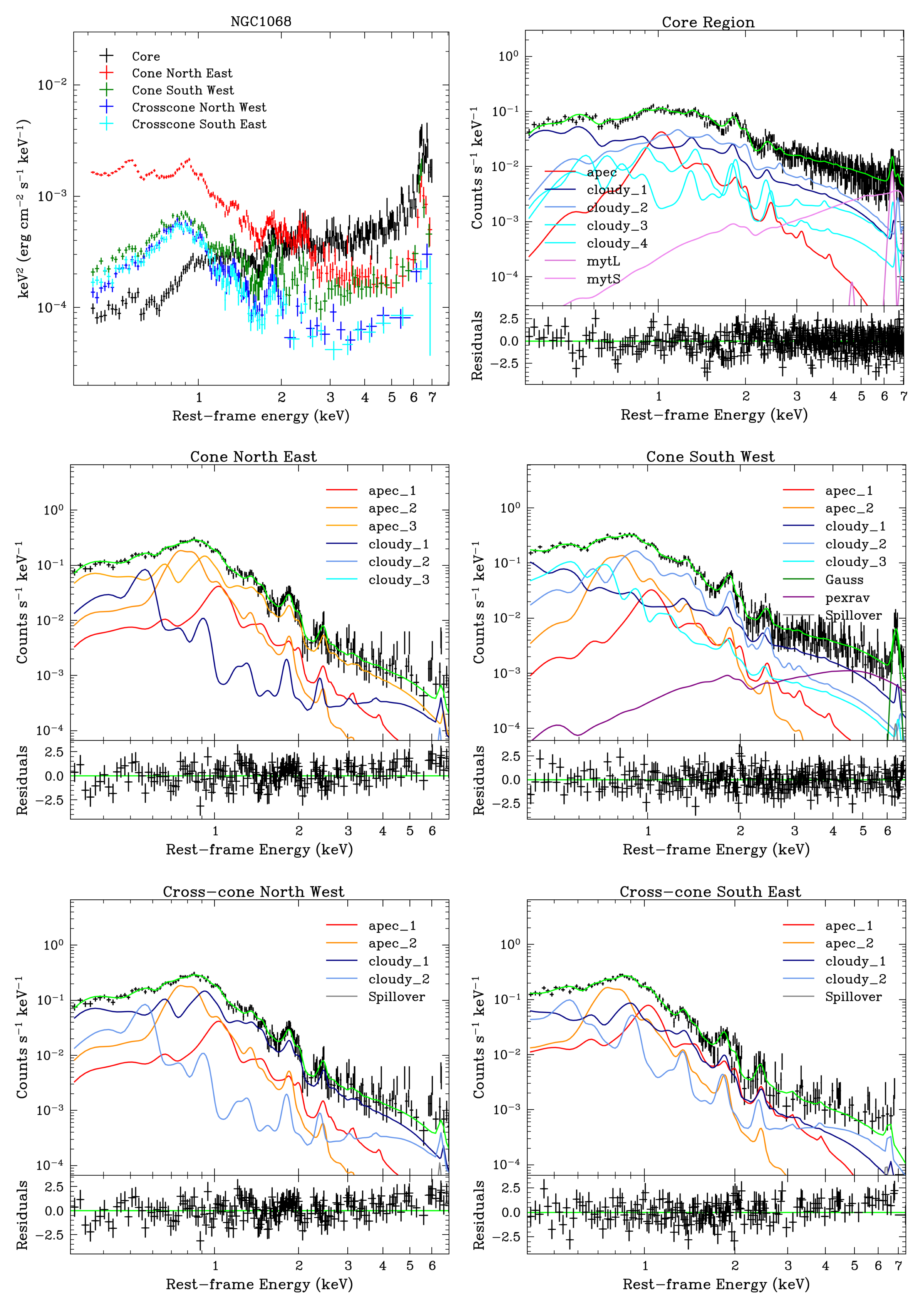}
\caption{Same as in Fig.~8, spectra for NGC~1068.}
\label{fig:ngc1068_spec}
\end{figure}

\begin{figure}[t]
\centering
\includegraphics[width=0.95\textwidth]{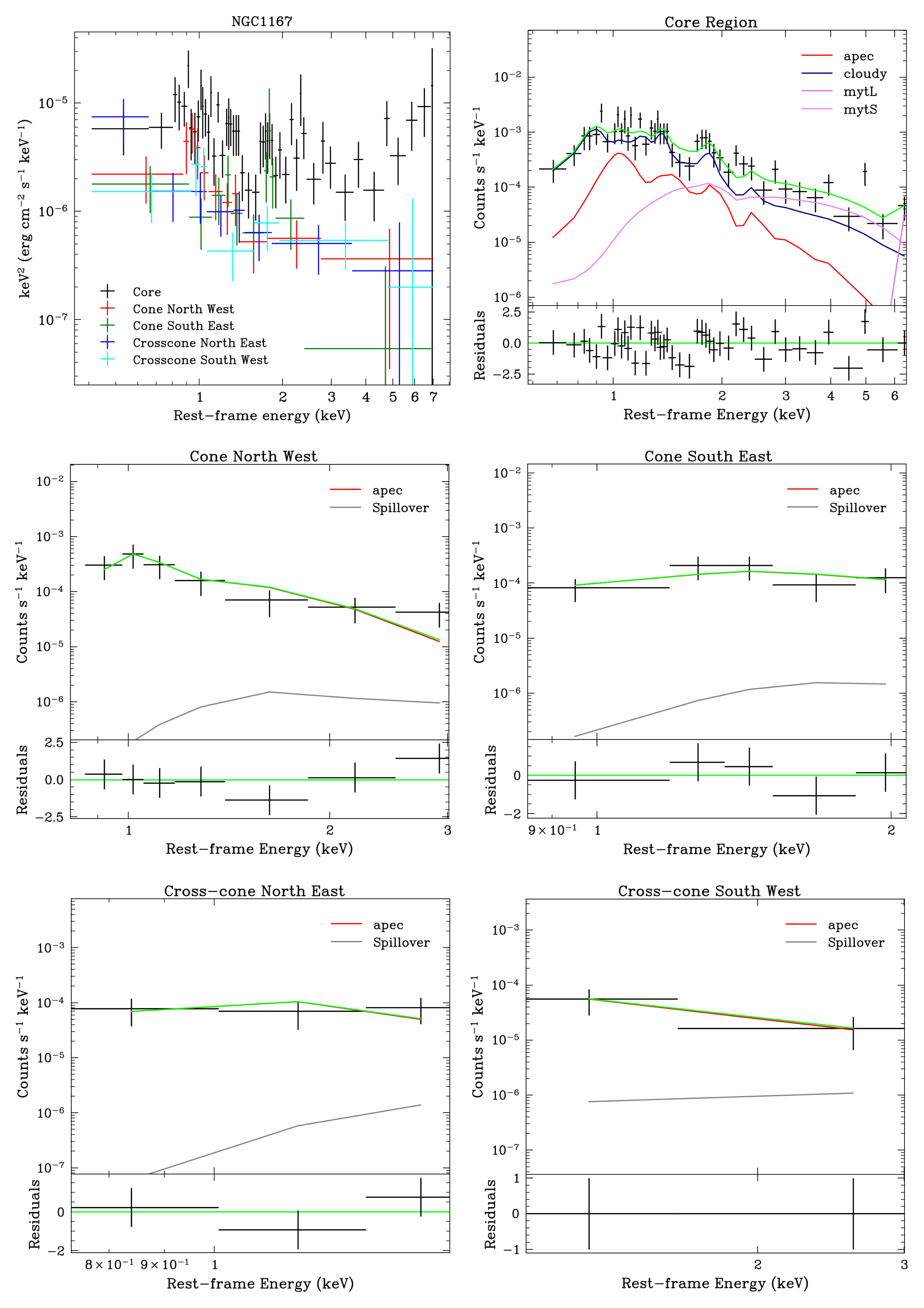}
\caption{Same as in Fig.~8, spectra for NGC~1167.}
\label{fig:ngc1167_spec}
\end{figure}

\begin{figure}[t]
\centering
\includegraphics[width=0.95\textwidth]{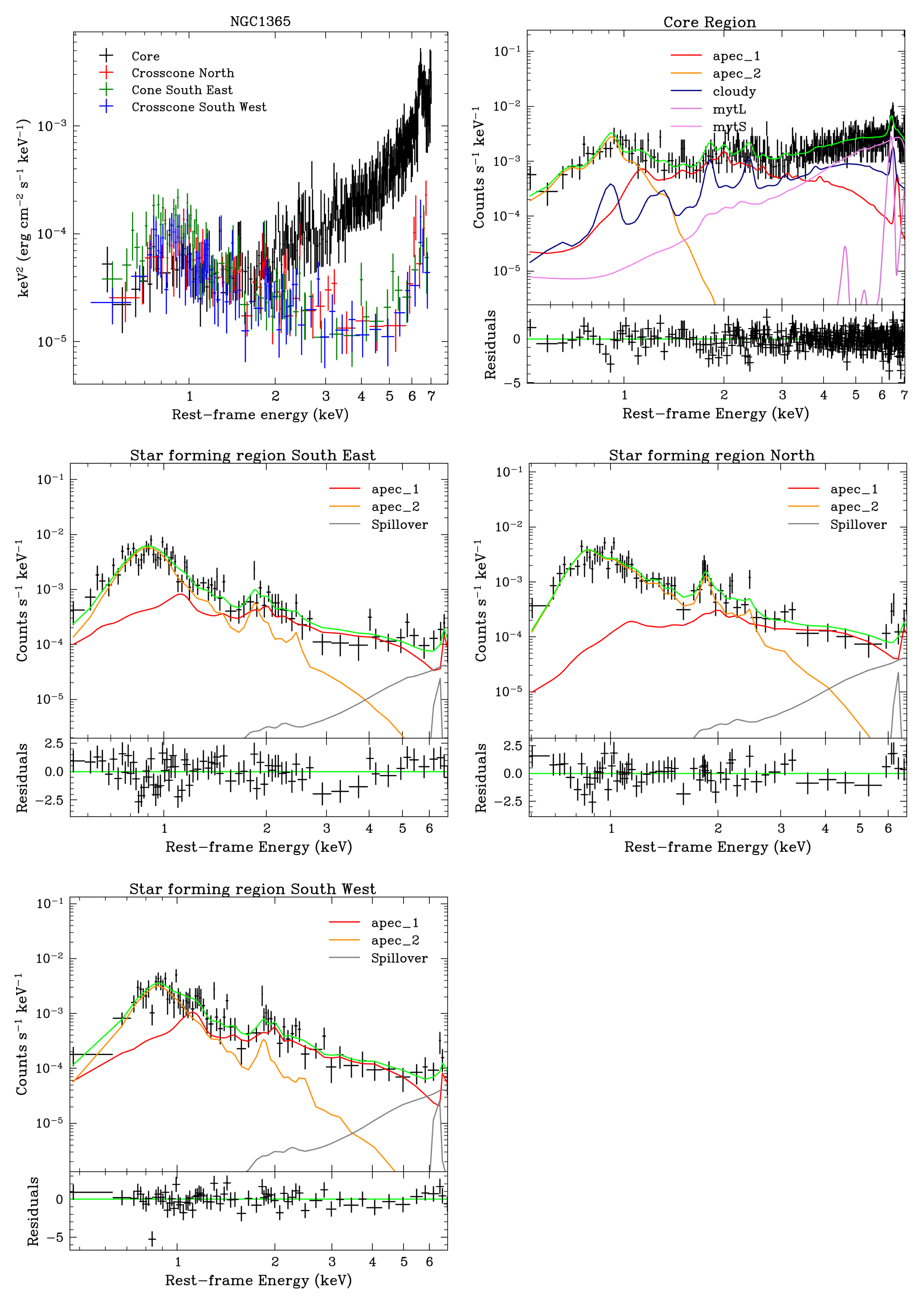}
\caption{Same as in Fig.~8, spectra for NGC~1365.}
\label{fig:ngc1365_spec}
\end{figure}

\begin{figure}[t]
\centering
\includegraphics[width=0.95\textwidth]{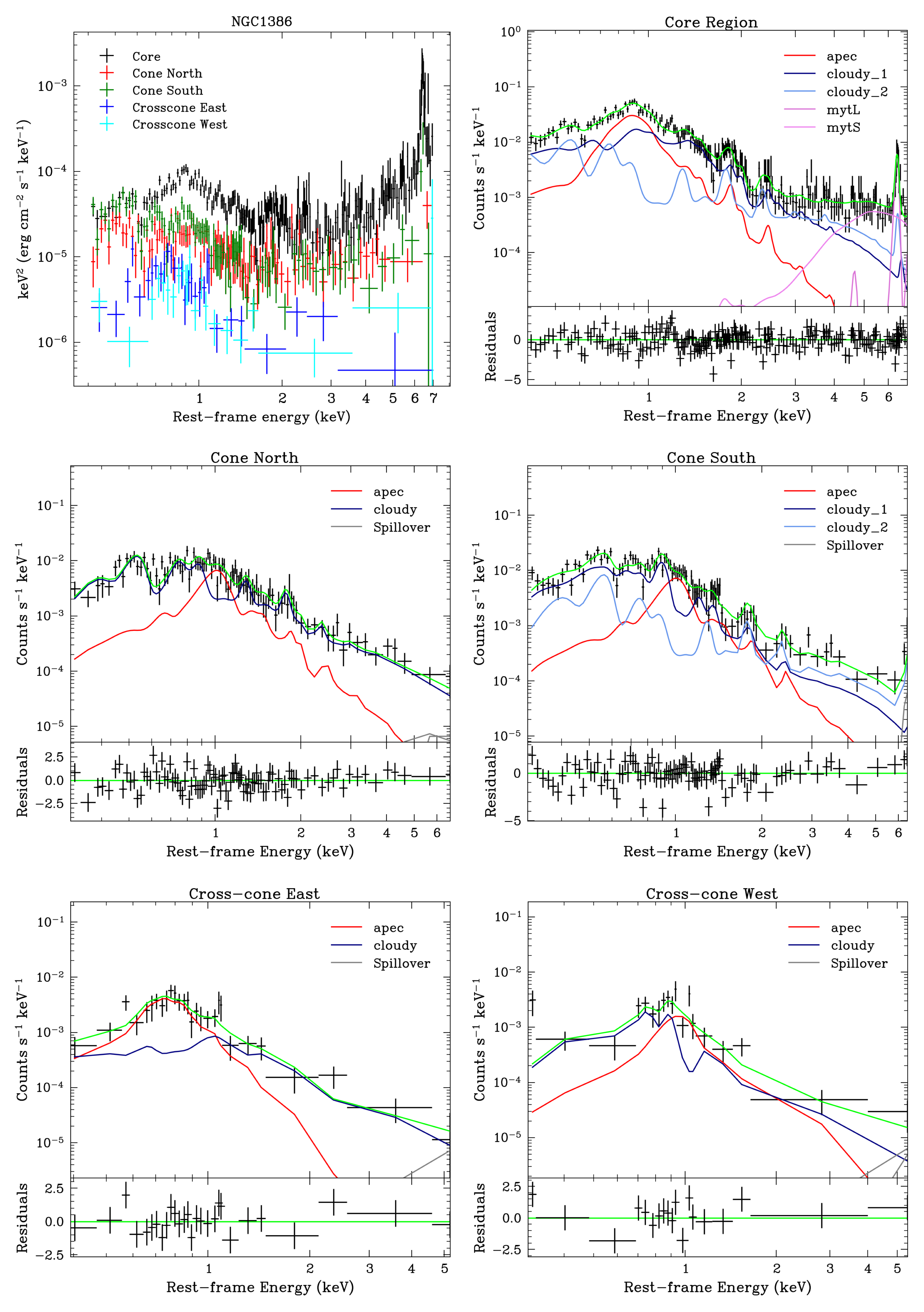}
\caption{Same as in Fig.~8, spectra for NGC~1386.}
\label{fig:ngc1386_spec}
\end{figure}

\begin{figure}[t]
\centering
\includegraphics[width=0.95\textwidth]{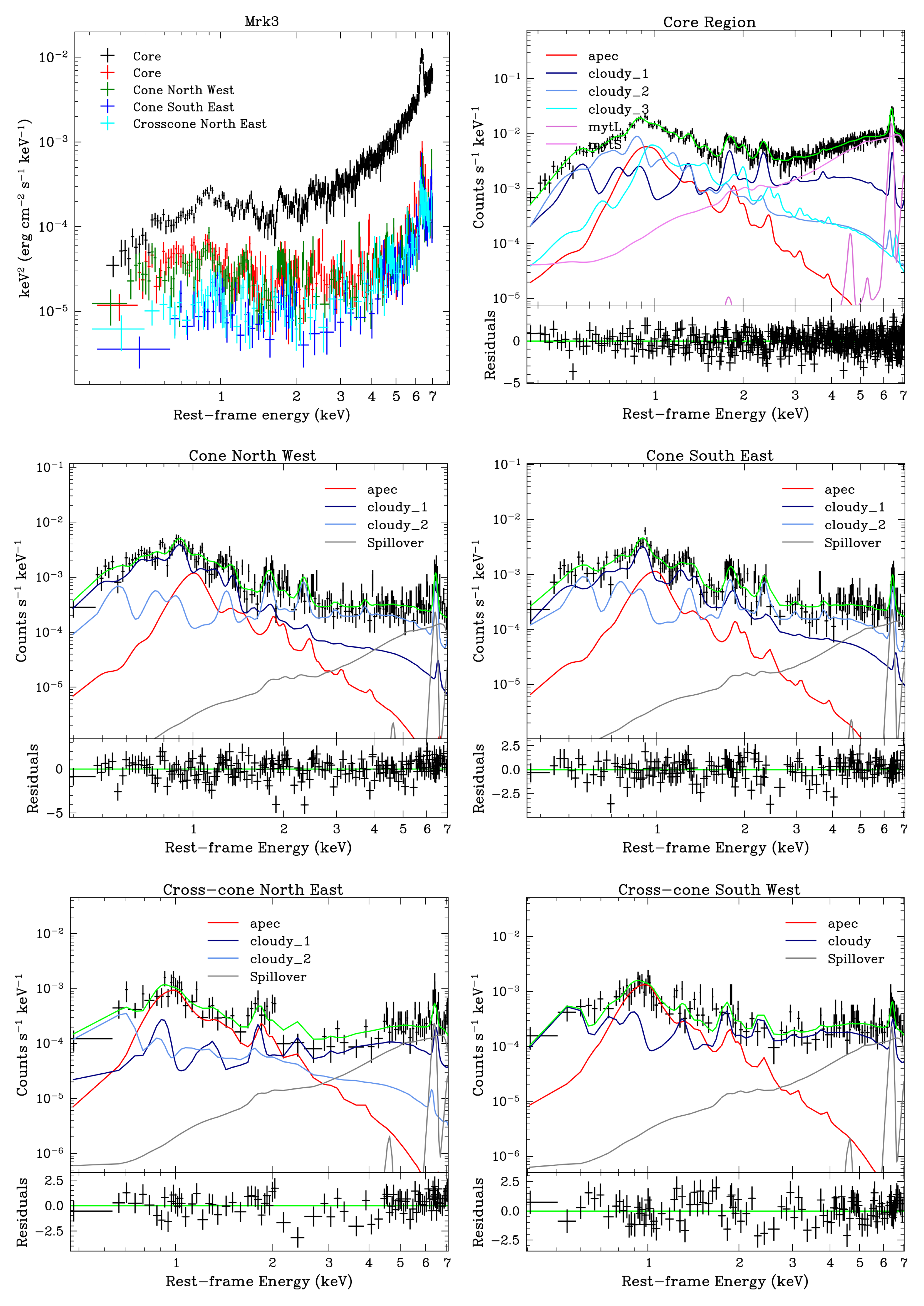}
\caption{Same as in Fig.~8, spectra for Mrk~3.}
\label{fig:mrk3_spec}
\end{figure}

\begin{figure}[t]
\centering
\includegraphics[width=0.95\textwidth]{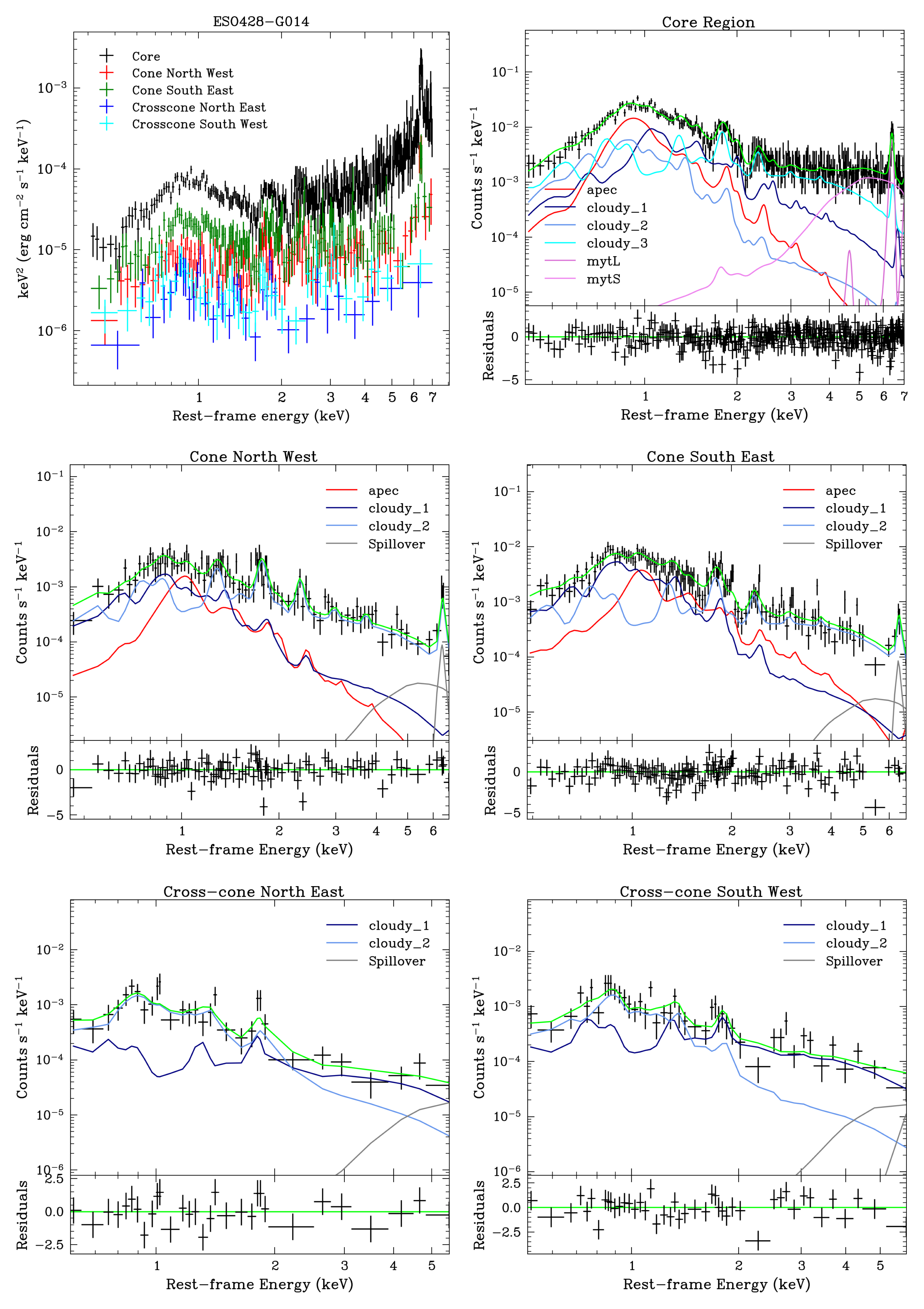}
\caption{Same as in Fig.~8, spectra for ESO~428-G014.}
\label{fig:eso428_spec}
\end{figure}

\begin{figure}[t]
\centering
\includegraphics[width=0.95\textwidth]{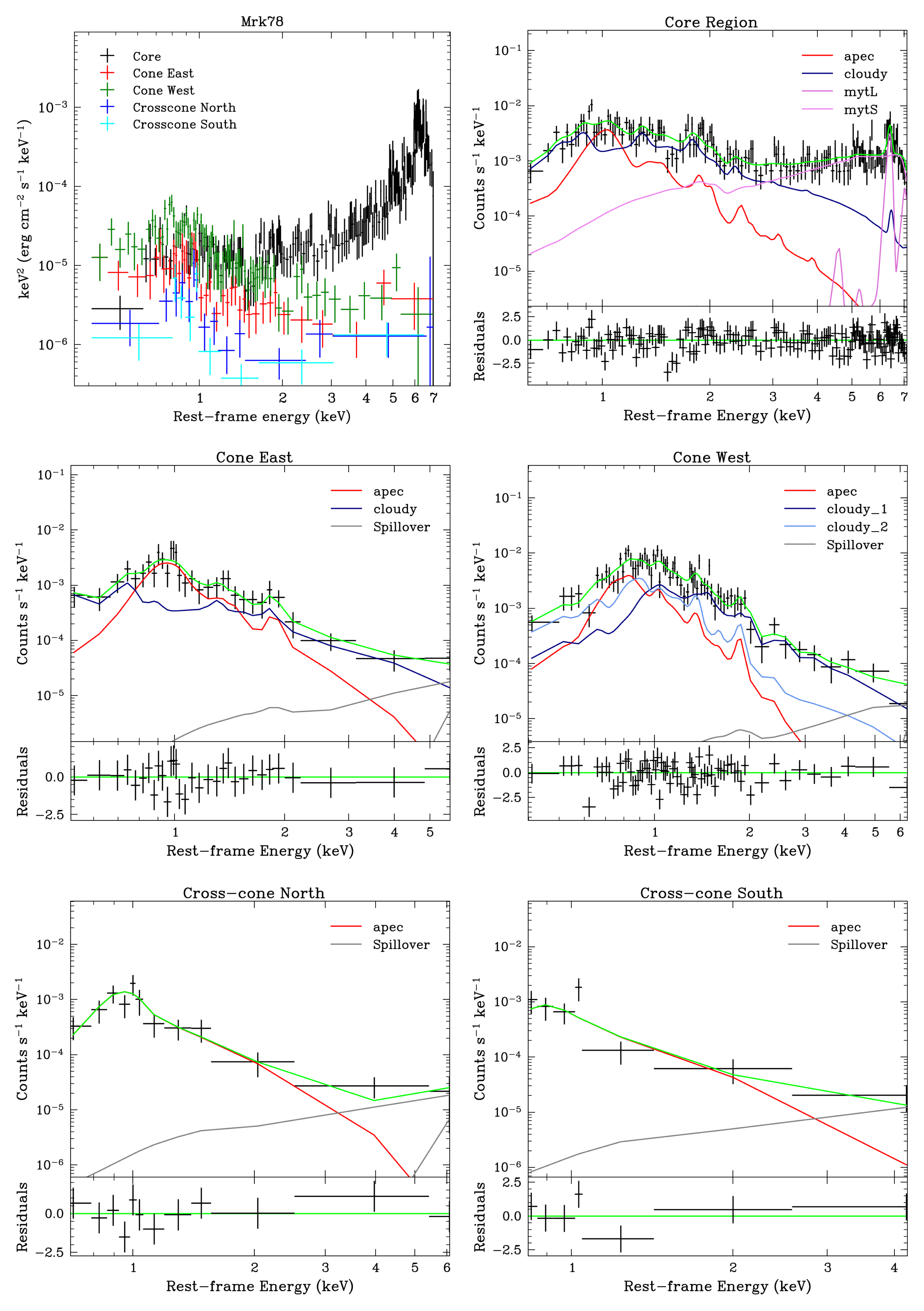}
\caption{Same as in Fig.~8, spectra for Mrk~78.}
\label{fig:mrk78_spec}
\end{figure}

\begin{figure}[t]
\centering
\includegraphics[width=0.95\textwidth]{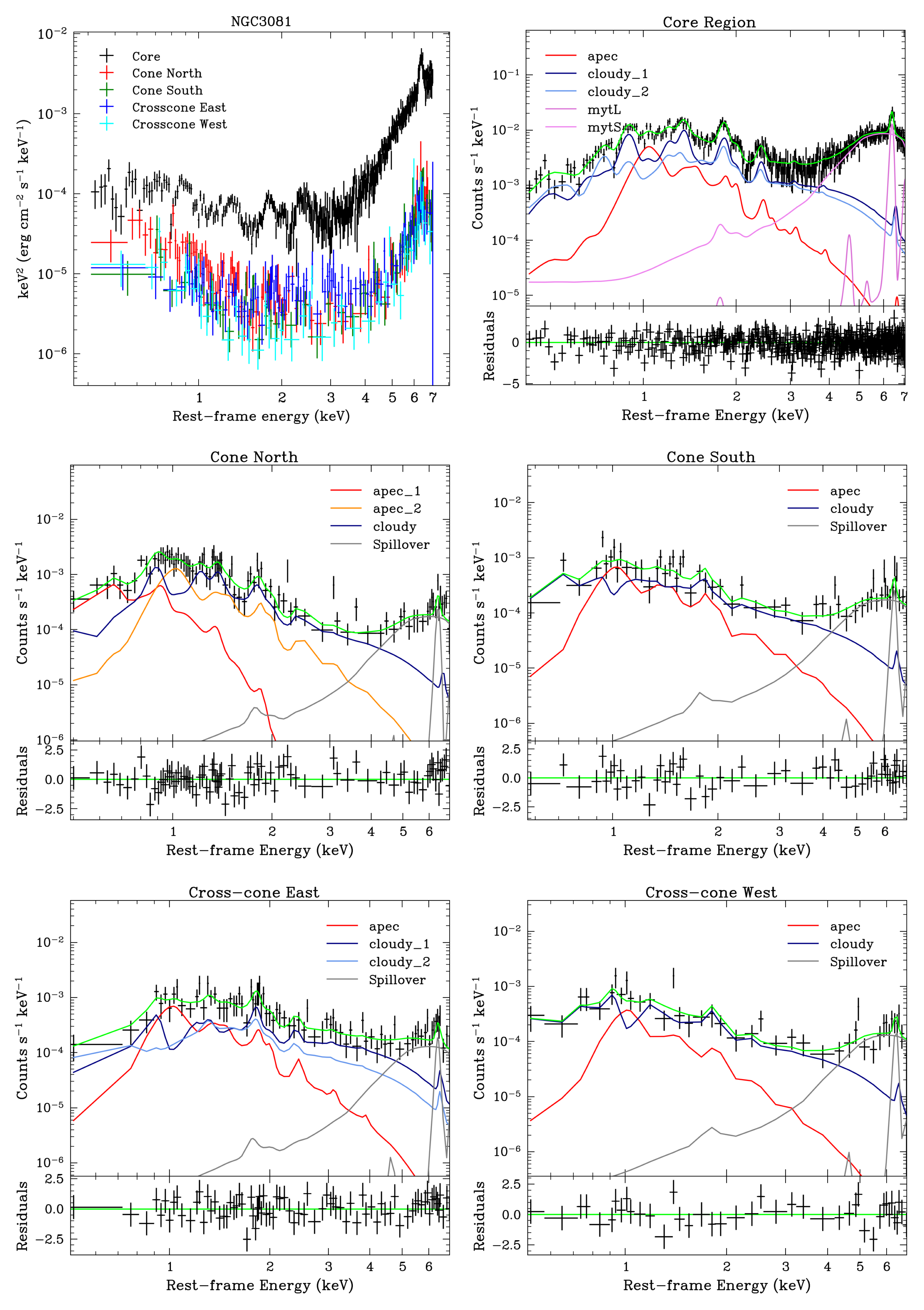}
\caption{Same as in Fig.~8, spectra for NGC~3081.}
\label{fig:ngc3081_spec}
\end{figure}

\begin{figure}[t]
\centering
\includegraphics[width=0.95\textwidth]{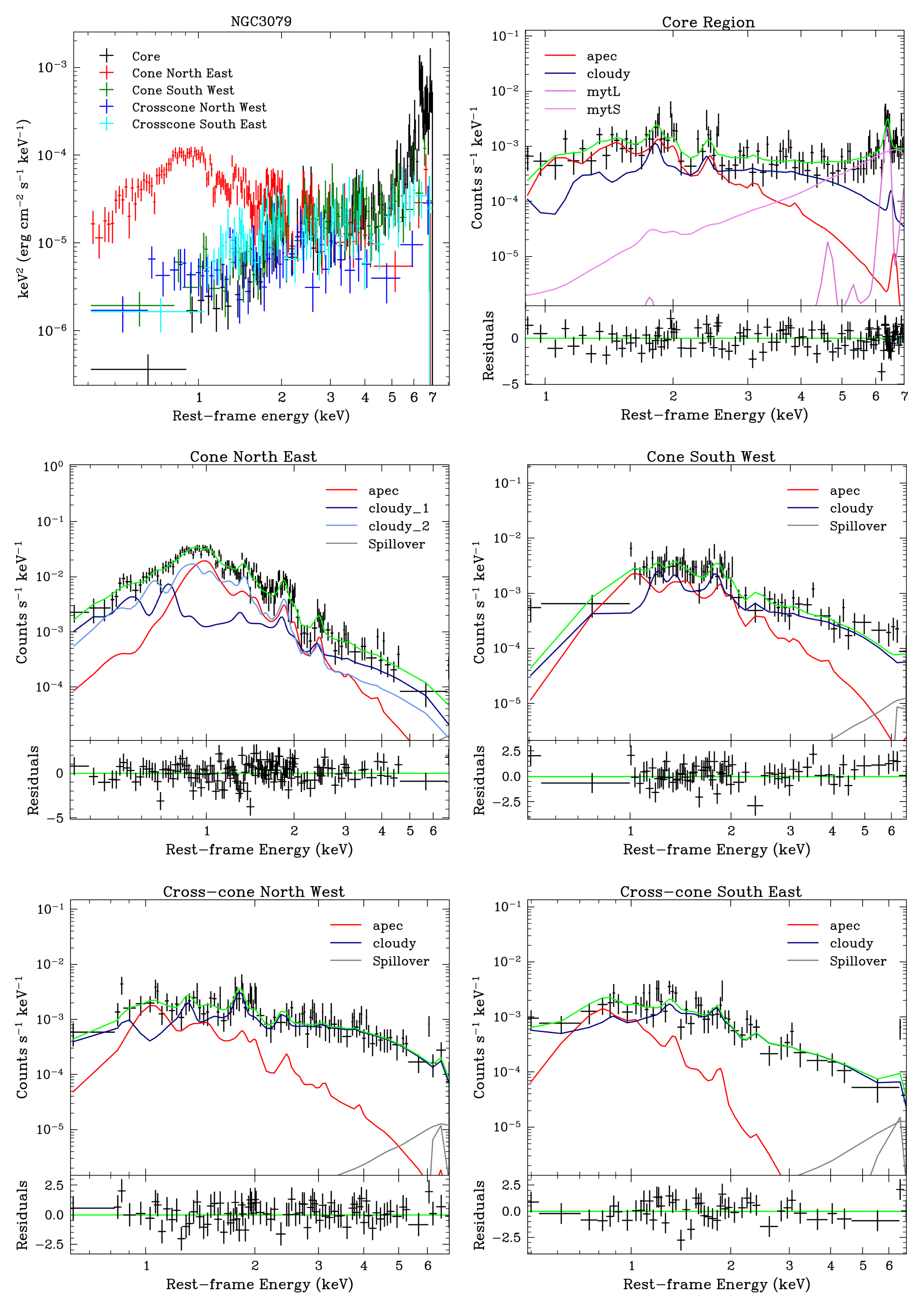}
\caption{Same as in Fig.~8, spectra for NGC~3079.}
\label{fig:ngc3079_spec}
\end{figure}

\begin{figure}[t]
\centering
\includegraphics[width=0.95\textwidth]{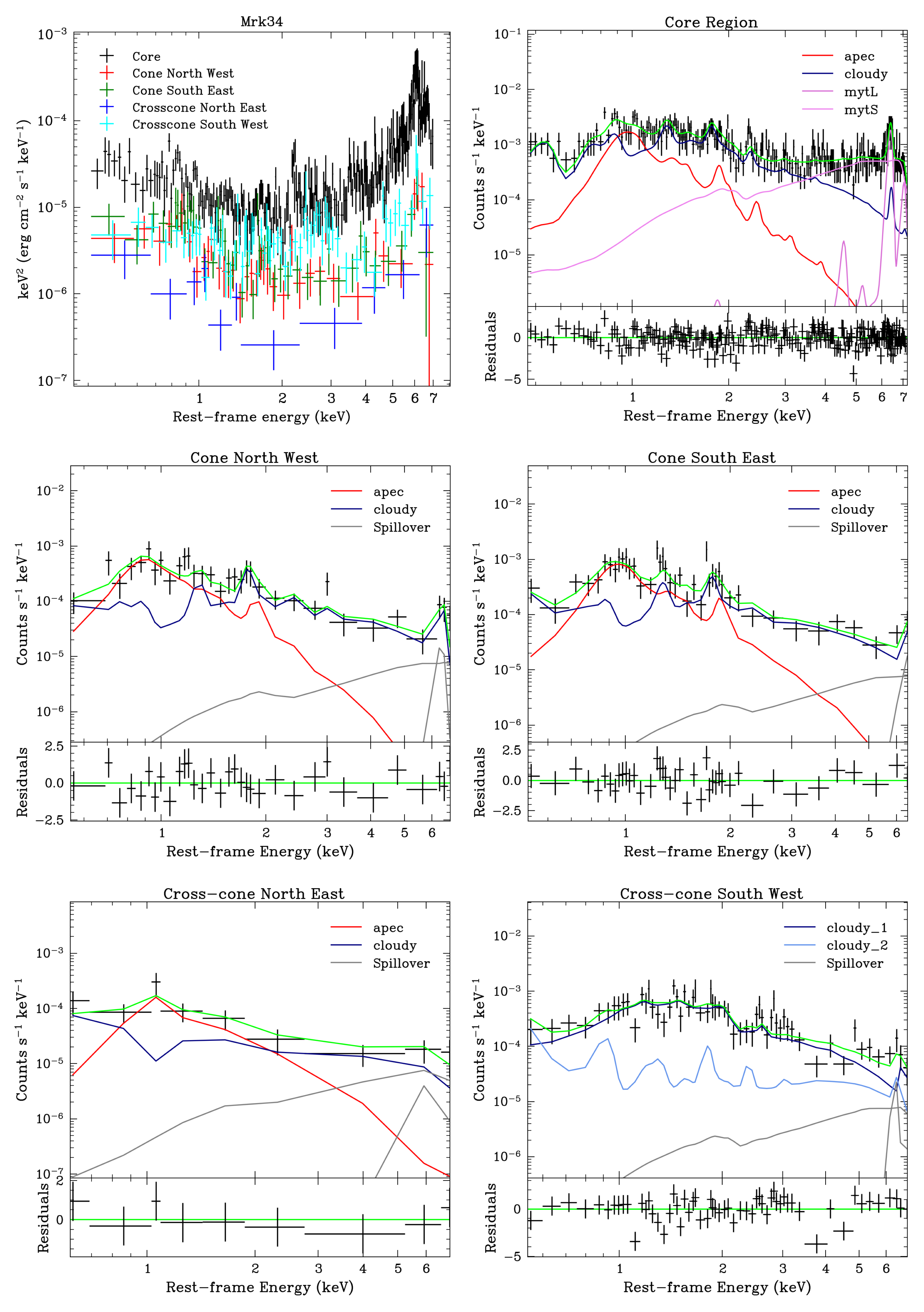}
\caption{Same as in Fig.~8, spectra for Mrk~34.}
\label{fig:mrk34_spec}
\end{figure}

\begin{figure}[t]
\centering
\includegraphics[width=0.95\textwidth]{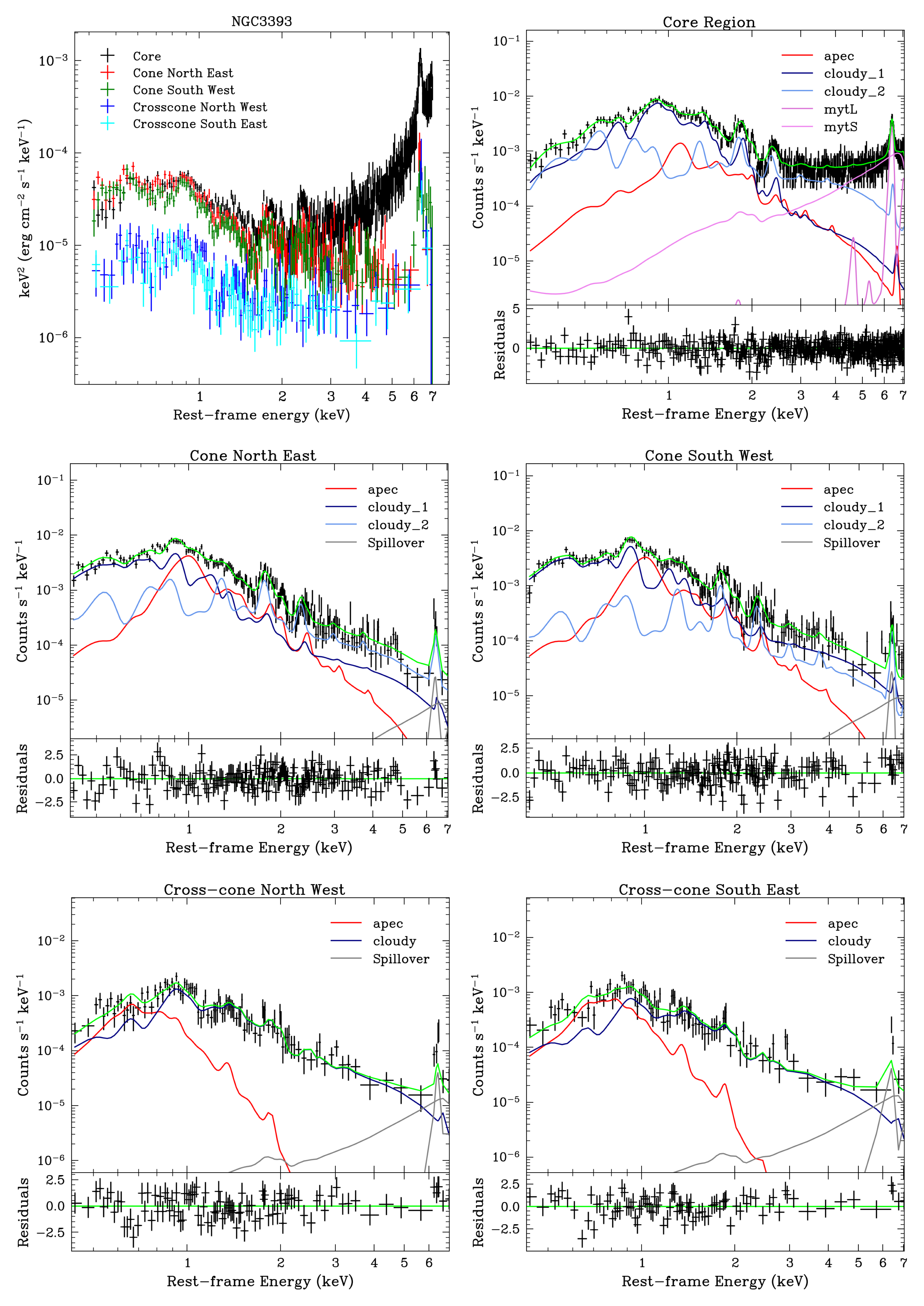}
\caption{Same as in Fig.~8, spectra for NGC~3393.}
\label{fig:ngc3393_spec}
\end{figure}

\begin{figure}[t]
\centering
\includegraphics[width=0.95\textwidth]{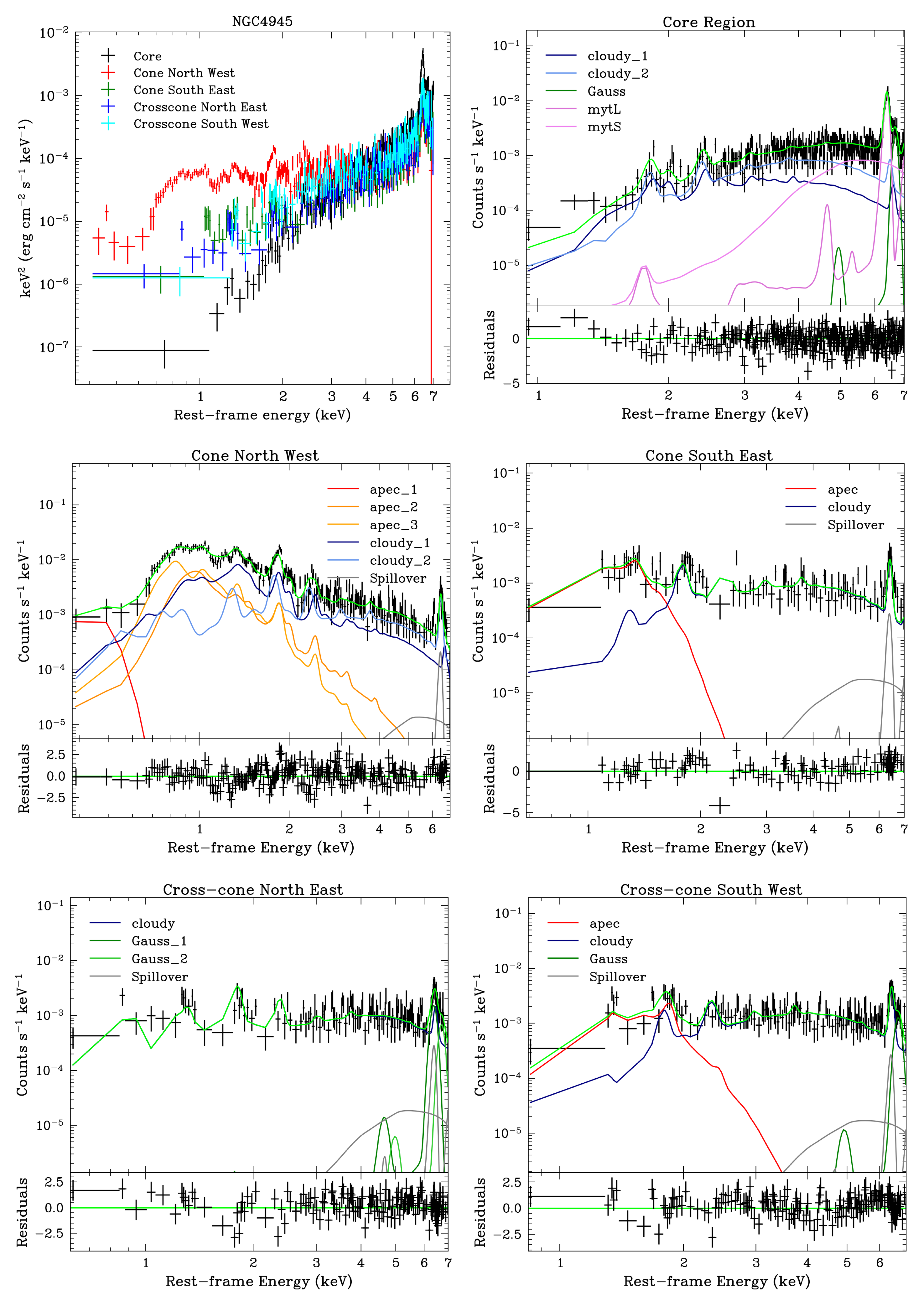}
\caption{Same as in Fig.~8, spectra for NGC~4945.}
\label{fig:ngc4945_spec}
\end{figure}

\begin{figure}[t]
\centering
\includegraphics[width=0.95\textwidth]{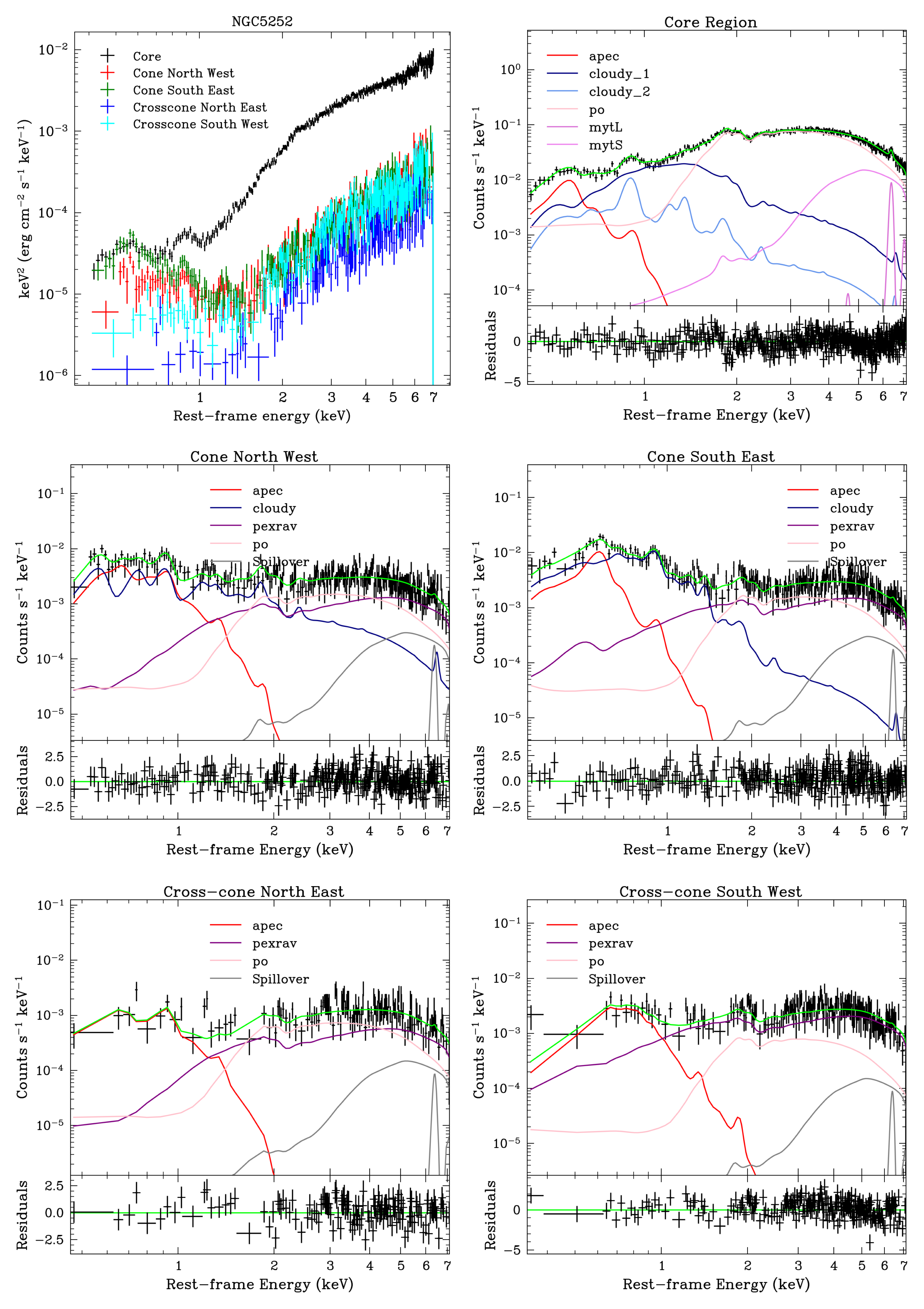}
\caption{Same as in Fig.~8, spectra for NGC~5252.}
\label{fig:ngc5252_spec}
\end{figure}

\begin{figure}[t]
\centering
\includegraphics[width=0.95\textwidth]{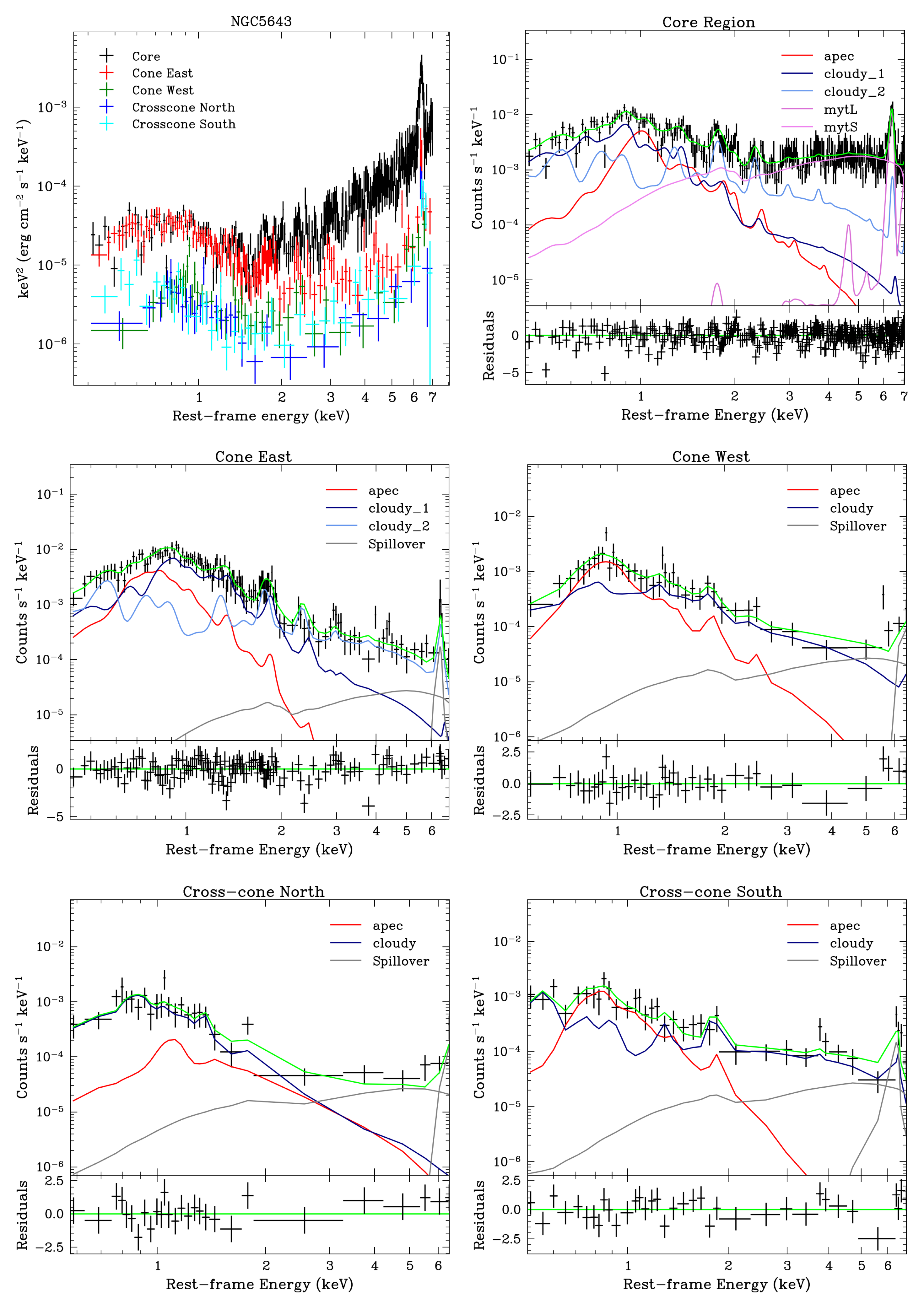}
\caption{Same as in Fig.~8, spectra for NGC~5643.}
\label{fig:ngc5643_spec}
\end{figure}

\begin{figure}[t]
\centering
\includegraphics[width=0.95\textwidth]{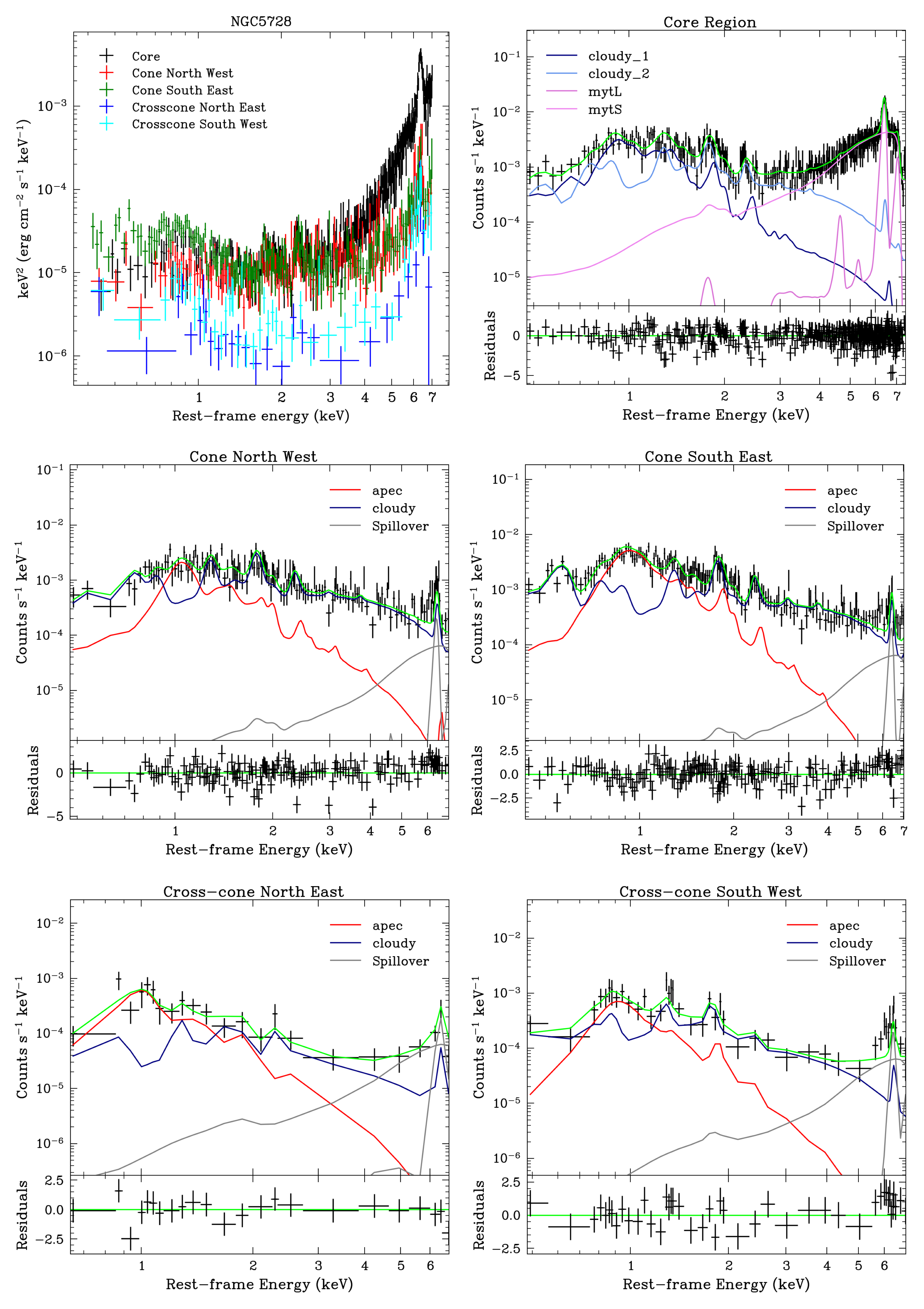}
\caption{Same as in Fig.~8, spectra for NGC~5728.}
\label{fig:ngc5728_spec}
\end{figure}

\begin{figure}[t]
\centering
\includegraphics[width=0.95\textwidth]{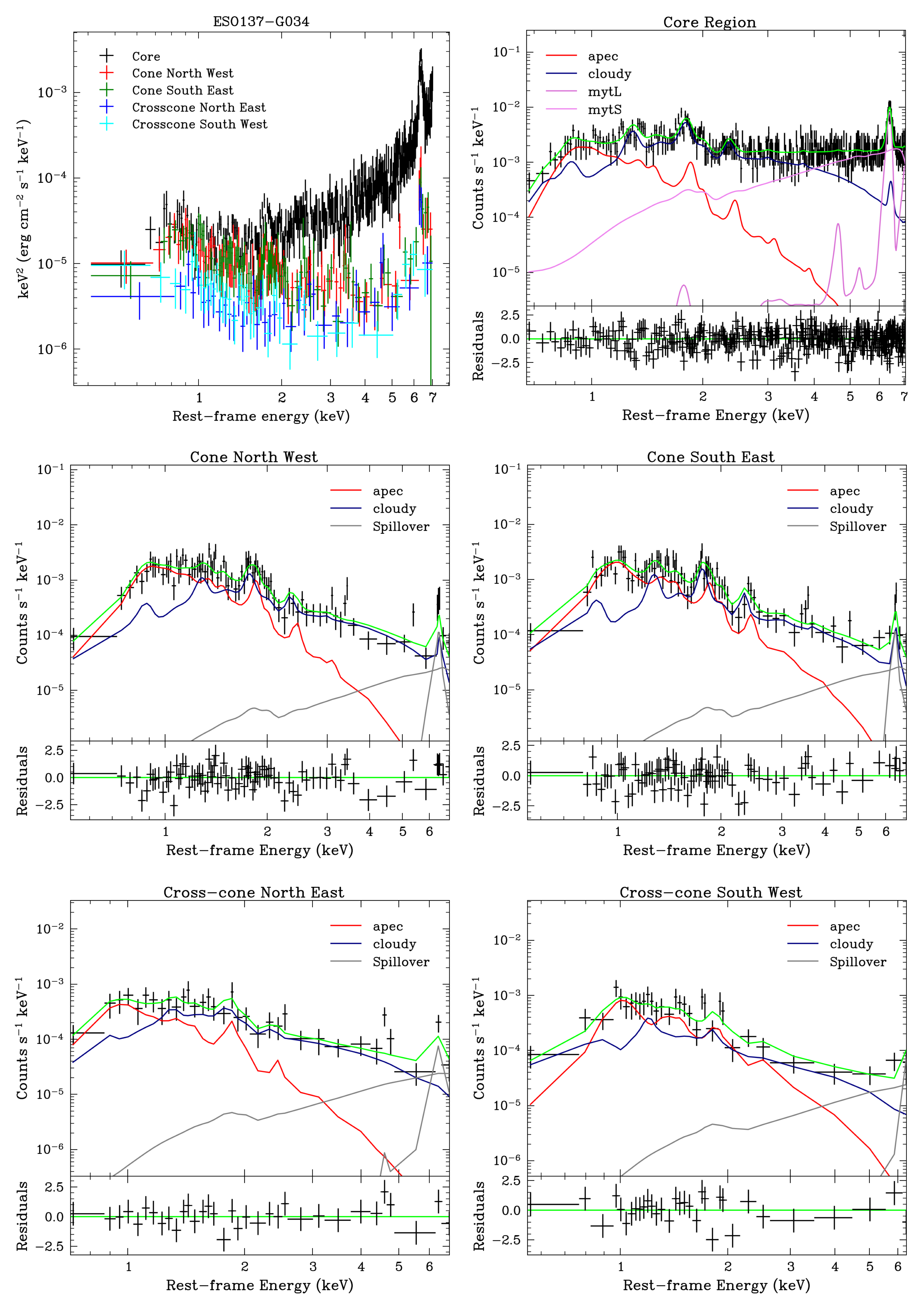}
\caption{Same as in Fig.~8, spectra for ESO~137-G034.}
\label{fig:eso137_spec}
\end{figure}

\begin{figure}[t]
\centering
\includegraphics[width=0.95\textwidth]{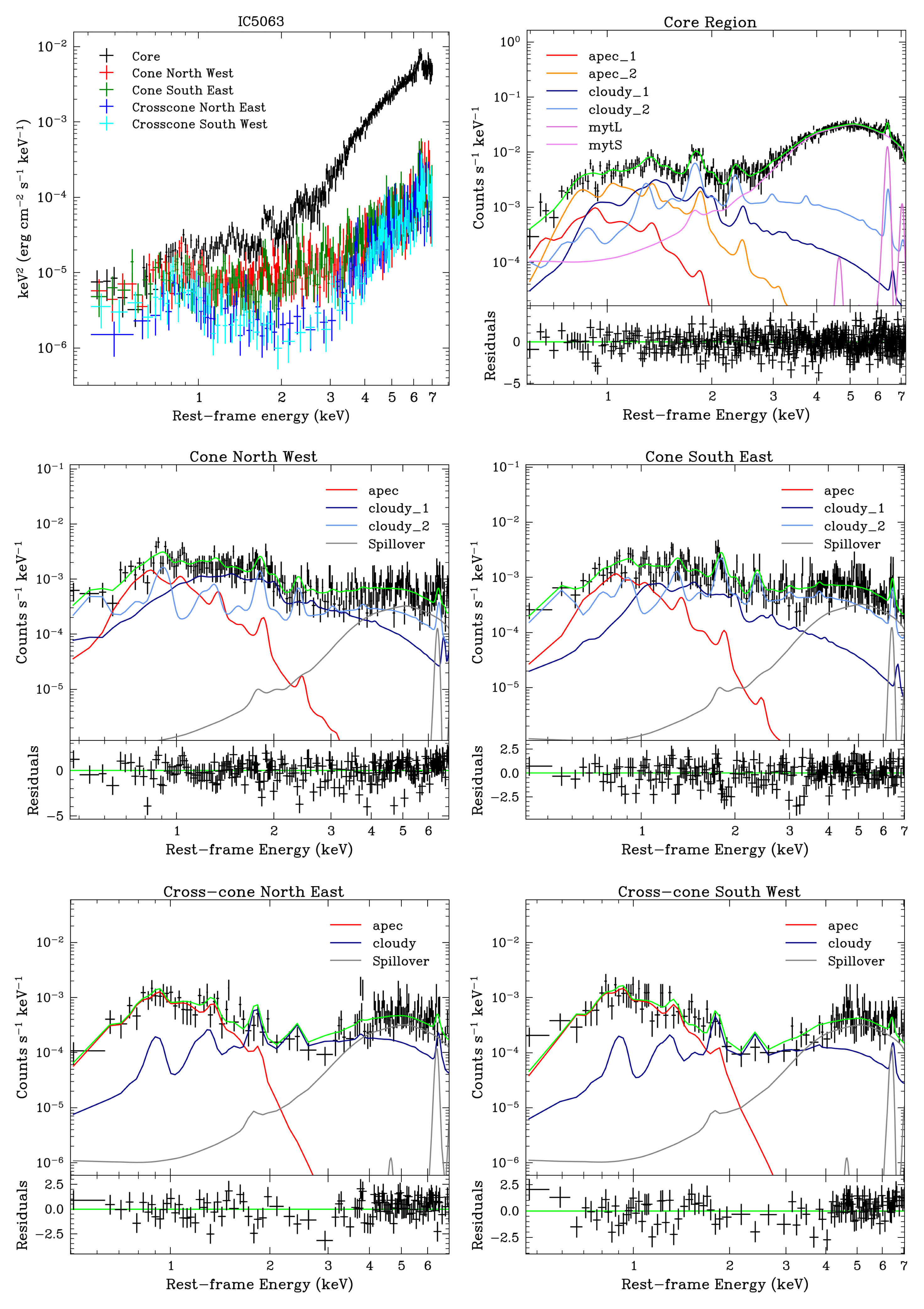}
\caption{Same as in Fig.~8, spectra for IC~5063.}
\label{fig:ic5063_spec}
\end{figure}

\begin{figure}[t]
\centering
\includegraphics[width=0.95\textwidth]{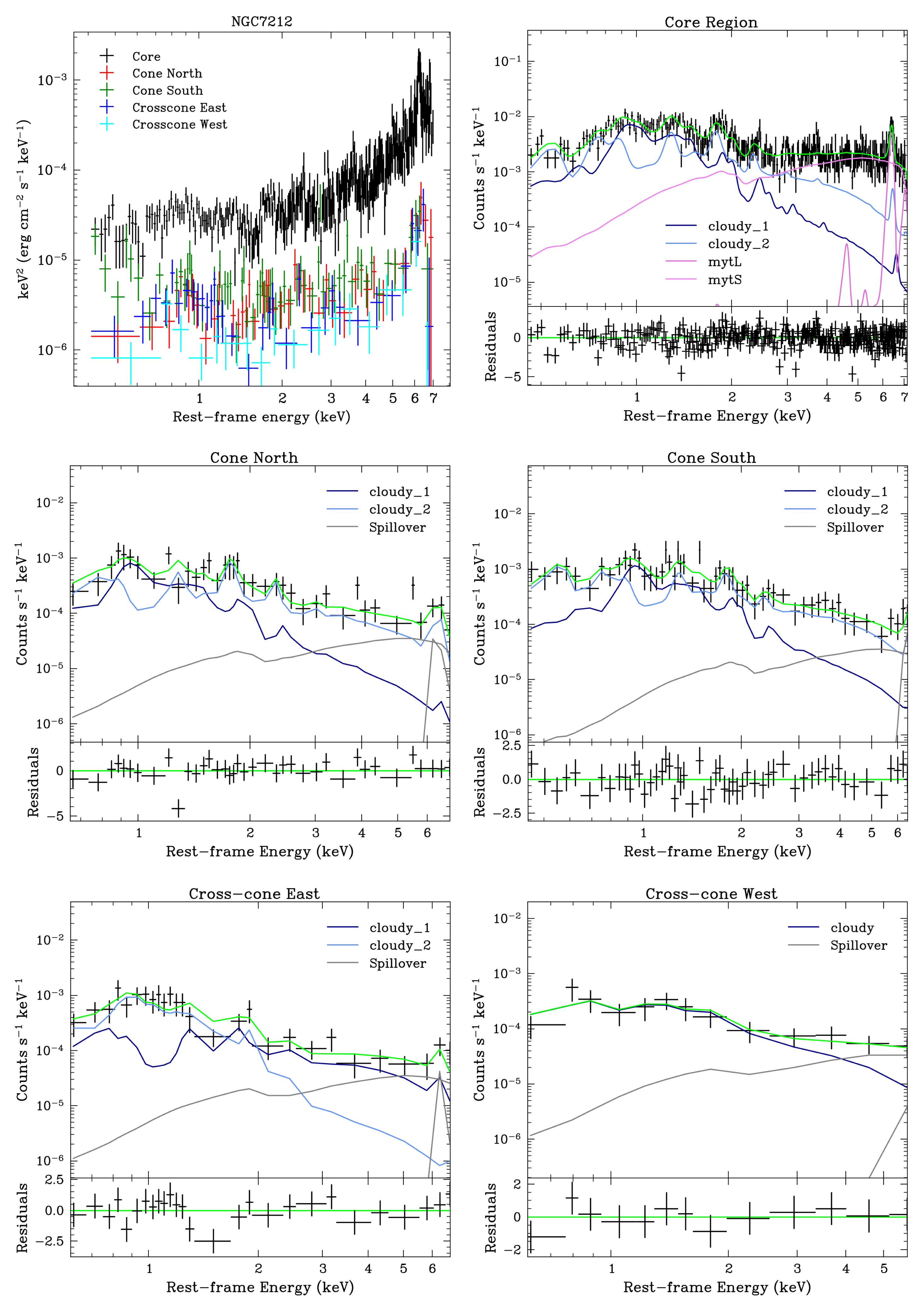}
\caption{Same as in Fig.~8, spectra for NGC~7212.}
\label{fig:ngc7212_spec}
\end{figure}

\begin{table}
\centering
\begin{minipage}[t]{0.48\textwidth} 
\caption{Best fit values for the different regions in \textbf{Mrk~573}.}
\label{tab:Mrk573}
\centering
    \vspace{-0.3cm}
\begin{tabular}{llcc}
        \toprule
        Component & Parameter & Value & log L$_{0.3-7~keV}$ \\
        \midrule
        \multicolumn{4}{c}{\textbf{core} (Cstat/d.o.f. = 1.00)} \\
        \midrule
        CLOUDY$_{1}$ & logU & 0.43$^{+0.05}_{-0.04}$ & - \\ 
        CLOUDY$_{1}$ & logN$_{\rm H}$ & 20.21$^{+0.04}_{-0.06}$ & - \\ 
        CLOUDY$_{1}$ & norm & 1.96$^{+0.14}_{-0.26}\times 10^{-15}$ & 41.02$^{+0.03}_{-0.04}$ \\ 
        CLOUDY$_{2}$ & logU & -0.71$^{+4.92e-03}_{-0.03}$ & - \\ 
        CLOUDY$_{2}$ & logN$_{\rm H}$ & 22.40$^{+0.21}_{-0.01}$ & - \\ 
        CLOUDY$_{2}$ & norm & 3.35$^{+0.18}_{-0.22}\times 10^{-15}$ & 40.94$^{+0.04}_{-0.04}$ \\
\bottomrule
    \end{tabular}
\end{minipage}
\hfill
\begin{minipage}[t]{0.48\textwidth}
\centering
\caption{Best fit values for the different regions in \textbf{NGC~835}.}
\label{tab:NGC835}
\vspace{-0.3cm}
    \begin{tabular}{llcc}
        \toprule
        \multicolumn{4}{c}{\textbf{Core} (Cstat/d.o.f. = 1.09)}\\
        Component & Parameter & Value & log $L_{0.3-7~\rm{keV}}$ \\
        \midrule
        apec & kT & 0.96$^{+0.06}_{-0.05}$ & - \\
        apec & norm & 2.34$^{+0.28}_{-0.24}\times 10^{-6}$ & 39.39$^{+0.04}_{-0.04}$ \\
        CLOUDY & logU & $<-1.47$ & - \\
        CLOUDY & logN$_{\rm H}$ & $>$22.55 & - \\
        CLOUDY & norm & 5.46$^{+4.59}_{-4.46}\times 10^{-15}$ & 39.11$^{+0.12}_{-0.15}$ \\
        MYTorusL & norm & 1.03$^{+0.36}_{-0.34}\times 10^{-3}$ & 39.79$^{+0.12}_{-0.16}$ \\
\bottomrule
    \end{tabular}
\end{minipage}
\end{table}

\begin{table}[h]
    \centering  
    \begin{minipage}[t]{0.48\textwidth} 
    \label{tab:NGC1068}
    \caption{Best-fit parameters for the different regions in \textbf{NGC\,1068}}
    \centering
    \begin{tabular}{llcc}
        \toprule
        \multicolumn{4}{c}{\textbf{Core} (Cstat/d.o.f. = 1.19)} \\
        Component & Parameter & Value & log$L_{0.3-7~\rm {keV}}$ \\
        \midrule
        TBabs & N$_{\rm H}$ & 0.30$^{+0.01}_{0.01}$ & - \\
        apec & kT & 1.21$^{+0.03}_{-0.04}$ & - \\
        apec & norm & 8.23$^{+0.92}_{-0.59}\times 10^{-5}$ & 39.76$^{+0.05}_{-0.05}$ \\
        CLOUDY$_{1}$ & logU & -1.88$^{+0.01}_{-0.01}$ & - \\
        CLOUDY$_{1}$ & logN$_{\rm H}$ & $<19.11$ & - \\
        CLOUDY$_{1}$ & norm & 3.43$^{+0.25}_{-0.12}\times 10^{-10}$ & 41.36$^{+0.03}_{-0.03}$ \\
\bottomrule
\end{tabular}
\end{minipage}
\hfill
\begin{minipage}[t]{0.48\textwidth}
\centering
\vspace{0.6cm}
    \begin{tabular}{llcc}
    \toprule
    \multicolumn{4}{c}{\textbf{Southwest Cone} (Cstat/d.o.f. = 1.21)} \\
    Component & Parameter & Value & $\log L_{\rm 0.3-7~keV}$ \\
    \midrule
    TBabs & N$_{\rm H}$ & 0.20$^{+0.02}_{-0.03}$ & - \\
        apec$_{1}$ & kT & 1.27$^{+0.20}_{-0.09}$ & - \\
        apec$_{1}$ & norm & 5.55$^{+1.51}_{-1.43}\times 10^{-5}$ & 39.56$^{+0.09}_{-0.11}$ \\
        apec$_{2}$ & kT & 0.50$^{+0.07}_{-0.04}$ & - \\
        apec$_{2}$ & norm & 1.60$^{+0.31}_{-0.11}\times 10^{-4}$ & 40.18$^{+0.03}_{-0.03}$ \\
        CLOUDY$_{1}$ & logU & -1.75$^{+0.11}_{-0.22}$ & - \\
\bottomrule
\end{tabular}
\end{minipage}
\end{table}

\begin{table}[h]
    \centering  
    \begin{minipage}[t]{0.48\textwidth} 
    \label{tab:NGC1167}
    \caption{Best fit values for different regions in \textbf{NGC\,1167}.}
    \centering
    \vspace{0.3cm}
    \begin{tabular}{llcc}
        \toprule
        \multicolumn{4}{c}{\textbf{Core} (Cstat/d.o.f. = 1.31)} \\
        Component & Parameter & Value & log$L_{0.3-7~\rm {keV}}$ \\
        \midrule
        apec & kT & $>$1.04 & - \\
        apec & norm & 1.62$^{+1.09}_{-0.59}\times 10^{-6}$ & 39.29$^{+0.21}_{-0.41}$ \\
        CLOUDY & logU & 1.25$^{+0.12}_{-0.21}$ & - \\
        CLOUDY & logN$_{\rm H}$ & 20.03$^{+0.30}_{-0.76}$ & - \\
        CLOUDY & norm & 3.11$^{+1.07}_{-1.64}\times 10^{-16}$ & 40.30$^{+0.08}_{-0.09}$ \\
        MYTorusL & norm & 6.50$^{+5.34}_{-4.04}\times 10^{-4}$ & 38.78$^{+0.26}_{-0.41}$ \\
\bottomrule
    \end{tabular}
    \end{minipage}
    \hfill
    \begin{minipage}[t]{0.48\textwidth}
    \centering
    \label{tab:NGC1365}
    \caption{Best fit values for the different regions in \textbf{NGC\,1365}.}
    \vspace{0.3cm}
    \begin{tabular}{llcc}
        \toprule
        \multicolumn{4}{c}{\textbf{Core} (Cstat/d.o.f. = 0.93)} \\
        Component & Parameter & Value & log$L_{0.3-7~\rm {keV}}$ \\
        \midrule
        TBabs & N$_{\rm H}$ & 0.98$^{+0.13}_{-0.13}$ & - \\
        apec$_{1}$ & kT & 2.77$^{+1.65}_{-0.72}$ & - \\
        apec$_{1}$ & norm & 1.90$^{+0.29}_{-0.29}\times 10^{-4}$ & 40.29$^{+0.04}_{-0.05}$ \\
        apec$_{2}$ & kT & 0.11$^{+0.02}_{-0.01}$ & - \\
        apec$_{2}$ & norm & 0.14$^{+0.47}_{-0.12}$ & 42.75$^{+0.04}_{-0.04}$ \\
        CLOUDY & logU & 0.36$^{+0.20}_{-0.24}$ & - \\
\bottomrule
    \end{tabular}
    \end{minipage}
\end{table}

\begin{table}[h]
    \centering  
    \begin{minipage}[t]{0.48\textwidth} 
    \label{tab:mrk3}
    \caption{Best fit values for the different regions in \textbf{Mrk~3}.}
    \centering
    \vspace{-0.3cm}
    \begin{tabular}{llcc}
        \toprule
        \multicolumn{4}{c}{\textbf{Core} (Cstat/d.o.f. = 1.12)} \\
        Component & Parameter & Value & log$L_{0.3-7~\rm {keV}}$ \\
        \midrule
        TBabs$_{1}$ & N$_{\rm H}$ & 0.26$^{+0.02}_{-4.50e-03}$ & - \\
        apec & kT & 0.88$^{+0.04}_{-0.04}$ & - \\
        apec & norm & 6.23$^{+0.77}_{-1.59}\times 10^{-5}$ & 40.89$^{+0.06}_{-0.07}$ \\
        CLOUDY$_{1}$ & logU & -0.31$^{+0.02}_{-0.03}$ & - \\
        CLOUDY$_{1}$ & logN$_{\rm H}$ & $>$23.42 & - \\
        CLOUDY$_{1}$ & norm & 5.13$^{+1.52}_{-0.31}\times 10^{-15}$ & 41.85$^{+0.02}_{-0.02}$ \\
\bottomrule
\end{tabular}
\end{minipage}
\hfill
\begin{minipage}[t]{0.48\textwidth}
\centering
\vspace{0.6cm}
    \begin{tabular}{llcc}
    \toprule
        \multicolumn{4}{c}{\textbf{Southwest Cross-Cone} (Cstat/d.o.f. = 1.07)} \\
        Component & Parameter & Value & log$L_{0.3-7~\rm {keV}}$ \\
        \midrule
        TBabs & N$_{\rm H}$ & 0.18$^{+0.03}_{-0.03}$ & - \\
        apec & kT & 0.97$^{+0.07}_{-0.06}$ & - \\
        apec & norm & 1.12$^{+0.14}_{-0.13}\times 10^{-5}$ & 40.11$^{+0.04}_{-0.04}$ \\
        CLOUDY & logU & -0.21$^{+0.18}_{-0.23}$ & - \\
        CLOUDY & logN$_{\rm H}$ & $>$22.90 & - \\
        CLOUDY & norm & 4.15$^{+5.03}_{-1.80}\times 10^{-16}$ & 40.89$^{+0.03}_{-0.03}$ \\
        \midrule
        \multicolumn{4}{c}{\textbf{Northeast Cross-Cone} (Cstat/d.o.f. = 1.26)} \\
        Component & Parameter & Value & log$L_{0.3-7~\rm {keV}}$ \\
        \midrule
        TBabs & N$_{\rm H}$ & 0.40$^{+0.06}_{-0.05}$ & - \\
        apec & kT & 0.95$^{+0.08}_{-0.07}$ & - \\
        apec & norm & 1.45$^{+0.27}_{-0.21}\times 10^{-5}$ & 40.23$^{+0.08}_{-0.09}$ \\
        CLOUDY$_{1}$ & logU & 0.25$^{+0.19}_{-0.17}$ & - \\
        CLOUDY$_{1}$ & logN$_{\rm H}$ & $>$23.39 & - \\
        CLOUDY$_{1}$ & norm & 5.76$^{+3.53}_{-2.56}\times 10^{-17}$ & 40.63$^{+0.09}_{-0.11}$ \\
        CLOUDY$_{2}$ & logU & -0.50$^{+0.26}_{-0.36}$ & - \\
        CLOUDY$_{2}$ & logN$_{\rm H}$ & $<19.32$ & - \\
        CLOUDY$_{2}$ & norm & 4.00$^{+7.18}_{-2.18}\times 10^{-13}$ & 41.05$^{+0.15}_{-0.26}$ \\
        \bottomrule
\end{tabular}
\end{minipage}
\end{table}

\begin{table}[h]
    \centering  
    \begin{minipage}[t]{0.48\textwidth} 
    \label{tab:NGC1386}
    \caption{Best fit values for the different regions in \textbf{NGC~1386}.}
    \centering
    \begin{tabular}{llcc}
        \toprule
        \multicolumn{4}{c}{\textbf{Core} (Cstat/d.o.f. = 1.12)} \\
        Component & Parameter & Value & log$L_{0.3-7~\rm {keV}}$ \\
        \midrule
        apec & kT & 0.84$^{+0.02}_{-0.02}$ & - \\
        apec & norm & 1.53$^{+0.09}_{-0.07}\times 10^{-5}$ & 39.00$^{+0.03}_{-0.03}$ \\
        CLOUDY$_{1}$ & logU & 2.05$^{+0.04}_{-0.04}$ & - \\
        CLOUDY$_{1}$ & logN$_{\rm H}$ & 22.27$^{+0.02}_{-0.08}$ & - \\
        CLOUDY$_{1}$ & norm & 2.37$^{+0.10}_{-0.15}\times 10^{-18}$ & 39.30$^{+0.03}_{-0.03}$ \\
        CLOUDY$_{2}$ & logU & -0.490$^{+0.003}_{-0.002}$ & - \\
\bottomrule
        \end{tabular}
\end{minipage}
\hfill
\begin{minipage}[t]{0.48\textwidth} 
\label{tab:NGC3081}
\caption{Best fit values for the different regions in \textbf{NGC~3081}.}
\centering
\vspace{-0.3cm}
\begin{tabular}{llcc}
        \toprule
        \multicolumn{4}{c}{\textbf{Core} (Cstat/d.o.f. = 1.02)} \\
        Component & Parameter & Value & log$L_{0.3-7~\rm {keV}}$ \\
        \midrule
        TBabs & N$_{\rm H}$ & 0.08$^{+0.03}_{-0.01}$ & - \\
        apec & kT & 1.25$^{+0.14}_{-0.07}$ & - \\
        apec & norm & 2.06$^{+0.72}_{-0.39}\times 10^{-5}$ & 39.80$^{+0.05}_{-0.05}$ \\
        CLOUDY$_{1}$ & logU & 1.280$^{+0.004}_{-0.010}$ & - \\
        CLOUDY$_{1}$ & logN$_{\rm H}$ & 22.78$^{+0.10}_{-0.10}$ & - \\
        CLOUDY$_{1}$ & norm & 2.53$^{+0.10}_{-0.69}\times 10^{-17}$ & 40.69$^{+0.04}_{-0.04}$ \\
\bottomrule
    \end{tabular}
\end{minipage}
\end{table}

\begin{table}
\centering
\begin{minipage}[t]{0.48\textwidth}
\centering
\label{tab:ESO428-G014}
\caption{Best fit values for the different regions in \textbf{ESO\,428-G014}.}
    \begin{tabular}{llcc}
        \toprule
        \multicolumn{4}{c}{\textbf{Core} (Cstat/d.o.f. = 1.01)} \\
        Component & Parameter & Value & log$L_{0.3-7~\rm {keV}}$ \\
        \midrule
        apec & kT & 0.85$^{+0.05}_{-0.03}$ & - \\
        apec & norm & 2.05$^{+0.23}_{-0.17}\times 10^{-5}$ & 39.66$^{+0.05}_{-0.05}$ \\
        CLOUDY$_{1}$ & logU & 2.52$^{+0.12}_{-0.04}$ & - \\
        CLOUDY$_{1}$ & logN$_{\rm H}$ & 19.09$^{+4.41}_{-0.09}$ & - \\
        CLOUDY$_{1}$ & norm & 9.31$^{+1.95}_{-0.55}\times 10^{-16}$ & 39.74$^{+0.04}_{-0.04}$ \\
        CLOUDY$_{2}$ & logU & 1.22$^{+0.10}_{-0.07}$ & - \\
\bottomrule
        \end{tabular}
\end{minipage}
\hfill
\begin{minipage}[t]{0.48\textwidth}
\centering
\vspace{0.6cm}
\begin{tabular}{llcc}        
        \toprule
        \multicolumn{4}{c}{\textbf{Southwest Cross-Cone} (Cstat/d.o.f. = 1.31)} \\
        Component & Parameter & Value & log$L_{0.3-7~\rm {keV}}$ \\
        \midrule
        CLOUDY$_{1}$ & logU & -0.83$^{+0.05}_{-0.08}$ & - \\
        CLOUDY$_{1}$ & logN$_{\rm H}$ & 22.40$^{+0.04}_{-0.27}$ & - \\
        CLOUDY$_{1}$ & norm & 8.25$^{+1.01}_{-1.29}\times 10^{-16}$ & 39.12$^{+0.06}_{-0.06}$ \\
        CLOUDY$_{2}$ & logU & 1.10$^{+0.11}_{-0.07}$ & - \\
        CLOUDY$_{2}$ & logN$_{\rm H}$ & 20.06$^{+0.34}_{-0.22}$ & - \\
        CLOUDY$_{2}$ & norm & 1.45$^{+1.04}_{-0.33}\times 10^{-16}$ & 39.22$^{+0.05}_{-0.06}$ \\
        \midrule
        \multicolumn{4}{c}{\textbf{Northeast Cross-Cone} (Cstat/d.o.f. = 1.08)} \\
        Component & Parameter & Value & log$L_{0.3-7~\rm {keV}}$ \\
        \midrule
        CLOUDY$_{1}$ & logU & -0.59$^{+0.60}_{-0.65}$ & - \\
        CLOUDY$_{1}$ & logN$_{\rm H}$ & $>$21.89 & - \\
        CLOUDY$_{1}$ & norm & 1.40$^{+24.49}_{-0.90}\times 10^{-16}$ & 39.05$^{+0.11}_{-0.13}$ \\
        CLOUDY$_{2}$ & logU & 1.51$^{+0.10}_{-0.05}$ & - \\
        CLOUDY$_{2}$ & logN$_{\rm H}$ & 19.13$^{+0.82}_{-0.02}$ & - \\
        CLOUDY$_{2}$ & norm & 5.99$^{+5.16}_{-5.72}\times 10^{-16}$ & 39.03$^{+0.05}_{-0.05}$ \\
        \bottomrule
    \end{tabular}
\end{minipage}
\end{table}

\begin{table}[h]
    \centering  
    \begin{minipage}[t]{0.48\textwidth} 
    \label{tab:mrk78}
    \caption{Best fit values for the different regions in \textbf{Mrk~78}.}
    \centering
    \begin{tabular}{llcc}
        \toprule
        \multicolumn{4}{c}{\textbf{Core} (Cstat/d.o.f. = 1.03)} \\
        Component & Parameter & Value & log$L_{0.3-7~\rm {keV}}$ \\
        \midrule
        apec & kT & 1.20$^{+0.13}_{-0.13}$ & - \\
        apec & norm & 6.13$^{+2.35}_{-1.54}\times 10^{-6}$ & 40.63$^{+0.08}_{-0.09}$ \\
        CLOUDY & logU & $<-1.46$ & - \\
        CLOUDY & logN$_{\rm H}$ & 21.31$^{+0.72}_{-0.29}$ & - \\
        CLOUDY & norm & 2.88$^{+2.80}_{-2.48}\times 10^{-13}$ & 41.34$^{+0.04}_{-0.05}$ \\
        MYTorusL & N$_{\rm H}$ & 3.96$^{+0.35}_{-0.25}$ & - \\
\bottomrule
        \end{tabular}
\end{minipage}
\hfill
\begin{minipage}[t]{0.48\textwidth}  
\label{tab:mrk34}
\caption{Best fit values for the different regions in \textbf{Mrk~34}.}
\centering
\vspace{-0.3cm}
\begin{tabular}{llcc}
        \toprule
        \multicolumn{4}{c}{\textbf{Core} (Cstat/d.o.f. = 1.06)} \\
        Component & Parameter & Value & log$L_{0.3-7~\rm {keV}}$ \\
        \midrule
        apec & kT & 0.95$^{+0.10}_{-0.07}$ & - \\ 
        apec & norm & 4.89$^{+0.51}_{-0.50}\times 10^{-6}$ & 40.90$^{+0.06}_{-0.06}$ \\ 
        CLOUDY & logU & $<-1.64$ & - \\ 
        CLOUDY & logN$_{\rm H}$ & 21.49$^{+0.07}_{-0.18}$ & - \\ 
        CLOUDY & norm & 3.24$^{+1.00}_{-0.15}\times 10^{-13}$ & 41.59$^{+0.03}_{-0.03}$ \\ 
        MYTorusL & N$_{\rm H}$ & 4.23$^{+0.60}_{-0.33}$ & - \\
\bottomrule
\end{tabular}
\end{minipage}
\end{table}

\begin{table}[h]
\centering 
\begin{minipage}[t]{0.48\textwidth}
\centering
\label{tab:NGC3079}
\caption{Best fit values for the different regions in \textbf{NGC~3079}.}
\vspace{-0.3cm}
\begin{tabular}{llcc}
        \toprule
        \multicolumn{4}{c}{\textbf{Core} (Cstat/d.o.f. = 1.23)} \\
        Component & Parameter & Value & log$L_{0.3-7~\rm {keV}}$ \\
        \midrule
        TBabs & N$_{\rm H}$ & 1.67$^{+0.18}_{-0.32}$ & - \\
        apec & kT & 1.20$^{+0.32}_{-0.17}$ & - \\
        apec & norm & 3.60$^{+1.35}_{-1.47}\times 10^{-5}$ & 39.38$^{+0.06}_{-0.07}$ \\
        CLOUDY & logU & 0.29$^{+0.27}_{-0.29}$ & - \\
        CLOUDY & logN$_{\rm H}$ & 22.38$^{+0.40}_{-0.38}$ & - \\
        CLOUDY & norm & 2.07$^{+4.67}_{-1.07}\times 10^{-16}$ & 39.99$^{+0.07}_{-0.08}$ \\
\bottomrule
        \end{tabular}
\end{minipage}
\hfill
\begin{minipage}[t]{0.48\textwidth}
\centering
\vspace{0.6cm}
\begin{tabular}{llcc}
        \toprule
        \multicolumn{4}{c}{\textbf{Southeast Cross-Cone} (Cstat/d.o.f. = 0.99)} \\
        Component & Parameter & Value & log$L_{0.3-7~\rm {keV}}$ \\
        \midrule
        TBabs & N$_{\rm H}$ & 0.39$^{+0.29}_{-0.12}$ & - \\
        apec & kT & 0.34$^{+0.35}_{-0.09}$ & - \\
        apec & norm & 1.05$^{+4.18}_{-0.74}\times 10^{-5}$ & 38.88$^{+0.09}_{-0.11}$ \\
        CLOUDY & logU & $<-1.15$ & - \\
        CLOUDY & logN$_{\rm H}$ & $<20.41$ & - \\
        CLOUDY & norm & 5.18$^{+37.89}_{-4.72}\times 10^{-12}$ & 39.97$^{+0.04}_{-0.04}$ \\
        \midrule
        \multicolumn{4}{c}{\textbf{Northwest Cross-Cone} (Cstat/d.o.f. = 0.86)} \\
        Component & Parameter & Value & log$L_{0.3-7~\rm {keV}}$ \\
        \midrule
        TBabs$_{2}$ & N$_{\rm H}$ & 0.60$^{+0.17}_{-0.10}$ & - \\
        apec & kT & 1.21$^{+0.23}_{-0.19}$ & - \\
        apec & norm & 9.51$^{+5.53}_{-2.67}\times 10^{-6}$ & 38.80$^{+0.08}_{-0.09}$ \\
        CLOUDY & logU & -0.25$^{+0.26}_{-0.19}$ & - \\
        CLOUDY & logN$_{\rm H}$ & 22.42$^{+0.17}_{-0.14}$ & - \\
        CLOUDY & norm & 1.21$^{+0.75}_{-0.64}\times 10^{-15}$ & 40.08$^{+0.03}_{-0.03}$ \\  
        \bottomrule
    \end{tabular}
\end{minipage}
\end{table}

\begin{table}
\centering
\begin{minipage}[t]{0.48\textwidth} 
\caption{Best fit values for the various regions in \textbf{NGC\,3393.}}
\label{tab:NGC3393}
\centering
\begin{tabular}{llcc}
            \toprule
        \multicolumn{4}{c}{\textbf{Core} (Cstat/d.o.f. = 0.88)} \\
        Component & Parameter & Value & log$L_{0.3-7~\rm {keV}}$ \\
        \midrule
        apec & kT & 1.940$^{+0.002}_{-0.300}$ & - \\
        apec & norm & 8.82$^{+0.90}_{-2.20}\times 10^{-6}$ & 39.69$^{+0.06}_{-0.07}$ \\
        CLOUDY$_{1}$ & logU & 1.55$^{+0.01}_{-0.02}$ & - \\
        CLOUDY$_{1}$ & logN$_{\rm H}$ & 19.19$^{+0.06}_{-0.01}$ & - \\
        CLOUDY$_{1}$ & norm & 1.70$^{+0.60}_{-0.17}\times 10^{-15}$ & 40.54$^{+0.01}_{-0.01}$ \\
        CLOUDY$_{2}$ & logU & -0.72$^{+0.02}_{-0.02}$ & - \\
\bottomrule
    \end{tabular}
\end{minipage}
\hfill
\begin{minipage}[t]{0.48\textwidth}
\centering
\vspace{0.6cm}
\begin{tabular}{llcc}
    \toprule
        \multicolumn{4}{c}{\textbf{Southeast Cross-Cone} (Cstat/d.o.f. = 1.32)} \\
        Component & Parameter & Value & log$L_{0.3-7~\rm {keV}}$ \\
        \midrule
        apec & kT & 0.30$^{+0.02}_{-0.03}$ & - \\
        apec & norm & 4.32$^{+0.39}_{-0.61}\times 10^{-6}$ & 39.54$^{+0.04}_{-0.05}$ \\
        CLOUDY & logU & 1.96$^{+0.01}_{-0.01}$ & - \\
        CLOUDY & logN$_{\rm H}$ & 19.400$^{+0.003}_{-0.004}$ & - \\
        CLOUDY & norm & 2.84$^{+0.19}_{-0.24}\times 10^{-16}$ & 39.65$^{+0.03}_{-0.03}$ \\
        \midrule
        \multicolumn{4}{c}{\textbf{Northwest Cross-Cone} (Cstat/d.o.f. = 1.14)} \\
        Component & Parameter & Value & log$L_{0.3-7~\rm {keV}}$ \\
        \midrule
        apec & kT & 0.24$^{+0.01}_{-0.02}$ & - \\
        apec & norm & 5.07$^{+0.81}_{-0.71}\times 10^{-6}$ & 39.59$^{+0.05}_{-0.05}$ \\
        CLOUDY & logU & 1.950$^{+0.030}_{-0.005}$ & - \\
        CLOUDY & logN$_{\rm H}$ & 19.40$^{+0.02}_{-0.01}$ & - \\
        CLOUDY & norm & 3.73$^{+0.13}_{-0.52}\times 10^{-16}$ & 39.81$^{+0.02}_{-0.02}$ \\  
        \bottomrule
    \end{tabular}
\end{minipage}
\end{table}

\begin{table}
\centering
\begin{minipage}[t]{0.48\textwidth} 
\caption{Best fit values for the various regions in \textbf{NGC\,4945.}}
\label{tab:NGC4945}
\centering
\begin{tabular}{llcc}
    \toprule
        \multicolumn{4}{c}{\textbf{Core} (Cstat/d.o.f. = 1.04)} \\
        Component & Parameter & Value & log$L_{0.3-7~\rm {keV}}$ \\
        \midrule
        TBabs & N$_{\rm H}$ & 4.67$^{+0.46}_{-0.14}$ & - \\
        CLOUDY$_{1}$ & logU & 2.27$^{+0.15}_{-0.09}$ & - \\
        CLOUDY$_{1}$ & logN$_{\rm H}$ & 23.50$^{+-0.93}_{-1.29}$ & - \\
        CLOUDY$_{1}$ & norm & 5.00$^{+6.19}_{-2.49}\times 10^{-19}$ & 39.33$^{+0.09}_{-0.10}$ \\
        CLOUDY$_{2}$ & logU & 0.25$^{+0.52}_{-0.29}$ & - \\
        CLOUDY$_{2}$ & logN$_{\rm H}$ & 22.98$^{+0.37}_{-0.22}$ & - \\
\bottomrule
        \end{tabular}
\end{minipage}
\hfill
\begin{minipage}[t]{0.48\textwidth}
\centering
\vspace{0.6cm}
    \begin{tabular}{llcc}
    \toprule
     \multicolumn{4}{c}{\textbf{Northeast Cross-Cone} (Cstat/d.o.f. = 1.16)} \\
        Component & Parameter & Value & log$L_{0.3-7~\rm {keV}}$ \\
        \midrule
        TBabs$_{1}$ & N$_{\rm H}$ & 0.85$^{+0.11}_{-0.09}$ & - \\
        CLOUDY & logU & 0.24$^{+0.07}_{-0.11}$ & - \\
        CLOUDY & logN$_{\rm H}$ & $>$23.27 & - \\
        CLOUDY & norm & 3.23$^{+0.99}_{-0.48}\times 10^{-16}$ & 39.67$^{+0.02}_{-0.02}$ \\
        zgauss$_{1}$ & LineE & 6.39$^{+0.02}_{-0.02}$ & - \\
        zgauss$_{1}$ & norm & 3.43$^{+0.78}_{-0.49}\times 10^{-6}$ & 38.44$^{+0.06}_{-0.06}$ \\
        zgauss$_{2}$ & LineE & 6.73$^{+0.04}_{-0.04}$ & - \\
        zgauss$_{2}$ & norm & 1.96$^{+0.48}_{-0.45}\times 10^{-6}$ & 38.21$^{+0.09}_{-0.11}$ \\ 
        \bottomrule
    \end{tabular}
\end{minipage}
\end{table}

\begin{table}
\centering
\begin{minipage}[t]{0.48\textwidth} 
\caption{Best fit values for the various regions in \textbf{NGC\,5252.}}
\label{tab:NGC5252}
\centering
\vspace{-0.3cm}
\begin{tabular}{llcc}
        \toprule
        \multicolumn{4}{c}{\textbf{Core} (Cstat/d.o.f. = 1.12)} \\
        Component & Parameter & Value & log$L_{0.3-7~\rm {keV}}$ \\
        \midrule
        TBabs$_{1}$ & N$_{\rm H}$ & 0.29$^{+0.02}_{-0.03}$ & - \\
        apec & kT & 0.09$^{+0.01}_{-0.02}$ & - \\
        apec & norm & 1.89$^{+2.18}_{-0.92}\times 10^{-3}$ & 41.95$^{+0.04}_{-0.04}$ \\
        CLOUDY$_{1}$ & logU & 2.61$^{+0.06}_{-0.05}$ & - \\
        CLOUDY$_{1}$ & logN$_{\rm H}$ & $>$23.40 & - \\
        CLOUDY$_{1}$ & norm & 4.19$^{+0.09}_{-0.50}\times 10^{-19}$ & 41.78$^{+0.02}_{-0.02}$ \\
\bottomrule
        \end{tabular}
    \end{minipage}
\hfill
\begin{minipage}[t]{0.48\textwidth}
\centering
\vspace{0.6cm}  

\begin{tabular}{llcc}
        \toprule
        \multicolumn{4}{c}{\textbf{Southwest Cross-Cone} (Cstat/d.o.f. = 1.12)} \\
        Component & Parameter & Value & log$L_{0.3-7~\rm {keV}}$ \\
        \midrule
        TBabs & N$_{\rm H}$ & $<0.07$ & - \\
        apec & kT & 0.28$^{+0.03}_{-0.02}$ & - \\
        apec & norm & 4.74$^{+3.12}_{-0.69}\times 10^{-6}$ & 40.09$^{+0.04}_{-0.05}$ \\
        pexrav & norm & 1.83$^{+0.07}_{-0.07}\times 10^{-3}$ & 41.40$^{+0.02}_{-0.02}$ \\
        \midrule
        \multicolumn{4}{c}{\textbf{Northeast Cross-Cone} (Cstat/d.o.f. = 1.18)} \\
        Component & Parameter & Value & log$L_{0.3-7~\rm {keV}}$ \\
        \midrule
        TBabs & N$_{\rm H}$ & 0.50$^{+0.11}_{-0.12}$ & - \\
        apec & kT & 0.15$^{+0.02}_{-0.02}$ & - \\
        apec & norm & 1.85$^{+4.64}_{-1.30}\times 10^{-4}$ & 41.45$^{+0.05}_{-0.07}$ \\
        pexrav & norm & 5.23$^{+0.44}_{-0.44}\times 10^{-4}$ & 40.86$^{+0.03}_{-0.04}$ \\
        \bottomrule
    \end{tabular}
\end{minipage}
\end{table}

\begin{table}
	\begin{minipage}[t]{0.48\textwidth} 
    \caption{Best fit values for the various regions in \textbf{NGC\,5643}.}
    \begin{tabular}{llcc}
        \toprule
        \multicolumn{4}{c}{\textbf{core} (Cstat/d.o.f. = 1.10)} \\
        Component & Parameter & Value & logL$_{0.3-7~keV}$ \\
        \midrule
        apec & kT & 1.13$^{+0.06}_{-0.06}$ & - \\ 
        apec & norm & 7.09$^{+0.94}_{-0.54}\times 10^{-6}$ & 38.79$^{+0.05}_{-0.06}$ \\ 
        CLOUDY$_{1}$ & logU & 0.86$^{+0.10}_{-0.05}$ & - \\ 
        CLOUDY$_{1}$ & logN$_{\rm H}$ & 19.80$^{+0.18}_{-0.25}$ & - \\ 
        CLOUDY$_{1}$ & norm & 1.56$^{+1.18}_{-0.12}\times 10^{-15}$ & 39.51$^{+0.04}_{-0.05}$ \\ 
        CLOUDY$_{2}$ & logU & -0.88$^{+0.07}_{-0.15}$ & - \\
\bottomrule
    \end{tabular}
    \label{tabNGC5643}
\end{minipage}
\end{table}
\begin{table}
\begin{minipage}[t]{0.48\textwidth} 
\caption{Best fit values for the various regions in \textbf{NGC\,5728.}}
\label{tab:NGC5728}
\centering
\begin{tabular}{llcc}
        \toprule
        \multicolumn{4}{c}{\textbf{Core} (Cstat/d.o.f. = 0.98)} \\
        Component & Parameter & Value & log$L_{0.3-7~\rm {keV}}$ \\
        \midrule
        CLOUDY$_{1}$ & logU & 1.65$^{+0.04}_{-0.04}$ & - \\
        CLOUDY$_{1}$ & logN$_{\rm H}$ & 20.03$^{+0.24}_{-0.03}$ & - \\
        CLOUDY$_{1}$ & norm & 1.08$^{+0.08}_{-0.62}\times 10^{-16}$ & 39.88$^{+0.03}_{-0.03}$ \\
        CLOUDY$_{2}$ & logU & -1.02$^{+0.02}_{-0.25}$ & - \\
        CLOUDY$_{2}$ & logN$_{\rm H}$ & 22.50$^{+0.01}_{-0.01}$ & - \\
        CLOUDY$_{2}$ & norm & 6.15$^{+0.52}_{-0.36}\times 10^{-15}$ & 40.06$^{+0.03}_{-0.03}$ \\
\bottomrule
    \end{tabular}
\end{minipage}
\end{table}

\begin{table}
\centering
\begin{minipage}[t]{0.48\textwidth}
\caption{Best fit values for the various regions in \textbf{IC\,5063.}}
\label{tab:IC5063}
\centering
\vspace{-0.3cm}
\begin{tabular}{llcc}
        \toprule
        \multicolumn{4}{c}{\textbf{Core} (Cstat/d.o.f. = 1.23)} \\
        Component & Parameter & Value & log$L_{0.3-7~\rm {keV}}$ \\
        \midrule
        TBabs & N$_{\rm H}$ & 0.74$^{+0.17}_{-0.13}$ & - \\
        apec$_{1}$ & kT & $<0.57$ & - \\
        apec$_{1}$ & norm & $<1.89\times 10^{-3}$ & 41.05$^{+0.18}_{-0.29}$ \\
        apec$_{2}$ & kT & 0.52$^{+0.16}_{-0.13}$ & - \\
        apec$_{2}$ & norm & 5.93$^{+1.89}_{-4.35}\times 10^{-5}$ & 40.71$^{+0.08}_{-0.10}$ \\
        CLOUDY$_{1}$ & logU & 1.77$^{+0.04}_{-0.08}$ & - \\
\bottomrule
        \end{tabular}
\end{minipage}
\hfill
\begin{minipage}[t]{0.48\textwidth}
\centering
\vspace{0.6cm}
\begin{tabular}{llcc}
        \toprule
        \multicolumn{4}{c}{\textbf{Southwest Cross-Cone} (Cstat/d.o.f. = 1.23)} \\
        Component & Parameter & Value & log$L_{0.3-7~\rm {keV}}$ \\
        \midrule
        TBabs & N$_{\rm H}$ & 0.62$^{+0.09}_{-0.10}$ & - \\
        apec & kT & 0.22$^{+0.04}_{-0.03}$ & - \\
        apec & norm & 1.43$^{+1.65}_{-0.81}\times 10^{-4}$ & 40.91$^{+0.03}_{-0.04}$ \\
        CLOUDY & logU & 0.25$^{+0.27}_{-0.12}$ & - \\
        CLOUDY & logN$_{\rm H}$ & $>$23.34 & - \\
        CLOUDY & norm & 2.87$^{+0.94}_{-1.41}\times 10^{-17}$ & 40.18$^{+0.05}_{-0.06}$ \\
        \midrule
        \multicolumn{4}{c}{\textbf{Northeast Cross-Cone} (Cstat/d.o.f. = 1.25)} \\
        Component & Parameter & Value & log$L_{0.3-7~\rm {keV}}$ \\
        \midrule
        TBabs & N$_{\rm H}$ & 0.76$^{+0.09}_{-0.10}$ & - \\
        apec & kT & 0.20$^{+0.03}_{-0.03}$ & - \\
        apec & norm & 2.24$^{+3.13}_{-1.24}\times 10^{-4}$ & 41.09$^{+0.04}_{-0.04}$ \\
        CLOUDY & logU & 0.25$^{+0.21}_{-0.08}$ & - \\
        CLOUDY & logN$_{\rm H}$ & $>$23.34 & - \\
        CLOUDY & norm & 4.30$^{+0.95}_{-1.77}\times 10^{-17}$ & 40.36$^{+0.04}_{-0.04}$ \\ 
        \bottomrule
    \end{tabular}
\end{minipage}
\end{table}

\begin{table}
\centering
\begin{minipage}[t]{0.48\textwidth} 
\caption{Best fit values for the various regions in \textbf{ESO\,137-G034.}}
\label{tab:ESO137}
\centering
    \vspace{-0.3cm}
    \begin{tabular}{llcc}
         \toprule
        \multicolumn{4}{c}{\textbf{Core} (Cstat/d.o.f. = 0.95)} \\
        Component & Parameter & Value & log$L_{0.3-7~\rm {keV}}$ \\
        \midrule
        apec & kT & 0.79$^{+0.10}_{-0.12}$ & - \\
        apec & norm & 1.37$^{+0.26}_{-0.22}\times 10^{-5}$ & 39.94$^{+0.05}_{-0.05}$ \\
        CLOUDY & logU & -1.13$^{+0.18}_{-0.11}$ & - \\
        CLOUDY & logN$_{\rm H}$ & 22.17$^{+0.20}_{-0.20}$ & - \\
        CLOUDY & norm & 2.63$^{+2.00}_{-1.21}\times 10^{-14}$ & 40.78$^{+0.02}_{-0.02}$ \\
        MYTorusL & norm & 0.05$^{+0.01}_{-0.01}$ & 40.29$^{+0.03}_{-0.03}$ \\
\bottomrule
    \end{tabular}
\end{minipage}
\hfill
\begin{minipage}[t]{0.48\textwidth}
\centering
\caption{Best fit values for the different regions in \textbf{NGC\,7212.}}
\label{tab:NGC7212}
\begin{tabular}{llcc}
        \toprule
        \multicolumn{4}{c}{\textbf{Core} (Cstat/d.o.f. = 1.11)} \\
        Component & Parameter & Value & log$L_{0.3-7~\rm {keV}}$ \\
        \midrule
        CLOUDY$_{1}$ & logU & 2.05$^{+0.05}_{-0.02}$ & - \\
        CLOUDY$_{1}$ & logN$_{\rm H}$ & 20.91$^{+0.14}_{-0.15}$ & - \\
        CLOUDY$_{1}$ & norm & 2.01$^{+0.51}_{-0.51}\times 10^{-17}$ & 40.95$^{+0.03}_{-0.03}$ \\
        CLOUDY$_{2}$ & logU & -1.000$^{+0.004}_{-0.0200}$ & - \\
        CLOUDY$_{2}$ & logN$_{\rm H}$ & 22.50$^{+0.01}_{-0.01}$ & - \\
        CLOUDY$_{2}$ & norm & 1.13$^{+0.07}_{-0.04}\times 10^{-14}$ & 41.28$^{+0.03}_{-0.03}$ \\
\bottomrule
    \end{tabular}
    \end{minipage}
\end{table}

\clearpage
	
	\section{Results}
	\label{sec:results}
	The results of the spectral analysis are presented in Figures~\ref{fig:mrk573_spec}–\ref{fig:ngc7212_spec} and Tables~\ref{tab:Mrk573}–\ref{tab:NGC7212}. For each AGN, the top-left panel of Figures~\ref{fig:mrk573_spec}–\ref{fig:ngc7212_spec} compares the observed spectra extracted from the different spatial regions. The remaining panels show, for each region individually, the observed spectrum, the best-fit model and its individual components, and the residuals. We also plot the expected nuclear ``spillover'' contribution (Section~\ref{sec:spillover}) for reference. A summary of these results is shown in Table~\ref{tab:summary_spec}.
	We note a caveat regarding our sample selection. Our detailed spectral analysis showed that the nucleus of NGC~1167 is essentially unobscured ($\log (N_{\rm H}/{\rm cm}^{-2}) \sim 21.5$), while NGC~5252 was confirmed to be only moderately obscured, and therefore neither source strictly satisfies our original selection criterion for heavily obscured AGN. Nevertheless, the weak AGN in NGC~1167 and the prominent diffuse emission in NGC~5252 provide valuable opportunities to investigate their extended soft X-ray emission. For this reason, we retain both objects in the paper, but identify them separately in the figures and distinguish them from the main sample in the subsequent analyses. We also verified that their inclusion does not significantly affect the observed trends or alter the overall physical interpretation of our results.
	
	\subsection{Nuclear Emission}
	
	The nuclear/core ($r<1.5''$) spectra differ markedly from the surrounding extended emission, exhibiting systematically harder continua. In most sources the hard component becomes dominant above $\sim$3~keV, although the transition occurs at somewhat lower energies in NGC~835, NGC~5643, and IC~5063 ($\sim$2~keV), and is most extreme in NGC~5252, where the hard continuum emerges near $\sim$1~keV. NGC~3081 also shows a pronounced hump beginning near $\sim$3~keV. In NGC~5252 the nuclear spectrum requires an additional absorbed power-law component, indicating the presence of a strong transmitted nuclear continuum in addition to the reflected emission described by the \texttt{MYTorus} model.
	
	All nuclei require a scattered hard AGN component modeled with \texttt{MYTorusS}, accompanied by neutral Fe~K$\alpha$ emission modeled with \texttt{MYTorusL}, as expected for heavily obscured systems \citep[e.g.][]{Ghisellini1994}. The nuclei also commonly show complex soft X-ray emission. Photoionized gas modeled with \texttt{CLOUDY} is required in all nuclei ($20/20$, Table~\ref{tab:summary_spec}), with between one and four components per source (median $=2$). At least one collisionally ionized plasma component (\texttt{APEC}) is statistically required in $14/20$ nuclei, with temperatures spanning $kT \sim 0.09$–2.8~keV. Thus, even within the central $r<1.5''$, the soft emission is often multiphase and cannot be described by a single photoionized component alone.
	
	To quantify the relative importance of the soft components in the nuclear spectra, we define
	\[
	f_{\rm soft,core} \equiv \frac{L_{\rm \texttt{CLOUDY}}+L_{\rm \texttt{APEC}}}{L_{\rm 0.3-7,tot}},
	\]
	where all luminosities are measured from the nuclear/core fits. In $12/20$ sources the core luminosity is dominated by photoionized plus thermal emission ($f_{\rm soft,core}>0.5$, Table~\ref{tab:summary_spec}). This strongly suggests that, despite the presence of an obscured AGN component in every nucleus, the observed 0.3–7~keV nuclear spectrum is often dominated by circumnuclear gas emission rather than by the scattered nuclear continuum.
	
	\subsection{Extended Emission}
	
	At energies $\lesssim$3~keV, spectra extracted from different extended regions within a given galaxy often show substantial variations, particularly in the highest S/N observations. The clearest example is NGC~1068, whose northeast cone requires six components (three \texttt{CLOUDY} and three \texttt{APEC}) to reproduce its line-rich spectrum. More generally, the extended emission is frequently multiphase. In the cone regions, a combination of photoionized and thermal components is required in $17/20$ sources ($34$ cone regions), while in the cross-cone regions the same combination is required in $14/20$ sources ($26$ cross-cone regions).
	
	The extended spectra are generally soft-dominated. In many systems, the continuum above $\sim$3~keV can still be reproduced by the same \texttt{CLOUDY} components that fit the soft band, as seen in sources such as Mrk~3, ESO~428-G014, NGC~3079, Mrk~34, NGC~3393, NGC~4945, and NGC~5728. The strength of hard emission above $\sim$3~keV varies significantly across the sample. No significant extended hard emission is detected in NGC~1167, while only weak hard-band emission is present in Mrk~573, IC~5063 and in the cross-cone regions of NGC~835, NGC~1386, and NGC~5728.
	
	Above $\sim$4-5~keV, a measurable high-energy contribution from nuclear spillover is seen in NGC~835, Mrk~78, NGC~3081, ESO~137-G034, IC~5063, and NGC~7212, and in some regions dominates the continuum at the highest energies covered by \textit{Chandra}. In NGC~1386, Mrk~3, ESO~428-G014, NGC~3393, and NGC~5728, the high-energy tail affected by spillover is important only in the cross-cone regions. Hard extended emission in excess of the level expected from nuclear PSF spillover is clearly detected in 73 regions across $19/20$ sources. 
	
	Only two galaxies require an additional reflected hard component modeled with \texttt{pexrav} to reproduce the continuum above $\sim$3~keV: NGC~1068 (southwest cone) and NGC~5252 (all extended regions). NGC~5252 is the most extreme case in the sample, with both \texttt{pexrav} and a power-law component required to reproduce the extended hard-band continuum, indicating a substantial contribution from reflected and/or scattered nuclear emission outside the core.
	
	\subsection{Neutral Fe~K$\alpha$ Emission}
	
	Neutral Fe~K$\alpha$ emission at 6.4~keV is detected in the nuclear spectra of all sources (Table~\ref{tab:summary_spec}).
	
	Outside the core, the line is detected in $12/20$ (30 extended regions) systems, while in the remaining $8/20$ systems it is confined to the nucleus. The latter group comprises Mrk~573, NGC~835, NGC~1167, Mrk~78, NGC~3081, NGC~3079, NGC~5252, and ESO~137-G034. Extended Fe~K$\alpha$ emission is clearly present in all regions of NGC~1068, NGC~1365, Mrk~3, and NGC~4945, while is detected solely in the cone regions of ESO~428-G014, NGC~3393, NGC~5728, and IC~5063. 
	
	In several sources, for example Mrk~3, Mrk~34, NGC~3393, and IC~5063, the extended Fe~K$\alpha$ feature is reproduced by the same \texttt{CLOUDY} components used to fit the continuum. In NGC~1068 and NGC~4945, additional Gaussian components are required in the Fe~K-band, reflecting the greater complexity of the iron-line emission and, in NGC~4945, the presence of Fe~XXV at 6.7~keV.
	
	These results are summarized in Table~\ref{tab:summary_spec}.
	\begin{table*}[t]
\caption{Summary of the main spectral properties of the 20 CT AGNs in our sample. Column (2) indicates whether photoionized (\texttt{CLOUDY}) and/or thermal (\texttt{APEC}) components are required to model the nuclear (core) spectra. Column (3) lists the soft-band flux fraction in the core, $f_{\rm soft,core}$, where values $>0.5$ indicate soft emission dominance. Columns (4) and (5) indicate whether photoionized and thermal components are required in the cone and cross-cone regions, respectively. Column (6) reports the presence of emission above 3~keV in the cones and cross-cones. Column (7) identifies the model component that primarily accounts for the continuum emission above 4~keV in these regions. Column (8) indicates the detection of the neutral Fe~K$\alpha$ line in the core, cones, and cross-cone spectra. The label \texttt{spill} denotes emission dominated by nuclear PSF spillover rather than intrinsic extended emission.}

\resizebox{\textwidth}{!}{%
\centering
\begin{tabular}{lccccccc}
\toprule
\textbf{Galaxy} & \textbf{Core} & \bm{$f_{\rm soft,core}$} & \textbf{Cones} & \textbf{Cross-Cones} & \textbf{Emission \bm{$>$}3~keV} & \textbf{Continuum \bm{$>$}4~keV} & \textbf{Fe~K$\alpha$ Line} \\
& (\textbf{\texttt{CLOUDY}}, \textbf{\texttt{APEC}}) &   & (\textbf{\texttt{CLOUDY}}, \textbf{\texttt{APEC}})   & (\textbf{\texttt{CLOUDY}}, \textbf{\texttt{APEC}}) & \textbf{(cones, cross-cones)}  & \textbf{(cones, cross-cones)}  & \textbf{(core, cones, cross-cones)} \\
\midrule

Mrk~573    &  (\cmark, \xmark) & 0.82 &(\cmark, \xmark) & (\cmark, \cmark) &(Weak, Weak) & (\texttt{CLOUDY}, \texttt{CLOUDY}/\texttt{APEC}) & (\cmark, \xmark, \xmark)\\

NGC~835  &  (\cmark, \cmark) & 0.02 &(\cmark, \cmark) & (\cmark, \cmark) &(\cmark, Weak) & (\texttt{spill}, \texttt{spill}) & (\cmark, \xmark, \xmark)\\

NGC~1068 &  (\cmark, \cmark) & 0.95 &(\cmark, \cmark) & (\cmark, \cmark) &(\cmark, \cmark) & (\texttt{CLOUDY}/\texttt{pexrav}, \texttt{CLOUDY}) & (\cmark, \cmark, \cmark)\\

NGC~1167 & (\cmark, \cmark) & 0.89 & (\xmark, \cmark) & (\xmark, \cmark) &(\xmark, \xmark) & (\texttt{APEC}, \texttt{APEC}) & (\cmark, \xmark, \xmark)\\

NGC~1365 & (\cmark, \cmark) & 0.99 & (\xmark, \cmark) & (\xmark, \cmark) &(\cmark, \cmark) & (\texttt{APEC}, \texttt{APEC}) & (\cmark, \cmark, \cmark)\\

NGC~1386 & (\cmark, \cmark) & 0.66 & (\cmark, \cmark) & (\cmark, \cmark) &(\cmark, Weak) & (\texttt{CLOUDY}, \texttt{CLOUDY}) & (\cmark, Weak, \xmark)\\

Mrk~3  & (\cmark, \cmark) & 0.63 & (\cmark, \cmark) & (\cmark, \cmark) &(\cmark, \cmark) & (\texttt{CLOUDY}, \texttt{CLOUDY}) & (\cmark, \cmark, \texttt{spill})\\

ESO~428-G014 & (\cmark, \cmark) & 0.69 & (\cmark, \cmark) & (\cmark, \xmark) &(\cmark, \cmark) & (\texttt{CLOUDY}, \texttt{CLOUDY}) & (\cmark, \cmark, \xmark)\\

Mrk~78 & (\cmark, \cmark) & 0.31 & (\cmark, \cmark) & (\xmark, \cmark) &(\cmark, \texttt{spill}) & (\texttt{CLOUDY}, \texttt{APEC}) & (\cmark, \xmark, \xmark)\\

NGC~3081 & (\cmark, \cmark) & 0.47 & (\cmark, \cmark) & (\cmark, \cmark) & (\cmark, \cmark) & (\texttt{spill}, \texttt{spill}) & (\cmark, \texttt{spill}, \texttt{spill})\\

NGC~3079 & (\cmark, \cmark) & 0.78 & (\cmark, \cmark) & (\cmark, \cmark) & (\cmark, \cmark) & (\texttt{CLOUDY}, \texttt{CLOUDY}) & (\cmark, \xmark, Weak)\\

Mrk~34 & (\cmark, \cmark) & 0.80 & (\cmark, \cmark) & (\cmark, \cmark) & (\cmark, \cmark) & (\texttt{CLOUDY}, \texttt{CLOUDY}) & (\cmark, \cmark, Weak)\\

NGC~3393 & (\cmark, \cmark) & 0.59 & (\cmark, \cmark) & (\cmark, \cmark) & (\cmark, \cmark) & (\texttt{CLOUDY}, \texttt{CLOUDY}) & (\cmark, \cmark, \texttt{spill})\\

NGC~4945 & (\cmark, \xmark) & 0.74 & (\cmark, \cmark) & (\cmark, \cmark) & (\cmark, \cmark) & (\texttt{CLOUDY}, \texttt{CLOUDY}) & (\cmark, \cmark, \cmark)\\

NGC~5252& (\cmark, \cmark) & 0.08 & (\cmark, \cmark) & (\xmark, \cmark) & (\cmark, \cmark) & (\texttt{pexrav+plaw}, \texttt{pexrav+plaw}) & (\cmark, \texttt{spill}, \texttt{spill})\\

NGC~5643 & (\cmark, \cmark) & 0.36 & (\cmark, \cmark) & (\cmark, \cmark) & (\cmark, Weak) & (\texttt{CLOUDY}, \texttt{spill}) & (\cmark, \cmark, \texttt{spill})\\

NGC~5728 & (\cmark, \xmark) & 0.18 & (\cmark, \cmark) & (\cmark, \cmark) & (\cmark, Weak) & (\texttt{CLOUDY}, \texttt{spill}) & (\cmark, \cmark, \texttt{spill})\\

ESO~137-G034 & (\cmark, \cmark) & 0.54 & (\cmark, \cmark) & (\cmark, \cmark) & (\cmark, \cmark) & (\texttt{CLOUDY}, \texttt{CLOUDY}) & (\cmark, \cmark, \texttt{spill})\\

IC~5063 & (\cmark, \cmark) & 0.26 & (\cmark, \cmark) & (\cmark, \cmark) & (\cmark, Weak) & (\texttt{spill}, \texttt{spill}) & (\cmark, \cmark, \texttt{spill})\\

NGC~7212  & (\cmark, \xmark) & 0.40 & (\cmark, \xmark) & (\cmark, \xmark) & (Weak, Weak) & (\texttt{CLOUDY}, \texttt{CLOUDY}/\texttt{spill}) & (\cmark, \cmark, \texttt{spill})\\

\bottomrule
\end{tabular}%
}
\label{tab:summary_spec}
\end{table*}

	\subsection{Ensemble Properties of the Soft X-ray Emission}
	
\begin{figure*}
    \centering
    \includegraphics[width=\textwidth]{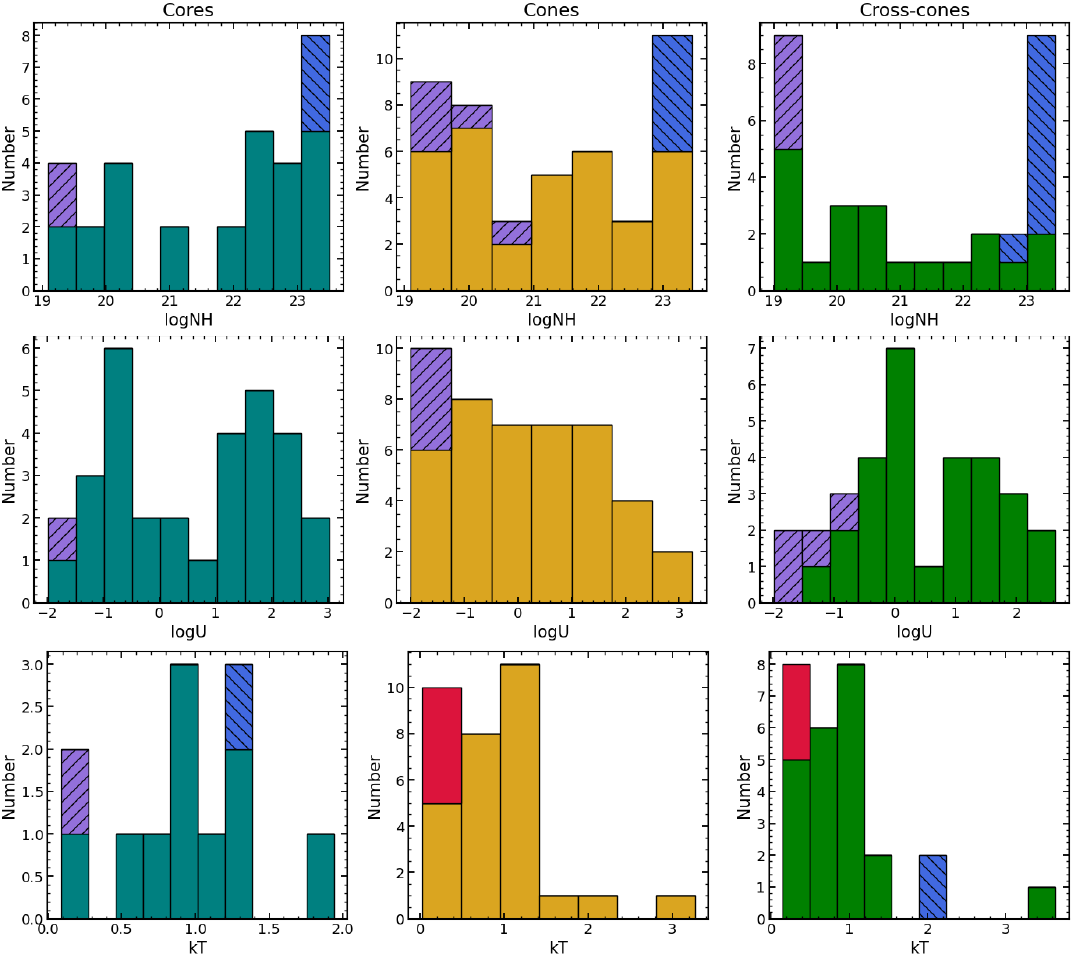}
    \caption{Distributions of best-fit spectral parameters for the core (left panels), cone (central panels), and cross-cone (right panels) regions. The top row shows the photoionized column density ($\log N_{\rm H}$), the middle row the ionization parameter ($\log U$), and the bottom row the thermal plasma temperature ($kT$). Core regions extend to higher $\log N_{\rm H}$ and $\log U$ values on average, but all regions exhibit broad and overlapping distributions, with no clear monotonic radial trend. Cone and cross-cone regions span similarly wide ranges in both parameters. Thermal plasma temperatures cluster around $kT \sim 0.3$-1~keV across all regions; however, cross-cone regions show a distribution that is more concentrated at lower temperatures but extends to higher $kT$ ($\gtrsim 5$~keV), while core temperatures are typically confined to $kT \lesssim 3$~keV. Red areas account for low signal to noise measurements, see final paragraph in Sect. 4.4, while upper/lower limits are in blue/purple, respectively.}
    \label{fig:isto}
\end{figure*}

\begin{figure*}
	\centering
	\includegraphics[width=\textwidth]{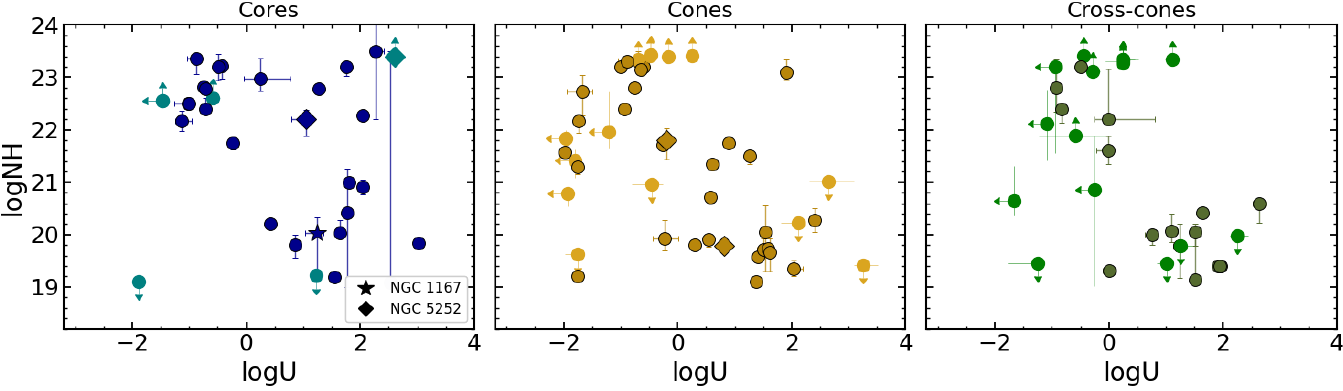}
	\caption{Distribution of photoionized column density versus ionization parameter for all spatial regions. Different colors identify the various spatial regions. Black-outlined markers represent measured values, while lighter-colored markers without black outlines indicate upper or lower limits. Core components extend to higher $\log N_{\rm H}$ and $\log U$ values on average, but all regions exhibit broad and overlapping distributions, with cone and cross-cone regions spanning similarly wide ranges in both parameters. No clear one-to-one correlation between $\log N_{\rm H}$ and $\log U$ is observed, consistent with a multiphase medium in which density, geometry, and illumination vary independently.}
	\label{fig:lognhlogu}
\end{figure*}

\begin{figure*}
    \centering
    \includegraphics[width=\textwidth]{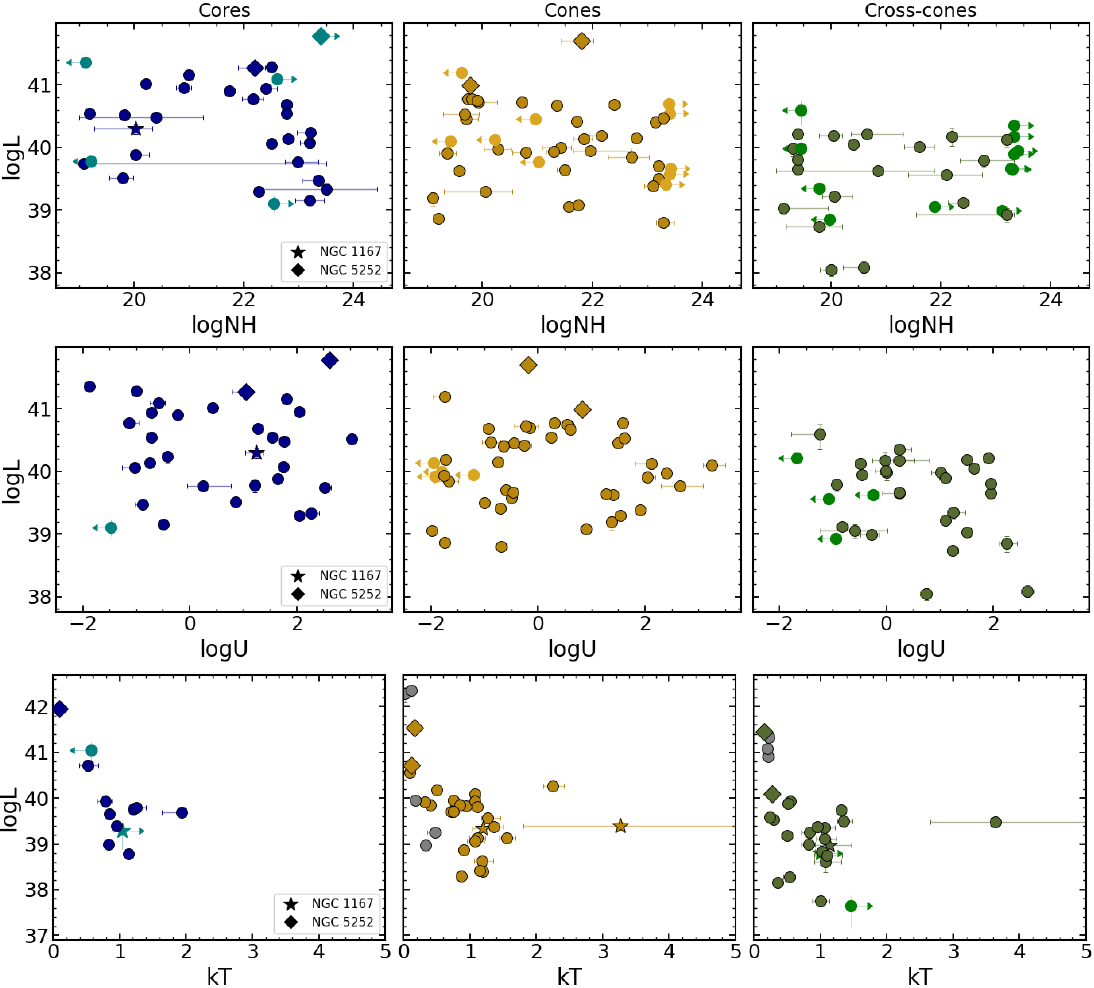}
    \caption{Intrinsic (unabsorbed) 0.3-7~keV X-ray luminosity of individual spectral components as a function of photoionized column density ($\log N_{\rm H}$), ionization parameter ($\log U$), and thermal temperature ($kT$), shown separately for the core (left), cone (middle), and cross-cone (right) regions. Across all regions, the luminosities span broad ranges with no strong or systematic correlations with $\log N_{\rm H}$, $\log U$, or $kT$. High-luminosity components are found across a wide range of ionization states and column densities. The absence of clear trends is consistent with a multiphase medium in which the emitting gas properties and energetics vary independently. Gray points identify measurements from low statistic spectra in the $0.3-0.5$ keV energy range.}
    \label{fig:lumvsparam}
\end{figure*}

	We now examine the ensemble properties of the spectral parameters characterizing the soft X-ray emission. The sample comprises 20 core spectra, 38 cone spectra, 38 cross-cone spectra, and 3 star-forming region spectra (in NGC~1365). For the purpose of constructing the ensemble distributions in Figures~\ref{fig:isto}-\ref{fig:lumvsparam}, and to maintain a uniform three-class comparison, we did not include any of the extended regions of NGC~1365. We emphasize that these ensemble distributions also include a number of censored measurements (upper and lower limits). In the following histograms and scatter plots, these limits are plotted at their nominal threshold values but are color-coded to differentiate them from the detections. Given the presence of these censored data, the comparisons presented in this section are intentionally kept qualitative and descriptive, focusing on the overall parameter space occupancy rather than formal statistical test inferences.
	
	The median number of photoionized and collisionally ionized components varies with spatial class (core: 2 and 1; cones: 3 and 2; cross-cones: 2 and 2, respectively), consistent with a decrease in the relative importance of photoionized components outside the nucleus and with cross-cone regions exhibiting a more comparable contribution from thermal and photoionized plasma. In contrast, the star-forming regions in NGC~1365 are well described by purely thermal emission, requiring only \texttt{APEC} components and no \texttt{CLOUDY} components, highlighting the distinct physical origin of the soft X-ray emission in those regions.
	
	Figure~\ref{fig:isto} shows the distributions of $\log\,N_{\rm H}$, $\log U$, and $kT$ for the core, cone, and cross-cone regions. The \texttt{CLOUDY} column densities span a wide range in all environments, from $\log (N_{\rm H}/\rm cm^{-2})\sim19$ up to $\gtrsim23$, confirming that the emitting gas is multiphase. The nuclear regions exhibit the broadest overall distribution, including a substantial population of high-column components ($\log (N_{\rm H}/\rm cm^{-2})\gtrsim22.5$) as well as a secondary population at low column densities ($\log (N_{\rm H}/\rm cm^{-2})\sim$19-20). In contrast, the cone and cross-cone regions show strong overlap with one another and are often populated by lower-column photoionized components ($\log (N_{\rm H}/\rm cm^{-2})\sim 19-21$), although high-column components are also present in both spatial classes.
	
	The ionization parameter distributions are similarly broad, extending from $\log U\sim-2$ to $\sim3$ in all regions. While the nuclei include many moderately and highly ionized components, the cone and cross-cone regions span comparable ranges and do not show a clean separation in $\log U$. Overall, there is no strong evidence for a single characteristic ionization state in any spatial class, further supporting the presence of multiple coexisting photoionized phases.
	
	Thermal plasma temperatures cluster primarily around $kT\sim0.2$-1~keV across all regions. The distributions are broadly similar between cones and cross-cones, with most measurements concentrated in the sub-keV to $\sim1$~keV range and a tail extending to higher temperatures. The nuclear regions span roughly $kT\sim0.1$-3~keV, while some of the highest-temperature components occur in the extended regions, especially in the cross-cone class, reaching $kT\sim5$-6~keV in a small number of cases. The NGC~1365 star-forming regions lie within the same broad thermal range, but are exclusively thermal, consistent with an origin in stellar processes rather than AGN photoionization.
	
	Figure~\ref{fig:lognhlogu} shows the distribution of $\log N_{\rm H}$ versus $\log U$. In all regions, the photoionized components occupy a broad locus characteristic of multiphase gas, and no strong one-to-one correlation is present between $\log N_{\rm H}$ and $\log U$. The inferred ionization parameters span a wide range, from $\log U \sim -2$ to $\sim 3$, while the column densities extend from $\log N_{\rm H} \sim 19$~cm$^{-2}$ to $\gtrsim 23.5$~cm$^{-2}$, with numerous measurements and limits clustering at both low and high column densities. This large dynamic range is consistent with the presence of multiple distinct photoionized phases coexisting within the same spatial regions.
	
	Despite the overall scatter, some systematic differences emerge between environments. In the nuclear regions (left panel), the distribution includes many high-column components ($\log (N_{\rm H}/\rm cm^{-2}) \gtrsim 23$), spanning a broad range in ionization parameter. At the same time, low-column ($\log (N_{\rm H}/\rm cm^{-2})\lesssim20$) components are also common, particularly among moderately and highly ionized gas, reinforcing the multiphase nature of the nuclear emission.
	
	In the cone regions (central panel), the distribution remains broad, with many components clustering around $\log (N_{\rm H}/\rm cm^{-2})\sim19$-21 over a wide range of ionization parameters. However, high-column components are still present in several sources, indicating that dense gas is not confined to the nucleus and can also be found along the ionization cones. The coexistence of low- and high-column phases at similar $\log U$ further supports a clumpy or stratified structure within these regions.
	
	The cross-cone regions exhibit a 2 clumps distribution, again with both low-column/high-ionization and high-column/lower-ionization solutions. There is some tendency for high-column cross-cone components to occur at low-to-moderate ionization parameter, while higher-ionization cross-cone components are often associated with lower columns, but the overlap with the cone population remains substantial. Overall, the cross-cone regions are not characterized by a single distinct locus; rather, they also host a broad multiphase distribution.


	Figure~\ref{fig:lumvsparam} shows the relation between the intrinsic 0.3-7~keV X-ray luminosity ($L_{\rm X}$) of the individual model components and their spectral parameters, separately for the core, cone, and cross-cone regions. No strong monotonic correlation is observed between component luminosity and $\log N_{\rm H}$, $\log U$, or $kT$. Instead, the luminosities span a broad range of $\sim$2 dex at nearly all parameter values, indicating that the X-ray output is distributed across multiple gas phases rather than being dominated by a single preferred ionization state, column density, or temperature.
	
	For the photoionized components, the $\log L$ versus $\log N_{\rm H}$ plane shows that both low-column ($\log (N_{\rm H} /\rm  cm^{-2})\sim19$-20) and high-column ($\log (N_{\rm H} /\rm  cm^{-2})\gtrsim22.5$) phases can contribute substantially to the extended X-ray emission. In the nuclear regions, the brightest \texttt{CLOUDY} components are found across much of this full range of column densities, including both low-column/high-ionization and high-column/lower-ionization solutions. The cone and cross-cone regions show a similarly broad distribution, although the most luminous photoionized components are more common in the cones than in the cross-cones.
	
	The $\log L$ versus $\log U$ relation is likewise highly scattered. Components with $\log U \lesssim -1$ and components with $\log U \gtrsim 1$ both reach substantial X-ray luminosities, particularly in the nuclei and cones. This indicates that the luminosity is not set primarily by ionization state alone. Instead, both low- and high-ionization phases contribute significantly to the observed X-ray emission in the 0.3-7~keV band. The core regions include the widest luminosity range, while the cone and cross-cone regions populate overlapping but somewhat narrower distributions. In particular, the cone regions contain several of the more luminous high-ionization components, whereas the cross-cone regions are more often characterized by moderate-luminosity components over a similarly broad range of $\log U$.
	
	For the thermal components, the $\log L$ versus $kT$ plane shows that most \texttt{APEC} components cluster at temperatures of $kT\sim0.2$-1.3~keV, but with luminosities extending over more than an order of magnitude. There is no clear evidence that more luminous thermal components are systematically hotter.
	Higher-temperature components ($kT\gtrsim2$-3~keV) are present in a subset of nuclei and extended regions, especially in the cross-cone class, but they are not uniformly associated with the highest thermal luminosities. The brightest thermal components are often found at relatively low or intermediate temperatures, particularly in systems such as NGC~5252 and NGC~1365, showing that the thermal soft X-ray output is likewise not governed by temperature alone.
	
	Taken together, these distributions show that the 0.3-7~keV X-ray luminosity is only weakly connected to any single fitted parameter. The lack of strong trends in $\log L$ with $\log N_{\rm H}$, $\log U$, or $kT$ supports a picture in which the observed emission arises from a complex superposition of photoionized and collisionally ionized phases with different physical conditions and spatial distributions.
	
	Finally, we note that best-fit model for at least one of the extended emission regions in NGC~4945, NGC~1068, IC~5063, NGC~5252, NGC~5643, and NGC~3081 includes an \texttt{APEC} component with a temperature $kT << 0.5$~keV. Since the $0.3$--$0.5$~keV energy range is significantly affected by the time-dependent degradation of the ACIS-S detector response, we performed a consistency check to assess the robustness of these soft components. Among these sources, only NGC~3081 was observed exclusively after 2019. However, the emission from NGC~4945, NGC~5643, and the cross-cone region of IC~5063 is background-dominated in this energy band. Consequently, the low-temperature \texttt{APEC} component in these specific cases likely lacks physical significance. In contrast, the soft thermal components in NGC~1068 and NGC~5252 appear reliable, as the extended emission in the $0.3$--$0.5$~keV range exceeds the background level and the data rely on pre-2019 observations. Nevertheless, it is important to note that the best-fit $kT$ values lie well below the effectively observed energy range; thus, the fit is constrained only by the high-energy tail of the thermal spectrum. In the relevant plots (i.e., Figures~\ref{fig:isto}--\ref{fig:lumvsparam}), parameters potentially unreliable due to low signal-to-noise ratio below $\sim0.5$~keV are displayed in red and gray, respectively. Lower and upper limits are represented by blue and purple shaded areas in the histograms (Fig.~\ref{fig:isto}), and by lighter colors with accompanying arrows in the scatter plots.
	
	\section{Discussion}
	\label{sec:discussion}
	
	This work provides a uniform, spatially resolved X-ray spectral characterization of the extended emission of 20 nearby AGN, observed with long {\it Chandra} ACIS-S exposures (Table 1., Fig. 1). Eighteen of these sources satisfy our original selection criterion for heavily obscured AGN, while we retain NGC~1167 and NGC\,5252 as additional sources because of their valuable extended soft X-ray emission. NGC\,1167 hosts a weak AGN with prominent extended emission, while NGC\,5252 is only moderately obscured. The key outcome is that the circumnuclear and kiloparsec-scale emission is rarely described by a single physical phase: across the sample, the soft band requires multiple photoionized and thermal components, and in systems the extended regions also show hard-band emission in excess of what is expected from nuclear PSF spillover. Below we discuss the physical implications of the ensemble trends, interpreting the spectral fit results presented in Section~\ref{sec:results}.
	While our spectral modeling successfully describes the extended emission using multiple discrete photoionized (CLOUDY) and, in some cases, thermal (APEC) components, we do not interpret these as strictly segregated physical phases. The circumnuclear and extended medium in these AGN is likely characterized by a complex, continuous distribution of ionization states, densities, and temperatures. However, given the relatively low spectral resolution of Chandra/ACIS-S ($\sim 100-150$ eV) and the limited photon statistics in the extended regions, cleanly resolving a continuous ionization distribution is unfeasible. Therefore, the multiple discrete phases inferred in our analysis are not uniquely required by the data as isolated zones; rather, they should be physically interpreted as a data-driven, coarse-grained approximation of the underlying continuous distribution, effectively representing the emission-weighted averages of the local gas conditions.
	
	\subsection{Multiphase Circumnuclear and Extended Emission}
	
	The ensemble parameter distributions (Figure~\ref{fig:isto}) reinforces the picture of a complex, spatially varying multiphase medium in both the nuclear and extended regions. The cores show the broadest overall spread in photoionized properties, including both a high-column population at $\log (N_{\rm H}/ \rm cm^{-2})\gtrsim22.5$ and a distinct lower-column population around $\log (N_{\rm H}/ \rm cm^{-2})\sim19-20$. The nuclear components also extend to relatively high ionization parameters, reaching $\log U\gtrsim2$ in several objects. These distributions imply that even within the central apertures the soft X-ray emission arises from multiple gas phases spanning a wide range of physical conditions, rather than from a single characteristic photoionized component.
	
	At larger radii, the cone and cross-cone regions remain broadly multiphase and overlap substantially in both $\log N_{\rm H}$ and $\log U$. The cones contain a large number of components at both low and high column density, and similarly span a wide ionization range from $\log U\lesssim-1.5$ up to $\log U\gtrsim2$. The cross-cones also cover a broad locus in parameter space, including both low-column/high-ionization and high-column/lower-ionization solutions. Thus, the plotted distributions do not suggest a clean separation between cones and cross-cones, nor a simple monotonic progression from core to cone to cross-cone. 
	
	This substantial overlap may indicate that anisotropic illumination is only part of the explanation. The cones are still consistent with representing gas that is more directly exposed to the AGN radiation field, but the cross-cone regions cannot be described as purely low-ionization shadows. Instead, the broad loci in $\log N_{\rm H}$-$\log U$ space (Figure~\ref{fig:lognhlogu}) indicate that local gas density, clumpiness, and line-of-sight geometry are also important in setting the observed spectral properties. In this sense, the extended soft X-ray emission appears to trace a complex distribution of clouds with different column densities and ionization states, rather than a single stratified structure. While the spectral fits formally result in a finite number of discrete ionization components, we  note that  the moderate spectral resolution of the ACIS-S CCD data does not allow us to rule out the possibility of a smoother, continuous distribution of physical properties. In any case, our results clearly indicate that the ISM and its ionization states are complex.

	In contrast, the star-forming regions in NGC~1365 are purely thermal, reinforcing that not all extended emission is AGN-driven.
	
	Thermal components are common in the extended regions and remain present in many cores (Table~\ref{tab:summary_spec}), indicating that collisional processes contribute significantly in addition to photoionization. The ensemble $kT$ values cluster most strongly at sub-keV to $\sim$1~keV temperatures, with a tail extending to several keV (Figure~\ref{fig:isto}). This behavior supports a multiphase thermal medium rather than a single characteristic plasma temperature. Interpreted as shock-heated gas, the sub-keV components are broadly consistent with $\sim$few$\times10^2$-$10^3$~km~s$^{-1}$ shocks, while the higher-$kT$ outliers would require faster shocks and/or additional heating channels. Several mechanisms may contribute, including interaction of AGN-driven outflows with the host ISM, jet-ISM coupling in systems with compact radio structures, and, in some targets, circumnuclear star formation.
	
	The thermal component distributions support the same conclusion reached from the photoionized gas: the soft X-ray-emitting medium is not uniform. Most \texttt{APEC} components in all spatial classes lie at sub-keV to $\sim$1~keV temperatures, but with tails to higher $kT$, including the most extreme values in some cross-cone regions. The cores and cones both span a broad range of temperatures, while the highest-temperature components are found in a subset of extended regions, particularly in the cross-cone class, where $kT$ reaches several keV in a few systems. The comparable $kT$ values found across cones and cross-cones further argue that the extraplanar soft X-ray emission is not produced by a single homogeneous medium, but instead reflects a mixture of photoionized gas and thermally emitting plasma whose relative contributions vary from source to source and from region to region. The $kT$ distribution of the cross-cone regions (Fig.~\ref{fig:isto}) however favors lower temperatures than the corresponding distribution observed for the cones. The presence of softer cross-cone distributions is in agreement with previous reports \citep[][]{fabbiano_interaction_2024}. This thermal cross-cone emission may be related to shock excitation and lateral outflows resulting from the interaction of radio jets embedded in the dense ISM of the host galaxy disk \citep[][]{Mukherjee2016,fabbiano_jet-ism_2022}.

	\subsection{Photoionization and Shocks Outside the Cones }
	
	Photoionized components are frequently required even in cross-cone spectra. In a strictly ``closed'' torus geometry, these regions would be largely shielded from the ionizing continuum. Instead, the cross-cone regions exhibit a wide range of ionization parameters and column densities (Figures~\ref{fig:isto} and \ref{fig:lognhlogu}), including clearly detected photoionized components spanning $\log U \sim -1$ to $\gtrsim2$ in several systems. Moreover, the cross-cone loci overlap substantially with those of the ionization cones, and do not form a distinct low-ionization population.
	
	This behavior indicates that ionizing photons reach a larger solid angle than predicted by simple axisymmetric obscuration models \citep[see e.g.,][]{antonucci_unified_1993}. A natural explanation is provided by clumpy or porous torus structures \citep[see review by][]{Nenkova2008,fabbiano_interaction_2024}, in which partial covering and low-density channels allow radiation to leak into nominally shadowed directions. Electron scattering by circumnuclear material may further redistribute ionizing photons on kiloparsec scales.
	
	Projection effects and large-scale ISM structure may contribute to the observed cone/cross-cone parameter overlap. The extraction regions can encompass gas spanning a range of true illumination angles and distances from the nucleus, mixing directly illuminated and partially shielded components. The absence of a clear separation between cones and cross-cones in $\log N_{\rm H}$--$\log U$ space therefore suggests that geometry alone does not set the observed ionization structure; instead, the combination of clumpiness, scattering, and environmental complexity plays a dominant role in shaping the extended photoionized emission.
	
	However, in many cases the photon statistics in cross-cone regions do not allow us to robustly distinguish between photoionization and shocked emission. Alternatively, the soft emission in the cross-cones can arise from shock excitation and lateral outflows driven by jet propagation through the dense ISM of the host galaxy disk \citep[][]{Mukherjee2016,fabbiano_jet-ism_2022}. This scenario seems to be favored by multi-wavelength studies. The shock vs. AGN photoionization contributions modeled based on optical separation models \citep[][2026 in preparation]{Zhu2025} show a significant presence of shock-driven emission in the bicone region. Additionally, the morphology of the extended X-ray emission in the cross-cone directions at $\sim1$~keV, an energy range dominated by thermal emission in multiple sources, seems to trace these shock-dominated regions (Zhu et al. 2026 in prep.).
	
	\subsection{Extended Hard Emission and Reflection}
	
	Hard-band excess emission is common, as suggested by previous studies \citep[see][]{fabbiano_interaction_2024}. In our sample, emission $>$3~keV appears in 38 cone regions, 3 star-forming regions (NGC~1365), and 33 cross-cone regions. This corresponds to 73 extended regions across 19/20 sources showing hard-band emission above the level expected from PSF spillover. Extended Fe~K$\alpha$ emission at 6.4~keV is detected in 20 cone regions, 3 star-forming regions, and 7 cross-cone regions. These detections indicate that genuinely extended hard emission is common in obscured AGN. Based on our spatial extraction regions, this hard-band excess is typically detected out to physical scales ranging from a few hundred parsecs up to $\sim 1-2$~kpc from the central engine, closely matching the extent of the soft X-ray components. Previous work has suggested an energy dependence on the extent, with harder components being less extended than the soft ones \citep[][]{jones_chandra_2020}.  A detailed and systematic characterization of the energy-resolved radial surface brightness profiles across the full sample will be presented in a forthcoming paper (Middei et al., in prep.).
	
	Reflection from cold material on galactic scales provides a natural explanation, as demonstrated in detailed studies of individual sources, and may trace dense clouds illuminated by the obscured nucleus \citep[see review by][]{fabbiano_interaction_2024}. The presence of Fe~K$\alpha$ emission outside the nucleus in particular is consistent with reflection and/or reprocessing by dense illuminated material distributed on circumnuclear to kiloparsec scales.
	
	The hard emission does not map cleanly onto a single geometric component. Although hard excess is more frequently identified in cone regions, a substantial number of cross-cone regions also show $>$3~keV emission, indicating that hard X-ray photons are not strictly confined to the ionization cones. This is consistent with a scenario in which reflection occurs off a wide distribution of dense clouds with a range of orientations, and where projection effects and inhomogeneous obscuration allow hard photons to be observed outside the nominal cone axis.
	
	In this context, the cases where cross-cone regions are consistent with spillover reflect a combination of lower surface brightness and reduced illumination, rather than a complete absence of hard emission. The lack of a clear separation between cones and cross-cones in the broader spectral properties (e.g., Figures~\ref{fig:isto} and \ref{fig:lumvsparam}) further supports the view that extended hard emission arises from a complex, multi-scale distribution of material rather than from a single well-defined geometric structure.

	\subsection{Luminosities and Geometric Modulation}
	
	The relation between X-ray luminosity and spectral parameters further emphasizes that the observed emission is not controlled by a single physical variable. Figure~\ref{fig:lumvsparam} shows that both photoionized and thermal components display a large dispersion in luminosity at fixed $\log N_{\rm H}$, $\log U$, and $kT$. In all spatial regions, components with similar best-fit parameters can differ in luminosity by more than an order of magnitude, demonstrating that the emitted power is not set uniquely by the local gas state.
	
	For the photoionized components, the strongest result is the broad scatter in $\log L$ at fixed $\log N_{\rm H}$ and $\log U$. The nuclear components reach the highest luminosities overall, extending to $\log (L/\rm erg~s^{-1})\gtrsim 41.5$, whereas cone and cross-cone components more commonly occupy the $\log (L/\rm erg~s^{-1})\sim 39$-41~erg~s$^{-1}$ range. However, the parameter space overlaps strongly across all three spatial regions: high- and low-luminosity components are found at both relatively low and relatively high $\log U$, and similarly across both low-column and high-column solutions. Thus, while the nuclei preferentially host the most luminous photoionized components, there is no tight global mapping between luminosity and either ionization parameter or column density.
	
	The $\log L$-$\log N_{\rm H}$ panels show in particular that both low-column and high-column photoionized phases can be either faint or luminous. This is especially clear in the cores, where components with $\log (N_{\rm H}/ \rm cm^{-2})\sim 19-20$ and components with $\log (N_{\rm H}/ \rm cm^{-2})\gtrsim 22$ both span a wide luminosity range. The same qualitative behavior is present in the cones and cross-cones, although with a lower upper envelope in luminosity. Likewise, the $\log L$-$\log U$ panels do not reveal a strong monotonic trend: more highly ionized components can be luminous, but comparably luminous components are also present at moderate or even low $\log U$. The figure therefore argues against a picture in which the non-nuclear emitted X-ray luminosity is determined primarily by ionization state alone.
	
	Instead, the large scatter strongly suggests an important role for geometric and structural effects. Variations in covering factor, filling factor, cloud number, and the three-dimensional distribution of emitting material can all change the emergent luminosity without requiring correspondingly large changes in $\log U$ or $\log N_{\rm H}$. In this sense, the spectral parameters trace the physical state of the gas, whereas the luminosity also encodes how much gas is present and how effectively it intercepts and reprocesses the AGN radiation field.
	
	The thermal components show a similarly broad behavior. Figure~\ref{fig:lumvsparam} shows no strong global correlation between thermal luminosity and $kT$ in any spatial class. Most \texttt{APEC} components cluster at sub-keV to $\sim$1~keV temperatures, yet at a given $kT$ the luminosities can vary by more than an order of magnitude. Conversely, relatively luminous and relatively faint components are found across much of the observed temperature range. Although a few high-$kT$ outliers are present, especially in some cone and cross-cone regions, they do not define a clear luminosity sequence. The thermal luminosity therefore appears to depend less on temperature itself than on the amount of emitting gas and the local efficiency of energy deposition into the ISM.
	
	A noteworthy aspect of Figure~\ref{fig:lumvsparam} is that cross-cone regions are not systematically segregated from cones in any of the three parameter-luminosity planes. Cross-cone components are often somewhat less luminous on average ($\sim$\,1 dex)), but they overlap substantially with cone components in both luminosity and parameter space. This is consistent with the picture emerging from Figures~\ref{fig:isto} and \ref{fig:lognhlogu}: the extended emission outside the nominal ionization cones is not drawn from a fundamentally different class of gas, but rather from material experiencing a different combination of illumination geometry, obscuration, and local environmental conditions.
	
	Overall, the ensemble behavior supports a structured, multiphase medium in which luminosity is modulated by geometry as much as by intrinsic gas properties. Photoionization dominates much of the line-rich soft emission, but its observed luminosity depends on how much gas subtends the ionizing source and how that gas is distributed spatially. Thermal plasma provides an additional, often substantial, contribution whose luminosity is likewise governed by local density, filling factor, and mechanical energy input. The broad distributions in Figure~\ref{fig:lumvsparam} therefore argue that the diversity of soft X-ray emission across cores, cones, and cross-cones is produced not by a single controlling parameter, but by the combined effects of anisotropic illumination, clumpy gas structure, and region-dependent interaction physics.
	
	\section{Conclusions}
	\label{sec:conclusions}
	
	We have presented a uniform, spatially resolved X-ray spectral analysis of nearby heavily obscured/weak AGN observed with deep {\it Chandra}/ACIS data. By separating nuclear and extended emission and modeling their spectra with photoionized and thermal components, we obtain the following main results:
	
	\begin{enumerate}
		
		
		\item \textbf{Nuclear regions are spectrally complex and often soft-dominated.}  
		Despite the presence of heavily obscured AGN continua in all nuclei, the observed 0.3-7~keV emission is frequently dominated by circumnuclear gas outside the obscuring torus. Photoionized components are required in all nuclei, and thermal plasma is present in the majority, indicating that even sub-kiloparsec regions host a mixture of gas phases.
		
		\item \textbf{Different spatial emission regions have different X-ray spectra}. The spatially resolved analyses presented support sheer complexity of the AGN interaction with the complex galaxy ISM.
		
		\item \textbf{Extended emission shows broad and overlapping parameter distributions.}  
		Cone and cross-cone regions span similarly wide ranges in photoionized column density ($\log (N_{\rm H}/ \rm cm^{-2})\sim19$ - $>23$), ionization parameter ($\log U\sim-2$ to $\sim3$), and temperature ($kT\sim0.3$ - few keV). No clear monotonic radial trend is observed, and the substantial overlap between regions indicates that the extended medium is structured and multiphase rather than stratified.
		
		\item \textbf{Photoionized gas is present outside the nominal ionization cones.}  
		Cross-cone regions frequently require photoionized components spanning a wide range of ionization states, often comparable to those in the cones. This suggests that ionizing radiation reaches a larger solid angle than predicted by simple axisymmetric obscuration, likely due to a clumpy or porous circumnuclear medium and scattering effects.
		
		\item  \textbf{Thermal emission is widespread across all spatial regions, with temperatures consistent with shock-heated plasma.} This supports a scenario in which mechanical processes, such as AGN-driven outflows and jet-ISM interactions, contribute alongside radiative photoionization in shaping the circumnuclear environment.
		
		\item \textbf{Extended hard X-ray emission is common.}  
		Hard X-ray emission in excess of nuclear PSF spillover is detected in 73 extended regions across 19/20 sources. Neutral Fe~K$\alpha$ emission at 6.4~keV is detected outside the nucleus in 12/20 systems. These results indicate that reflection and/or scattering from dense material occurs on circumnuclear and kiloparsec scales. We here note that non detections are possibly due to low S/N in the hard band.
		
		\item \textbf{Extended X-ray luminosity is not set by a single physical parameter.}  
		The luminosities of both photoionized and thermal components show large scatter at fixed photoionized $\log N_{\rm H}$, $\log U$, and $kT$. This indicates that the observed emission depends not only on the physical state of the gas but also on geometric factors such as covering fraction, filling factor, and spatial distribution.
		
	\end{enumerate}
	
	Taken together, these results support a picture in which the extended X-ray emission in heavily obscured/weal AGN arises from a complex, multiphase medium shaped by anisotropic illumination, clumpy gas structure, and interactions between AGN-driven outflows and the host galaxy interstellar medium. The lack of a simple separation between cone and cross-cone regions suggests that the classical picture of sharply defined ionization cones must be supplemented by a more realistic description that includes leakage, scattering, and environmental effects.
	
	In conclusion, this work establishes a uniform baseline for investigating the spectral properties of the spatially resolved X-ray emission of nearby AGN, where the nuclear source does not totally dominate the X-ray emission in Chandra observations. Our sample of 20 AGN was selected because of the previous detection of extended X-ray emission attributable to the effect of the AGN on the host ISM. All AGN, except NGC\,1167 and NGC\,5252, are heavily obscured. In NGC\,1167, the nuclear emission is instead intrinsically weak \citep{fabbiano_jet-ism_2022}, allowing the extended emission to be studied. Note that, because of the light-travel time, the extended emission in this case traces a previous high-luminosity state of the AGN. NGC\,5252, while only moderately obscured, provides a valuable case for investigating the properties and origin of the extended emission. In forthcoming papers of this series, we will present a comprehensive analysis of the ensemble properties of the sample, examining the spatial distribution of the X-ray emission across different energy bands together with complementary multi-wavelength observations. We will also investigate the scaling relations between the X-ray and host-galaxy properties, with the ultimate goal of placing quantitative constraints on the role of AGN feedback in galaxy evolution.
	\\
	
	{\it \textbf{Acknowledgments:}}This work was made possible by support from the Smithsonian Institution Combined Call for Research Award "Fingerprints of Black hole Feedback: X-ray Answers to a Cosmic Mistery' (PI: G. Fabbiano; co-PI Elvis, M.). Fabbiano acknowledges partial support from the Chandra X-ray Center (NASA contract NAS8-03060). Fabbiano and Elvis have benefited from discussions at the Aspen Center for Physics, which is supported by National Science Foundation grant PHY-2210452. We thank Guido Risaliti for useful discussion on this project and Peixin Zhu for comments on the manuscript. RM acknowledges financial support from the INAF Scientific Directorate. ATF was supported by an appointment to the NASA Postdoctoral Program at the NASA Goddard Space Flight Center, administered by Oak Ridge Associated Universities under contract with NASA.
	
	\facilities{Chandra}
	"This paper employs a list of Chandra datasets, obtained by the Chandra X-ray Observatory, contained in the Chandra Data Collection \url{[DOI:https://doi.org/10.25574/cdc.577]}." 
	
	\newpage
	\appendix
	\section{Imaging Parameters}
	
	\begin{table}[h]
\centering
\caption{Parameters used to generate the adaptively smoothed images shown in Figure~\ref{fig:imagescool}, obtained with the \texttt{dmadapt} routine.}
\begin{tabular}{l c c l}
\hline
\textbf{Galaxy} & \textbf{Binning} & \textbf{Counts/kernel} & \textbf{Contours (counts/pixel)} \\
\hline
Mrk\,573 & 1/8 & 9   & 0.015, 0.037, 0.102, 0.243, 0.545, 1.196, 2.599, 5.621, 12.132, 26.159 \\

NGC\,835 & 1/8 & 9   & 0.020, 0.039, 0.081, 0.172, 0.366, 0.786, 1.689, 3.635, 7.828, 16.861 \\

NGC\,1068 & 1/2 & 9   & 0.040, 0.328, 0.950, 5.000, 50.000, 150.000 \\

NGC\,1068-center & 1/8 & 9   & 0.437, 0.972, 2.125, 4.608, 9.959, 21.487, 46.322 \\

NGC\,1167 & 1/16 & 5   & 0.002, 0.003, 0.005, 0.010, 0.021, 0.043, 0.091, 0.195, 0.419, 0.902 \\

NGC\,1365 & 1/8 & 5   & 0.006, 0.013, 0.028, 0.061, 0.132, 0.285, 0.615, 1.326, 2.856, 6.153 \\

NGC\,1386 & 1/8 & 9   & 0.016, 0.038, 0.085, 0.186, 0.405, 0.876, 1.891, 4.077, 8.786, 18.933 \\

Mrk\,3 & 1/8 & 9 & 0.039, 0.122, 0.303, 0.691, 1.527, 3.330, 7.212, 15.577, 33.599\\

ESO\,428-G014 & 1/8 & 9   & 0.007, 0.0181, 0.0419, 0.093, 0.204, 0.443, 0.957, 2.065, 4.453, 9.597 \\

Mrk\,78 & 1/8 & 9 & 0.025, 0.039, 0.122, 0.303, 0.691, 1.527, 3.330, 7.212, 15.579, 33.599\\

Mrk\,78-center & 1/8 & 9 & 0.080, 0.103, 0.134, 0.161, 0.255, 0.335, 0.451, 0.616, 0.854, 1.197, 1.689, 2.398, 3.417, 4.883, 6.993\\

NGC\,3081 & 1/16 & 9   & 0.005, 0.030, 0.084, 0.201, 0.453, 0.996, 2.165, 4.683, 10.109, 21.799 \\

NGC\,3079 & 1/4 & 9   & 0.020, 0.030, 0.06, 0.084, 0.201, 0.453, 0.996, 3.500, 10.109 \\

NGC\,3079-center & 1/8 & 9   & 0.005, 0.008, 0.013, 0.025, 0.050, 0.104, 0.220, 0.472, 1.013, 2.178 \\

Mrk\,34 & 1/8 & 9 & 0.0303, 0.039, 0.122, 0.303, 0.691, 1.527, 3.330, 7.212, 15.578, 33.599\\

NGC\,3393 & 1/8 & 9   & 0.018, 0.051, 0.122, 0.276, 0.607, 1.321, 2.859, 6.172, 13.310, 28.689 \\

NGC\,4945 & 1/4 & 9   & 0.100, 0.154, 0.294, 0.593, 1.239, 2.629, 5.625, 17.000, 55.944 \\

NGC\,4945-center & 1/8 & 9   & 0.025, 0.050, 0.100, 0.350, 0.593, 1.239, 2.629, 5.625, 12.079, 25.985, 55.944 \\

NGC\,5252 & 1/16 & 9   & 0.006, 0.011, 0.021, 0.042, 0.088, 0.187, 0.501, 1.556, 3.996, 7.000, 15.000 \\

NGC\,5643 & 1/8 & 9   & 0.006, 0.017, 0.042, 0.095, 0.210, 0.457, 0.989, 2.136, 4.606, 9.927 \\

NGC\,5728 & 1/16 & 9   & 0.006, 0.011, 0.022, 0.046, 0.097, 0.208, 0.447, 0.962, 2.071, 4.459 \\

ESO 137-G034 & 1/8 & 9   & 0.007, 0.019, 0.043, 0.097, 0.213, 0.461, 0.998, 2.152, 4.641, 10.002 \\

IC\,5063 & 1/8 & 9   & 0.040, 0.102, 0.236, 0.526, 1.148, 2.490, 5.382, 11.611, 25.032, 53.946 \\

NGC\,7212 & 1/16 & 9   & 0.003, 0.011, 0.027, 0.063, 0.139, 0.304, 0.659, 1.425, 3.074, 6.628 \\
\hline
\end{tabular}
\label{tab:dmadapt_params}
\end{table}

	\section{Evaluation of Multi-Component Fits and Overfitting}
	
\begin{figure*}
		\centering
		\includegraphics[width=0.85\textwidth]{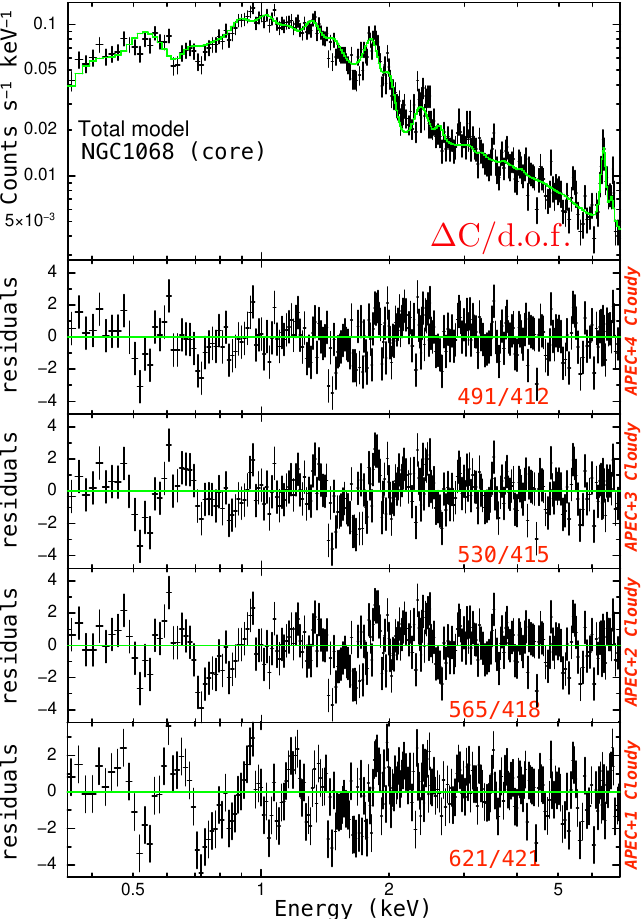}
		\caption{Stepwise assessment of the spectral complexity required to model the nuclear spectrum of NGC\,1068. Starting from a model composed of one \texttt{APEC} and one \texttt{CLOUDY} component, additional photoionized components are introduced sequentially. The residuals show that the correlated structures in the soft X-ray band are progressively removed, while the corresponding $C$-\text{stat}/d.o.f. values improve from 621/421 to 491/412.}
		\label{goodfit}
	\end{figure*}

	To demonstrate the need for multiple photoionized components and to show that the adopted procedure does not introduce unnecessary model complexity, we performed a step-by-step fitting test on the nuclear spectrum of NGC\,1068. Starting from a baseline model consisting of one thermal (\texttt{APEC}) and one photoionized (\texttt{CLOUDY}) component, we progressively added one additional \texttt{CLOUDY} component at a time, refitting the spectrum after each step and inspecting both the residuals and the fit statistic. As shown in Figure~\ref{goodfit}, the simplest configurations leave significant, correlated residuals in the soft X-ray band, indicating that the spectral complexity is not adequately described. The sequential addition of photoionized components progressively removes these systematic residuals while producing substantial improvements in the fit statistic (C-stat/d.o.f. = 621/421, 565/418, 530/415, and 491/412 for one to four \texttt{CLOUDY} components, respectively). Beyond the fourth photoionized component, an additional photoionized component results only in marginal improvements in C-stat and did not remove any remaining systematic residuals; therefore, more complex models were not retained. We adopted the same empirical criterion throughout the analysis, selecting the simplest model that adequately reproduces the observed spectra.
	
	\bibliography{references1}
	\bibliographystyle{aasjournalv7}

\end{document}